# Inner and Outer Bounds for the Gaussian Cognitive Interference Channel and New Capacity Results

Stefano Rini, Daniela Tuninetti, and Natasha Devroye
Department of Electrical and Computer Engineering
University of Illinois at Chicago, IL USA,
Email: {srini2, danielat, devroye}@uic.edu

**Abstract**

The capacity of the Gaussian cognitive interference channel, a variation of the classical two-user interference channel where one of the transmitters (referred to as *cognitive*) has knowledge of both messages, is known in several parameter regimes but remains unknown in general. In this paper we provide a comparative overview of this channel model as we proceed through our contributions: we present a new outer bound based on the idea of a broadcast channel with degraded message sets, and another series of outer bounds obtained by transforming the cognitive channel into channels with known capacity. We specialize the largest known inner bound derived for the discrete memoryless channel to the Gaussian noise channel and present several simplified schemes evaluated for Gaussian inputs in closed form which we use to prove a number of results. These include a new set of capacity results for the a) "primary decodes cognitive" regime, a subset of the "strong interference" regime that is not included in the "very strong interference" regime for which capacity was known, and for the b) "S-channel" in which the primary transmitter does not interfere with the cognitive receiver. Next, for a general Gaussian cognitive interference channel, we determine the capacity to within one bit/s/Hz and to within a factor two regardless of channel parameters, thus establishing rate performance guarantees at high and low SNR, respectively. We also show how different simplified transmission schemes achieve a constant gap between inner and outer bound for specific channels. Finally, we numerically evaluate and compare the various simplified achievable rate regions and outer bounds in parameter regimes where capacity is unknown, leading to further insight on the capacity region of the Gaussian cognitive interference channel.

## I. Introduction

A well studied channel model inspired by the newfound abilities of cognitive radio technology and its potential impact on spectral efficiency in wireless networks is the *cognitive radio channel* [4]. The cognitive radio channel is also referred to as the *interference channel with unidirectional cooperation* [5], the *interference channel with*

Parts of this work were presented at [1], [2], [3]. The work of S. Rini and D. Tuninetti was partially funded by NSF under award 0643954. The contents of this article are solely the responsibility of the authors and do not necessarily represent the official views of the NSF.

*degraded message sets* [6], or the *cognitive interference channel* [7]. This channel consists of a two-user interference channel, where one transmitter-receiver pair is referred to as the *primary* user and the other as the *cognitive*. The primary transmitter has knowledge of one of the two independent messages to be sent, while the cognitive transmitter has full, non-causal knowledge of both messages, thus idealizing the cognitive user's ability to detect transmissions taking place in the network. Since the cognitive transmitter can "broadcast" information to both receivers, the capacity of the cognitive interference channel contains features of both the interference and the broadcast channel.

Although the assumption of full, non-causal knowledge of the primary user's message at the cognitive transmitter might not be practical, the simplicity of the resulting model leads to closed form results and provides powerful insights on the role of unilateral cooperation among the users. The more practical scenario of causal unilateral cooperation may be studied in the framework of *interference channels with generalized feedback* of [8] (and references therein), but is outside the scope of this work.

### A. Past work

**Capacity results.** The cognitive interference channel was first posed in an information theoretic framework in [4], where an achievable rate region (for general discrete memoryless channels) and broadcast-channel-based outer bound (in Gaussian noise only) were proposed. The first capacity results were determined in [9], [6] for a class of channels with "weak interference" at the primary receiver. In this regime, capacity is achieved by having the cognitive transmitter pre-code against the interference created at its receiver, while the primary receiver treats the interference from the cognitive transmitter as noise. Capacity is also known in the "very strong interference" regime [10]. In this regime, capacity is achieved by having both receivers decode both messages as in a compound multiple access channel. In [11, Th. 7.1], we show that the outer bound of [6] is achievable in the "better cognitive decoding" regime, which includes both the "very weak interference" and the "very strong interference" regimes.

**Outer bounds.** A general outer bound was derived in [12, Th. 4] using a technique developed for broadcast channels in [13]. Both the "weak interference" outer bound of [6] and the "strong interference" outer bound of [10] may be derived by loosening [12, Th. 4]. Although the outer bound in [12, Th. 4] is the tightest known, it is difficult to evaluate because it contains three auxiliary random variables for which no cardinality bounds are given on the corresponding alphabets. Moreover, for the Gaussian channel, the "Gaussian maximizes entropy" property alone does not suffice to show that Gaussian inputs exhaust the outer bound. For these reasons, in [11, Th. 4.1] we proposed an outer bound that exploits the fact that the capacity region only depends on the conditional marginal distributions of the outputs given the inputs (as for broadcast channels [14], since the receivers do not cooperate). The resulting outer bound does not include any auxiliary random variable and every mutual information term involves all the inputs (like in the cut-set bound) and thus may be evaluated for both discrete memoryless and Gaussian channels. The bound in [11, Th. 4.1] was shown to be tight for a class of semi-deterministic cognitive interference channels with a noiseless output at the primary receiver.

**Achievable rate regions.** Different achievable schemes have been proposed for the cognitive interference channel which include features originally devised for the interference channel and for the broadcast channel, such as rate splitting, superposition coding, binning and simultaneous decoding. The scheme of [15] generalizes the "weak interference" capacity achieving scheme of [6] by making part of the cognitive message common. The same rate splitting idea is used in [12], [16] along with a more elaborated binning operation. The region in [17] introduces a binning scheme inspired by Marton's achievable rate region for a general broadcast channel [18]. This feature is further generalized in [19] and in [1] where more refined binning and superposition steps are added in the cognitive encoding process. Given the different encoding choices, a comparison of the different achievable schemes is often not straightforward. In particular, despite possible simplifications of the original scheme in [4] as described in [20], no region was shown to conclusively encompass [4], or the larger region of [21], until recently. A comparison of all the transmission schemes proposed in the literature was presented in [11], in which we show that our region in [11, Th. 5.1] is provably the largest known achievable rate region to date.

**Constant gap results.** While the capacity region remains unknown for a general channel, in [7] we demonstrated achievable rate regions which lie within 1.87 bits/s/Hz for any real-valued Gaussian cognitive interference channel. We derived this constant gap result by using insights from the high SNR deterministic approximation of the cognitive interference channel [22], a deterministic model that captures the behavior of a Gaussian network for large transmit powers [23].

**Z channel.** The special case where the cognitive transmitter does not create interference to the primary receiver is called the *Z cognitive interference channel*; inner and outer bounds when the cognitive-primary link is noiseless are obtained in [24], [25]. The Gaussian causal case is considered in [25], and is related to the general causal cognitive channel in [26].

### B. Contributions

In this work we focus on the *Gaussian* cognitive interference channel in a comprehensive and comparative manner. In particular, our main contributions are:

A) **We evaluate the outer bound of [11, Th. 4.1] for the Gaussian channel.** We show that it unifies the previously proposed outer bounds for the "weak interference" and the "strong interference" regimes of [6] and [12], respectively.

B) **We derive a new outer bound based on the broadcast channel and inspired by [4]**. The capacity region of the Gaussian MIMO (multi input multi output antenna) broadcast channel with degraded message sets is an outer bound for a channel in "strong interference". Interestingly, we show that the new bound may be strictly tighter than the "strong interference" outer bound of [12].

C) **Derive new outer bounds by transformation / inclusion into channels with known capacity**. We determine the conditions under which the capacity region of a Gaussian channel is contained in that of a channel with known capacity. The capacity of the latter channel thus provides an outer bound for the former.

D) **We specialize the largest known inner bound of [11, Th. 5.1] to the Gaussian channel.** We utilize it as a unified framework to derive and compare various achievable schemes in this and prior work.

E) **We prove a new capacity result for the "primary decodes cognitive" regime.** This regime is a subset of the "strong interference" regime that is not included in the "very strong interference" regime for which capacity was known [10]. In this regime capacity is achieved by having the primary receiver decode the message of the cognitive user in addition to its own message.

F) **We prove a new capacity result for the S-channel, a channel in which the primary transmission does not interfere with the cognitive receiver.** For this channel we show the achievability of the outer bound based on the capacity of the broadcast channel with degraded message sets.

G) **We show capacity to within half a bit/s/Hz per real dimension and to within a factor two regardless of channel parameters.** These two results characterize the capacity region of the Gaussian channel both at high and low SNR, respectively. To this end, we use a transmission scheme inspired by the capacity achieving scheme for the semi-deterministic cognitive interference channel of [11, Th. 8.1], where capacity is achieved by having the cognitive transmitter perform partial interference pre-coding for both decoders. The multiplicative gap is shown by using a simple time sharing argument between achievable points.

H) **We provide insights on the capacity region of the Gaussian channel for the regimes in which capacity is still unknown.** We do so by showing that very simple transmission strategies can achieve capacity to within a constant gap for large sets of parameters. We conclude by showing that a constant gap result may alternatively be proved by trading off interference pre-coding at the cognitive encoder and interference decoding at the primary receiver.

### *C. Organization*

The rest of the paper is organized as follows. Section II formally defines the cognitive interference channel model and summarizes known results for the Gaussian channel. Section III presents new outer bounds for the Gaussian channel. Section IV lists the achievable schemes used in the rest of the paper and shows how they may be obtained from the largest known inner bound of [11]. Section V proves the two new capacity results. Section VI characterizes the capacity of the Gaussian channel to within half a bit/s/Hz per real dimension and to within a factor two. Section VII shows some relevant numerical results. Section VIII concludes the paper. Most proofs may be found in the Appendices.

## II. Gaussian channel model and known results

### *A. Notation*

We use the following convention:

- The symbol $X \sim \mathcal{N}_{\mathbb{C}}(\mu, \sigma^2)$ indicates that the random variable (RV) $X$ is a complex-valued proper Gaussian RV with mean $\mu$ and covariance $\sigma^2$.
- We define $\mathcal{C}(x) := \log(1+x)$ for $x \in \mathbb{R}^+$.
- We define $\overline{x} := 1-x$ for $x \in [0,1]$.
- For any two RVs $X$ and $Y$, the symbol $X|_Y$ denotes the conditional distribution of $X$ given $Y$.

- We use $[1:n]$ to denote the set of natural numbers from 1 to $n$.
- The notation $A \stackrel{(n)}{=} B$ to indicate that the expression $B$ is obtained from $A$ with the assignment given in equation number $n$.
- For an integer $N$, the symbol $X^N$ indicates the length-$N$ vector $(X_1, ..., X_N)$.
- For the plots, the logarithms are in base 2, i.e., rates are expressed in bits/s/Hz.
- $\boldsymbol{C}(a, |b|, P_1, P_2)$ indicates the capacity of a Gaussian cognitive interference channel with channel parameters $a$ and $|b|$ and powers $P_1$ and $P_2$.
- $X^*$ denotes the complex conjugate of the complex number (or vector) $X$.

### B. General memoryless cognitive interference channel

A two-user InterFerence Channel (IFC) is a multi-terminal network with two input alphabets $\mathcal{X}_1$ and $\mathcal{X}_2$, two output alphabets $\mathcal{Y}_1$ and $\mathcal{Y}_2$, and a channel transition probability $P_{Y_1 Y_2|X_1 X_2}(y_1, y_2|x_1, x_2) : \mathcal{Y}_1 \times \mathcal{Y}_2 \to [0, 1]$ for all $(x_1, x_2) \in \mathcal{X}_1 \times \mathcal{X}_2$. Each transmitter $i$, $i \in \{1, 2\}$, wishes to communicate a message $W_i$, uniformly distributed on $[1 : 2^{NR_i}]$, to receiver $i$ in $N$ channel uses at rate $R_i$. The two messages are independent. In the classical IFC, the two transmitters operate independently having no knowledge of each others' messages. Here we consider a variation of this setup assuming that transmitter 1, in addition to its own message, also knows the message of transmitter 2 prior to transmission. We refer to transmitter/receiver 1 as the *cognitive* pair and to transmitter/receiver 2 as the *primary* pair. This model is commonly known as the Cognitive InterFerence Channel (CIFC).

The CIFC is an idealized model for the *unilateral source cooperation* of transmitter 1 with transmitter 2. The receivers however do not cooperate. This implies that the capacity region of the CIFC, similar to the broadcast channel (BC) [14], only depends on the output conditional marginals $P_{Y_1|X_1 X_2}$ and $P_{Y_2|X_1 X_2}$, and not on the output joint marginal $P_{Y_1 Y_2|X_1 X_2}$.

A rate pair $(R_1, R_2)$ is achievable if there exists a sequence of encoding functions

$$X_1^N = X_1^N(W_1, W_2),$$
$$X_2^N = X_2^N(W_2),$$

and a sequence of decoding functions

$$\widehat{W}_i = \widehat{W}_i(Y_i^N), \quad i \in \{1, 2\},$$

such that

$$\max_{i \in \{1,2\}} \mathbb{P}\left[\widehat{W}_i \neq W_i\right] \to 0, \quad N \to \infty.$$

The capacity region is defined as the convex closure of the region of achievable $(R_1, R_2)$-pairs [27].

### C. Gaussian CIFC

A Gaussian CIFC (G-CIFC) in *standard form* (see Appendix A) is described by the input/output relationship

$$Y_1 = X_1 + aX_2 + Z_1,$$
$$Y_2 = |b|X_1 + X_2 + Z_2,$$

where the channel gains $a$ and $b$ are complex-valued, constant, and known to all terminals, the channel inputs are subject to the power constraint

$$\mathbb{E}[|X_i|^2] \le P_i, \qquad P_i \in \mathbb{R}^+, \qquad i \in \{1,2\},$$

and the channel noise $Z_i \sim \mathcal{N}_{\mathbb{C}}(0,1)$, $i \in \{1,2\}$. Since the capacity only depends on the output conditional marginals, the correlation coefficient of $Z_1$ and $Z_2$ is irrelevant. A graphical representation of a G-CIFC is found in Fig. 1.

A G-CIFC is said to be a:

- **Z-channel** if $|b| = 0$; we refer to it as a Z-G-CIFC. In this case the primary decoder does not experience interference from the cognitive transmitter. Capacity is trivially given by

$$R_1 \le \mathcal{C}(P_1), \quad R_2 \le \mathcal{C}(P_2).$$

- **S-channel** if $a = 0$; we refer to it as a S-G-CIFC. In this channel the cognitive decoder does not experience interference from the primary transmitter. For this channel capacity is only known for $|b| \le 1$ [6].
- **Degraded channel** if $a|b| = 1$. In this case one channel output is a degraded version of the other. In particular, for $|b| > 1$, $Y_1$ is a degraded version of $Y_2$ since

$$Y_1 = X_1 + \frac{1}{|b|}X_2 + Z_1 \sim \frac{1}{|b|}Y_2 + Z_0,$$

  for $Z_0 \sim \mathcal{N}_{\mathbb{C}}(0, |b|^2 - 1)$ independent of everything else. Similarly, when $|b| \le 1$, $Y_2$ is a degraded version of $Y_1$. Capacity is known in the case $|b| \le 1$ [6].

### D. Known results for the G-CIFC

The capacity of the G-IFC is not known in general. However several capacity results exist, as summarized next.

**Theorem II.1.** ***"Weak interference" capacity of [6, Lemma 3.6] and [9, Th. 4.1].*** *If*

$$|b| \le 1, \quad \textit{(the "weak interference" regime/condition)} \tag{1}$$

*the capacity of the G-CIFC is:*

$$R_1 \le \mathcal{C}(\alpha P_1), \tag{2a}$$
$$R_2 \le \mathcal{C}\left(|b|^2 P_1 + P_2 + 2\sqrt{\bar{\alpha}|b|^2 P_1 P_2}\right) - \mathcal{C}(|b|^2 \alpha P_1), \tag{2b}$$

*taken over the union of all* $\alpha \in [0,1]$.

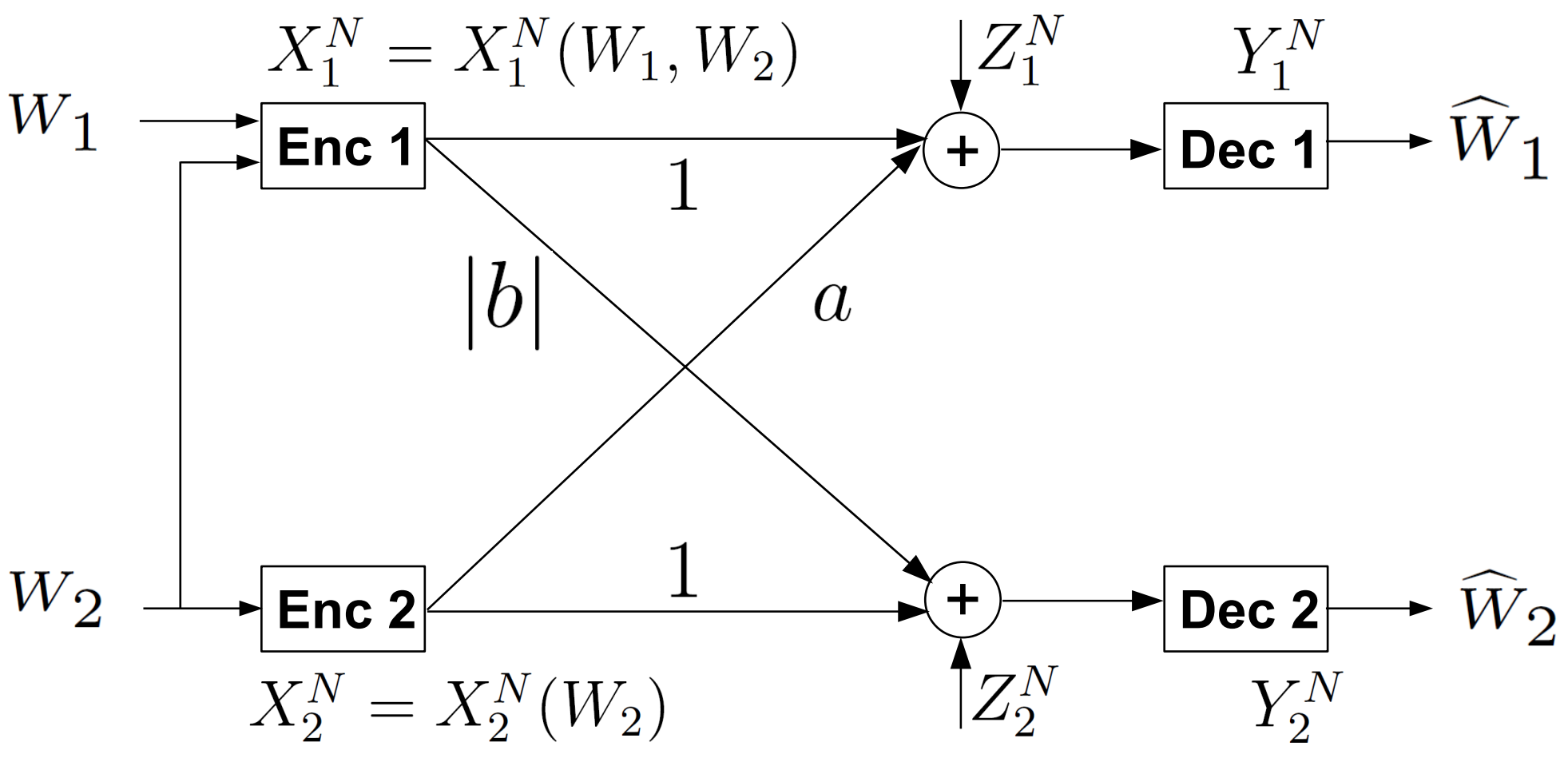

Fig. 1. The Gaussian cognitive interference channel (G-CIFC).

**Theorem II.2.** ***"Strong interference" outer bound of [12, Th. 4].*** *When*

$$|b| > 1, \quad \textit{(the "strong interference" regime/condition)} \tag{3}$$

*the capacity region of the G-CIFC is included in the region* $\mathcal{R}^{(\mathrm{SI})}$ *defined as:*

$$R_1 \leq \mathcal{C}(\alpha P_1), \tag{4a}$$

$$R_1 + R_2 \leq \mathcal{C}\left(|b|^2 P_1 + P_2 + 2\sqrt{\bar{\alpha}|b|^2 P_1 P_2}\right), \tag{4b}$$

*taken over the union of all* $\alpha \in [0, 1]$.

**Theorem II.3.** ***"Very strong interference" capacity of [10, Thm. 6]*** *extended to complex-valued channels (see Appendix B). When*

$$(|a|^2 - 1)P_2 - (|b|^2 - 1)P_1 - 2\big|a - |b|\big|\sqrt{P_1 P_2} \geq 0,$$

$$\textit{and } |b| > 1 \textit{ ("very strong interference" regime/condition)} \tag{5}$$

*the outer bound* $\mathcal{R}^{(\mathrm{SI})}$ *of Th.II.2 is tight.*

A plot of the capacity results of Th.II.1 and Th.II.3 for $a \in \mathbb{R}$ and $P_1 = P_2$ is depicted in Fig. 2. The channel gains $a$ and $|b|$ for which capacity is known are shaded, while those for which capacity is unknown are white.

## III. Outer bounds

In this section we prove several outer bounds:

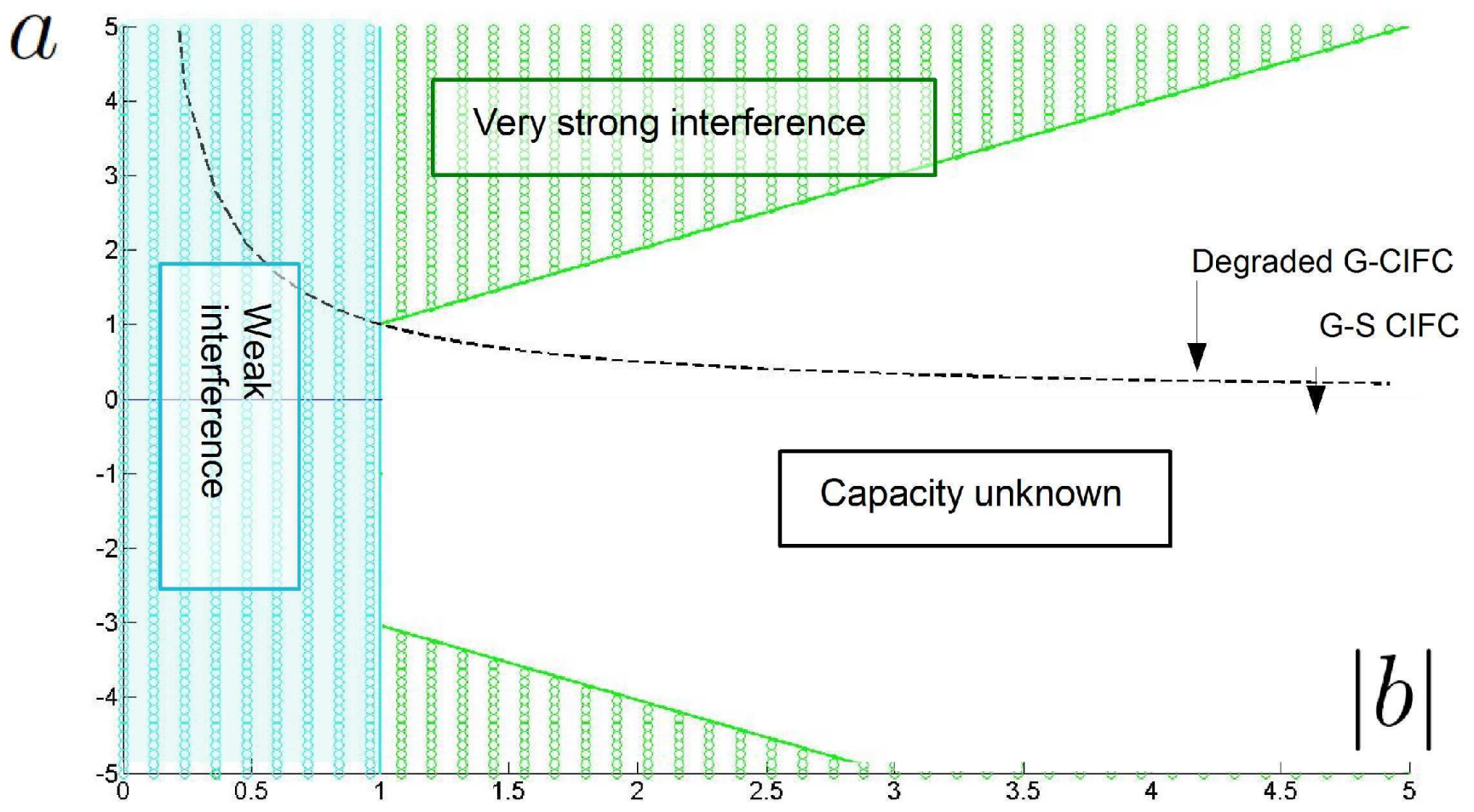

Fig. 2. A representation of the capacity results Th.II.1 and Th.II.3 for $P_1 = P_2$ and $(a, |b|) \in [-5, 5] \times [0, 5]$. The regions for which capacity is known are shaded, while those for which capacity is unknown are white.

A) First we evaluate the outer bound of [11, Th. 4.1] for the Gaussian channel and show that it coincides with the outer bounds of Th. II.1 and Th. II.2 in "weak" and "strong interference" respectively.

B) Then we tighten it by using the observation of [4] that the capacity region of a G-CIFC is included into the capacity region of the Gaussian MIMO BC obtained by allowing full cooperation among the transmitters. We further tighten the outer bound in "strong interference", where we show that the capacity region of a Gaussian broadcast channel with degraded message sets forms an outer bound to the capacity of the G-CIFC.

C) Finally, we propose outer bounds based on enhancing the original channel so as to transform it into a channel for which capacity is known.

### *A. A unifying framework for Th.II.1 and Th.II.2*

Our objective is to obtain an outer bound for the G-CIFC with $|b| > 1$ that improves on the "strong interference" outer bound of Th.II.2. Although the following theorem does not result in such a bound, it is of interest because it provides a simple unifying framework for Th.II.1 and Th.II.2. The proof of Th.II.1 and the proof of Th.II.2 use very different techniques. On the one hand, the bound in Th.II.1 is valid for a general channel under the "weak interference" condition in [6, Thm. 3.7] and is inspired by the converse for "less noisy BC". On the other hand, the bound in Th.II.2 is valid for Gaussian channels with "strong interference" only and is inspired by the converse of "strong interference IFC". We will show next that both results may be derived within the framework proposed in [11]. The proof of [11, Thm. 4.1] uses the argument originally devised by Sato for the BC [14] that, for channels without receiver cooperation, the capacity only depends on the output conditional marginals. The bound in [11, Thm. 4.1] is valid for *a general CIFC.*

**Theorem III.1. Unifying outer bound.** *The capacity region of the G-CIFC is contained in the region*

$$R_1 \le \mathcal{C}\left(\alpha P_1\right), \tag{6a}$$

$$R_2 \le \mathcal{C}\left(|b|^2 P_1 + P_2 + 2\sqrt{\bar{\alpha}|b|^2 P_1 P_2}\right), \tag{6b}$$

$$\begin{aligned} R_1 + R_2 \le\ & \mathcal{C}\left(|b|^2 \bar{\alpha} P_1 + P_2 + 2\sqrt{\bar{\alpha}|b|^2 P_1 P_2}\right) \\ & + \left[\mathcal{C}\left(\alpha P_1\right) - \mathcal{C}\left(|b|^2 \alpha P_1\right)\right]^+ \end{aligned} \tag{6c}$$

*taken over the union of all $\alpha \in [0,1]$. In "strong interference" ($|b| > 1$) the region in (6) reduces to Th.II.2, and in "weak interference" ($|b| \le 1$) to Th.II.1.*

*Proof:* In [11, Thm. 4.1], we showed that the capacity of a general CIFC is contained in the region

$$R_1 \le I(Y_1; X_1|X_2), \tag{7a}$$

$$R_2 \le I(X_1, X_2; Y_2), \tag{7b}$$

$$R_1 + R_2 \le I(X_1, X_2; Y_2) + I(Y_1; X_1|Y'_2, X_2), \tag{7c}$$

taken over the union of all joint distributions $P_{X_1,X_2}$ and where $Y'_2$ has the same conditional marginal distribution as $Y_2$, i.e., $P_{Y'_2|X_1,X_2} = P_{Y_2|X_1,X_2}$. The result in (7) specialized to the G-CIFC amounts to optimizing the correlation coefficient over the Gaussian additive noises, that is, optimizing with respect to $\gamma :\ |\gamma| \le 1$ in

$$\begin{bmatrix} Z_1 \\ Z_2 \end{bmatrix} \sim \mathcal{N}_{\mathbb{C}}\left(\mathbf{0}, \begin{bmatrix} 1 & \gamma \\ \gamma^* & 1 \end{bmatrix}\right).$$

First we show that a proper-complex Gaussian input exhausts the region in (7). For any $\alpha \in [0,1]$, let $\mathbf{S}$ be a covariance matrix defined as

$$\mathbf{S} \triangleq \begin{bmatrix} P_1 & \rho\sqrt{P_1 P_2} \\ \rho^*\sqrt{P_1 P_2} & P_2 \end{bmatrix} : \quad \rho = \sqrt{1-\alpha}\, \mathrm{e}^{\mathrm{j}\theta},\ \theta \in \mathbb{R}, \tag{8}$$

and let $(X_{1G}, X_{2G}) \sim \mathcal{N}_{\mathbb{C}}\left(\mathbf{0}, \mathbf{S}\right)$. By using the "Gaussian maximizes entropy" principle (see also [28, Eq.(3.29)]), we conclude that for a given input covariance constraint $\mathbf{S}$ in (8) for $P_{X_1,X_2}$, the regime in (7c) is upper bounded

by

$$\text{(7a)} \leq I(Y_1; X_{1G}|X_{2G}) = \text{(6a)}, \tag{9}$$

$$\begin{aligned}\text{(7b)} &\leq I(Y_2; X_{1G}, X_{2G}) \\ &= \log(1 + P_2 + |b|^2 P_1 + 2|b|\text{Re}\{\rho\}\sqrt{P_1\, P_2}) \leq \text{(6b)},\end{aligned} \tag{10}$$

$$\begin{aligned}\text{(7c)} &\leq I(Y_2; X_{1G}, X_{2G}) + I(Y_1; X_{1G}|Y_2, X_{2G}) \\ &\leq \text{(6b)} + \log\left(\frac{1 + (1-|\rho|^2)P_1 \dfrac{|b|^2 + 1 - 2|b|\text{Re}\{\gamma\}}{1-|\gamma|^2}}{1 + |b|^2(1-|\rho|^2)P_1}\right).\end{aligned} \tag{11}$$

Since the bound in (11) is valid for any $|\gamma| \leq 1$, the minimizing $\gamma$ is

$$\underset{\gamma:\ |\gamma|\leq 1}{\text{argmin}} \frac{|b|^2 + 1 - 2|b|\text{Re}\{\gamma\}}{1-|\gamma|^2} = \min\left\{|b|, \frac{1}{|b|}\right\}. \tag{12}$$

After substituting the optimal value of $\gamma$ given by (12) in (11) we obtain that the sum-rate in (7c) is bounded by (6c). This shows that a Gaussian input is optimal in (7) and that the worst conditional marginal is such that one of $Y_1|_{X_2}$ and $Y_2|_{X_2}$ is the degraded version of the other.

Finally, in "strong interference" the region in (6) reduces to Th.II.2 because the bound in (6b) is redundant due to (6c), while in "weak interference" it reduces to Th.II.1 because the closure of the region is determined by the rates pairs for which (6a) and (6c) are met with equality as argued in [29, Ex. 4.3]. ■

### *B. BC- based outer bounds*

In this subsection we propose an outer bound that is tighter than the "strong interference" outer bound of Th.II.2 in the "strong interference" regime. The following observation is key: if we provide the primary transmitter with the cognitive message, the G-CIFC becomes a Gaussian MIMO BC (with two antennas at the transmitter and one antenna at each receiver) where the input is subject to a per-antenna power constraint, as originally used in [4, page 1819]. Thus, our proposed outer bound, valid for a fully general C-IFC is:

**Theorem III.2. BC-based outer bound.** *The capacity of a general CIFC is contained in the following region*

$$\mathcal{R}^{(\text{BC}-\text{PR})} \cap \mathcal{R}^{(\text{[11, Thm. 4.1]})}, \tag{13}$$

*where* $\mathcal{R}^{(\text{BC}-\text{PR})}$ *is the capacity region (or an outer bound) for the* BC with private rates only *obtained by allowing the transmitters to fully cooperate and where* $\mathcal{R}^{(\text{[11, Thm. 4.1]})}$ *is the outer bound in [11, Thm. 4.1] given in (7).*

*Proof:* The theorem follows from the fact that allowing transmitter cooperation enlarges the capacity region of the CIFC and results in a BC. The closed form expression of $\mathcal{R}^{(\text{BC}-\text{PR})}$ is provided in Appendix C. ■

and can actually be capacity. Consider the G-CIFC with "strong interference" $|b| > 1$ and where the primary user is silent, i.e., $P_2 = 0$. This channel is equivalent to a (degraded) BC with input $X_1$ whose capacity $\boldsymbol{C}(a, |b|, P_1, 0)$

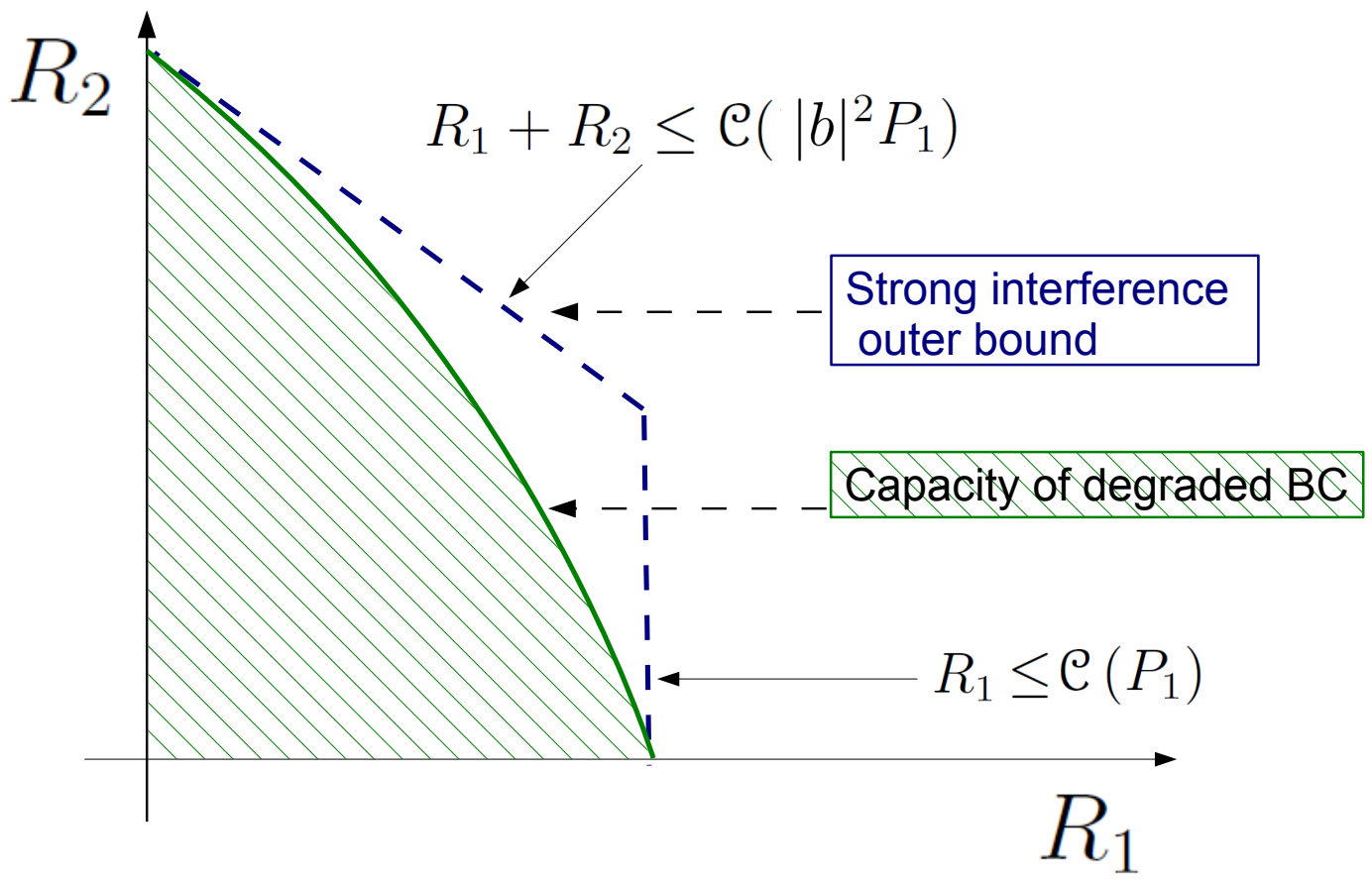

Fig. 3. The “strong interference” outer bound of Th.II.2 and the capacity region of the G-CIFC with $P_2 = 0$ and $|b| > 1$ (when the channel reduces to degraded BC).

is given by [30]

$$R_1 \leq \mathcal{C}\left(\frac{\alpha P_1}{\bar{\alpha} P_1 + 1}\right),$$

$$R_2 \leq \mathcal{C}(\alpha |b|^2 P_1),$$

taken over the union of all $\alpha \in [0, 1]$. For $P_2 = 0$, the “strong interference” outer bound of Th.II.2 reduces to

$$R_1 \leq \mathcal{C}(P_1),$$

$$R_1 + R_2 \leq \mathcal{C}(|b|^2 P_1).$$

These two regions are shown in Fig. 3 where it is clear that the “strong interference” outer bound of Th.II.2 fully contains the outer bound of the BC of Lemma VI.2. The two regions only coincide at the two Pareto optimal points A and B in (39).

Th. III.2 is valid for a general channel. It may be further tightened for the Gaussian channel in the “strong interference” regime. As previously noted in [12, Sec. 6.1], in the “strong interference” regime there is no loss of optimality in having the primary receiver decode the cognitive message in addition to its own message. Indeed, after decoding $W_2$, receiver 2 can reconstruct $X_2^N(W_2)$ and compute the following estimate of the receiver 1 output

$$\widetilde{Y}_1^N \triangleq \frac{Y_2^N - X_2^N}{|b|} + a X_2^N + \sqrt{1 - \frac{1}{|b|^2}} Z_0^N \sim Y_1^N, \tag{14}$$

where $Z_0^N \sim \mathcal{N}_{\mathbb{C}}(0, \mathbf{I})$ and independent of everything else. Hence, if receiver 1 can decode $W_1$ from $Y_1^N$, so can receiver 2 from $\widetilde{Y}_1^N$. For this reason the capacity region of the G-CIFC for $|b| > 1$ is unchanged if receiver 2 is required to decoded both messages. If we further allow the two transmitters to fully cooperate, the resulting channel is a *Gaussian MIMO BC with degraded message sets*, with per-antenna power constraint, where message $W_2$ is to be decoded at receiver 2 only and message $W_1$ at both receivers. This implies that the bound in Th.III.2 may be tightened for G-CIFC with $|b| > 1$ by using the capacity of the *Gaussian MIMO BC with degraded message sets* (BC-DMS) instead of the capacity of the *Gaussian MIMO BC with private rates only* (BC-PR):

**Theorem III.3. BC-DMS-based outer bound.** *The capacity of a G-CIFC in "strong interference" ($|b| > 1$) satisfies*

$$\boldsymbol{C}(a, |b|, P_1, P_2) \subseteq \mathcal{R}^{(\mathrm{BC-DMS})} \cap \mathcal{R}^{(\mathrm{SI})}, \tag{15}$$

*where $\mathcal{R}^{(\mathrm{BC-DMS})}$ is the capacity of the* MIMO BC with degraded message sets *determined in [31], [32] and $\mathcal{R}^{(\mathrm{SI})}$ is the "strong interference" outer bound of Th. II.2.*

*Remark* III.4. The capacity of the of the general BC-DMS is derived in [31] and it is an outer bound for a general CIFC in "strong interference". This observation was also pointed out in the independent work of [33]. It is possible to obtain the same outer bound by loosening the outer bound in [12, Th. 4], in particular by dropping [12, eq. (33)] and letting $U = [V, U_1]$. Our contribution is to determine a simpler expression for the capacity region of the Gaussian MIMO BC-DMS, in particular by proving the optimality of Gaussian inputs in the region of [31]; see Appendix D-B.

The analytical evaluation of the outer bound region in (13) of Th.III.2 (or in (15) of Th.III.3) is quite involved in general. For the special cases of degraded G-CIFC and of S-G-CIFC a closed form expression may be obtained as follows.

**Corollary III.5. BC-based outer bound for the degraded G-CIFC.** *For a degraded G-CIFC with $1/a = |b| \geq 1$, Th.III.2 and Th.III.3 coincide and reduce to*

$$R_1 \leq \mathcal{C}\left(\alpha P_1\right), \tag{16a}$$

$$R_2 \leq \mathcal{C}\left(\frac{P_2 + \bar{\alpha}|b|^2 P_1 + 2\sqrt{|b|^2 P_1 P_2}}{1 + \alpha P_1}\right), \tag{16b}$$

$$R_1 + R_2 \leq \mathcal{C}\left(P_2 + |b|^2 P_1 + 2\sqrt{\bar{\alpha}|b|^2 P_1 P_2}\right). \tag{16c}$$

*Moreover, the $R_2$-bound from the MIMO BC capacity region (in (16b)) is more stringent than the $R_2$-bound from the "strong interference" outer bound (from the difference of (16c) and (16a)) if*

$$|b| \geq \sqrt{\frac{P_2}{P_1}} + \sqrt{1 + \frac{P_2}{P_1}}.$$

*Proof:* See Appendix D. ■

**Corollary III.6. BC-DMS-based outer bound for the S-G-CIFC.** *For a S-G-CIFC with $a = 0$ and $|b| \geq 1$ the*

*outer bound of Th.III.3 is contained in the region*

$$R_1 \le \mathcal{C}\left(\alpha P_1\right), \tag{17a}$$

$$R_2 \le \mathcal{C}\left(P_2 + \frac{|b|^2 P_1 \bar{\alpha}}{1+\alpha P_1} + 2\sqrt{\frac{\bar{\alpha}|b|^2 P_1 P_2}{1+\alpha P_1}}\right), \tag{17b}$$

$$R_1 + R_2 \le \mathcal{C}\left(P_2 + |b|^2 P_1 + 2\sqrt{\bar{\alpha}|b|^2 P_1 P_2}\right). \tag{17c}$$

*Moreover, the $R_2$-bound from the MIMO BC capacity region (from (17b)) is more stringent than the $R_2$-bound from the "strong interference" outer bound (from the difference of (17c) and (17a)) if*

$$|b| \ge \sqrt{P_2 + 1}.$$

*Proof:* See Appendix D. ∎

## C. Outer bounds by transformation

Further outer bounds for the G-CIFC may be obtained by transforming the original G-CIFC into a different channel for which capacity is known. In the transformed channel the transmitters can reproduce the channel outputs of the original channel: this ensures that the transformation enlarges the capacity region thus providing an outer bound for the original channel.

**Theorem III.7. Outer bound by channel transformations.** *For the capacity region $\boldsymbol{C}(a, |b|, P_1, P_2)$ we have*

$$\boldsymbol{C}(a, b, P_1, P_2) \subseteq \bigcap_{A,B,C\ :|A|\ge 1, |\frac{C}{1-B|b|}|\ge 1} \boldsymbol{C}\left(\frac{aA-B}{C}, \frac{C|b|}{1-B|b|}, (\sqrt{|A|^2 P_1} + \sqrt{|B|^2 P_2})^2, |C|^2 P_2\right).$$

*Proof:* See Appendix E. ∎

**Corollary III.8. Special cases of Th. III.7.** *The capacity of the G-CIFC, $\boldsymbol{C}(a, |b|, P_1, P_2)$ is contained in the capacity region of the following channels:*

- **S-G-CIFC:**

$$\boldsymbol{C}(a, |b|, P_1, P_2) \subseteq \boldsymbol{C}\left(0, |b|, |\sqrt{P_1} + a\sqrt{P_2}|^2, |1 - a|b||^2 P_2\right),$$

- **G-CIFC in "weak interference":**

$$\boldsymbol{C}(a, |b|, P_1, P_2) \subseteq \boldsymbol{C}\left(a, 1, \left|\sqrt{|b|^2 P_1} + \frac{a(1-|b|)}{a-1}\sqrt{P_2}\right|^2, \left|\frac{a|b|-1}{a-1}\right|^2 P_2\right),$$

- **G-CIFC in "very strong interference":**

$$\boldsymbol{C}(a, |b|, P, P) \subseteq \boldsymbol{C}\left(|b|, |b|, P', P'\right),\ P' = \frac{P}{(|b|^2-1)^2}\max\{||b|^2 - 1 + |b| - a|^2, |1 - a|b||^2\}$$

*Proof:* See Appendix E. ∎

*Remark* III.9. The G-IFC with conferencing encoders of [34] encompasses the G-CIFC as a special case when $C_{12} = 0$ and $C_{21} = \infty$. The outer bound in [34, Lemma 4.1] with $C_{12} = 0$ and $C_{21} = \infty$ is an outer bound for the G-CIFC. This outer bound reduces to the "strong interference" outer bound of Th. II.2 when the channel is a G-CIFC. In particular we notice that for a CIFC, unlike for a classical IFC and the IFC with conferencing encoders, no bounds of the form $2R_1 + R_2$ are known. In [34] the authors provide an interesting interpretation of this type of bound for a channel with and without conferencing transmitters. With regard to this interpretation we point out that, with full a priori knowledge of the primary message, the cognitive transmitter can always pre-code its message against the interference from the primary user and thus the strategy of the primary encoder never limits the rate of the cognitive receiver.

## IV. Inner bounds

In [11] we introduce a new inner bound for the Discrete Memoryless CIFC (DM-CIFC) and show that this scheme encompasses all previously proposed achievable schemes as special cases; it is thus the largest known achievable rate region to date. This achievable scheme also introduces new transmission features that were crucial in proving capacity for the semi-deterministic DM-CIFC [11, Sec. VIII]. Here we use the inner bound of [11] as a unified framework to present the achievable schemes used in the remainder of the paper. In this section we introduce the general achievable scheme in [11, Th. V.1] and use it to obtain six simple sub-schemes that will be used in the following sections to prove capacity and constant gap results.

As the Gaussian CIFC encompasses classical interference, multiple-access and broadcast channels, the achievable rate region of [11] incorporates a combination of the transmission techniques devised for these channels.

- **Rate-splitting.** Both the primary and the cognitive message are split into private and common parts, as in the Han and Kobayashi scheme [35] for the IFC. Although rate-splitting was shown to be unnecessary in the "weak interference" [6] and "very strong interference" [5] regimes of (1) and (5), respectively, it allows significant rate improvement in the "strong interference" regime.
- **Superposition-coding.** The cognitive common message is superposed to the primary common message and parts of the cognitive message are superposed to parts of the primary message. Useful in multiple-access and broadcast channels [27], a simple superposition of the primary and cognitive messages (all common) is capacity achieving in the "very strong interference" regime [5].
- **Pre-coding.** Gel'fand-Pinsker coding [36], often referred to as binning or Dirty Paper Coding (DPC), allows a transmitter to pre-code (portions of) the interference known to be experienced at the receiver. Binning is also used by Marton in [18] to derive the largest known achievable rate region for the BC. In the scheme of [11], binning is performed at the cognitive encoder for both the common and the private message and it allows for the cancellation of interference from the primary transmitter.
- **Broadcasting.** In [11] we introduced the idea of having the cognitive encoder transmit part of the primary message. This is made possible by the perfect knowledge of the primary message at the cognitive transmitter, which is specific to this channel model. The additional primary message is superposed to the cognitive common message and also pre-coded against the cognitive private message. The incorporation of the broadcast feature at the cognitive transmitter was initially motivated by the fact that in certain regimes, this strategy was shown to be capacity achieving for the high-SNR linear deterministic approximation of the CIFC [37].

The achievable scheme may be described as follows:

- **Rate-splitting.** The independent messages $W_1$ and $W_2$, uniformly distributed on $\mathcal{M}_1 = [1 : 2^{nR_1}]$ and $\mathcal{M}_2 = [1 : 2^{nR_2}]$ respectively, are rate split into the messages $W_i$, $i \in \{1c, 2c, 1pb, 2pb, 2pa\}$, all independent and uniformly distributed on $[1 : 2^{nR_i}]$, each encoded using the RV $U_i$.
- **Primary encoder.** Transmitter 2 sends $X_2$ that carries the private message $W_{2pa}$ ("p" for private, "a" for alone) **superposed** to the common message $W_{2c}$ carried by $U_{2c}$ ("c" for common).
- **Cognitive encoder.** The common message of transmitter 1, encoded by $U_{1c}$, is **binned** against $X_2$ conditioned on $U_{2c}$. The private message of transmitter 2, $W_{2pb}$, encoded by $U_{2pb}$ ("b" for broadcast) and a portion of the private message of transmitter 1, $W_{1pb}$, encoded as $U_{1pb}$, are **binned** against each other as in Marton's region [18] conditioned on $U_{1c}, U_{2c}, X_2$. Transmitter 1 sends $X_1$, which is a function of all the RVs.
- **Primary decoder.** Receiver 2 jointly decodes $U_{2c}$ (carrying $W_{2c}$), $U_{1c}$ (carrying $W_{1c}$), $U_{2pb}$ (carrying $W_{2pb}$), and $X_2$ (carrying $W_{2pa}$).
- **Cognitive decoder.** Receiver 1 jointly decodes $U_{1c}$ (carrying $W_{1c}$), $U_{2pb}$ (carrying $W_{2pb}$), and $U_{1pb}$ (carrying $W_{1pb}$).
- **Analysis.** The codebook generation, encoding, decoding and the error analysis are provided in [11].

**Corollary IV.1. Achievable region $\mathcal{R}^{(\mathrm{RTD})}$ in [11, Th.4.1].**

*A rate pair $(R_1, R_2)$ such that*

$$R_1 = R_{1c} + R_{1pb}, \tag{18a}$$

$$R_2 = R_{2c} + R_{2pa} + R_{2pb} \tag{18b}$$

*is achievable for a general DM-CIFC if* $(R'_{1c}, R'_{1pb}, R'_{2pb}, R_{1c}, R_{1pb}, R_{2c}, R_{2pa}, R_{2pb}) \in \mathbb{R}_+^8$ *satisfies:*

$$
\begin{aligned}
{R'}_{1c} &\geq I(U_{1c}; X_2|U_{2c}) && \text{(19a)}\\
{R'}_{1c} + {R'}_{1pb} &\geq I(U_{1pb}; X_2|U_{1c}, U_{2c}) + I(U_{1c}; X_2|U_{2c}) && \text{(19b)}\\
{R'}_{1c} + {R'}_{1pb} + {R'}_{2pb} &\geq I(U_{1pb}; X_2, U_{2pb}|U_{1c}, U_{2c}) + I(U_{1c}; X_2|U_{2c}) && \text{(19c)}\\
R_{2c} + R_{2pa} + (R_{1c} + {R'}_{1c}) + (R_{2pb} + {R'}_{2pb}) &\leq I(Y_2; U_{2pb}, U_{1c}, X_2, U_{2c}) + I(U_{1c}; X_2|U_{2c}) && \text{(19d)}\\
R_{2pa} + (R_{1c} + {R'}_{1c}) + (R_{2pb} + {R'}_{2pb}) &\leq I(Y_2; U_{2pb}, U_{1c}, X_2|U_{2c}) + I(U_{1c}; X_2|U_{2c}) && \text{(19e)}\\
R_{2pa} + (R_{2pb} + {R'}_{2pb}) &\leq I(Y_2; U_{2pb}, X_2|U_{1c}, U_{2c}) + I(U_{1c}; X_2|U_{2c}) && \text{(19f)}\\
(R_{1c} + {R'}_{1c}) + (R_{2pb} + {R'}_{2pb}) &\leq I(Y_2; U_{2pb}, U_{1c}|X_2, U_{2c}) + I(U_{1c}; X_2|U_{2c}) && \text{(19g)}\\
(R_{2pb} + {R'}_{2pb}) &\leq I(Y_2; U_{2pb}|U_{1c}, X_2, U_{2c}) && \text{(19h)}\\
R_{2c} + (R_{1c} + {R'}_{1c}) + (R_{1pb} + {R'}_{1pb}) &\leq I(Y_1; U_{1pb}, U_{1c}, U_{2c}), && \text{(19i)}\\
(R_{1c} + {R'}_{1c}) + (R_{1pb} + {R'}_{1pb}) &\leq I(Y_1; U_{1pb}, U_{1c}|U_{2c}), && \text{(19j)}\\
(R_{1pb} + {R'}_{1pb}) &\leq I(Y_1; U_{1pb}|U_{1c}, U_{2c}), && \text{(19k)}
\end{aligned}
$$

*for some input distribution*

$$P_{Y_1,Y_2,X_1,X_2,U_{1c},U_{2c},U_{2pa},U_{1pb},U_{2pb}} = P_{U_{1c},U_{2c},U_{2pa},U_{1pb},U_{2pb},X_1,X_2} P_{Y_1,Y_2|X_1,X_2}.$$

We now present six different sub-schemes obtained from the achievable scheme of Corollary IV.1 by reducing the number of rate splits to at most three rather than five. By setting some rates to zero we may drop the corresponding RVs and simplify the region in (19). The resulting transmission schemes are used in the rest of the paper for achievability proofs (for capacity and constant gap results) and numerical evaluations. Tables I and II help illustrate the different schemes: Table I shows, for each scheme, which rate splits in the $\mathcal{R}^{(\mathrm{RTD})}$ are set to zero (the corresponding RV is in gray) and which ones are not (the corresponding RV is in black), while Table II indicates which result will be proved with the corresponding scheme.

### *A. Achievable scheme with $U_{2pb}$ and $U_{1pb}$: capacity achieving for the degraded broadcast channel.*

*Motivation: Achieve the capacity to within a finite gap in some parameter regime by having transmitter 2 silent.*

Consider the case where transmitter 2 is silent and transmitter 1 transmits to both decoders. In this case, the G-CIFC with $P_2 = 0$ reduces to a degraded BC with input $X_1$ [38]. When $|b| < 1$, $Y_2$ is a degraded version of $Y_1$ and the maximum achievable rate region when transmitter 2 is silent is

$$
\begin{aligned}
R_1 &\leq I(Y_1; U_{1pb}) - I(U_{1pb}; U_{2pb}) \overset{[38]}{=} \mathcal{C}(\alpha P_1), && \text{(20a)}\\
R_2 &\leq I(Y_2; U_{2pb}) \overset{[38]}{=} \mathcal{C}\left(\frac{\bar{\alpha}|b|^2 P_1}{1 + \alpha |b|^2 P_1}\right), && \text{(20b)}
\end{aligned}
$$

TABLE I

THE ACHIEVABLE SCHEMES OF SECTION IV.

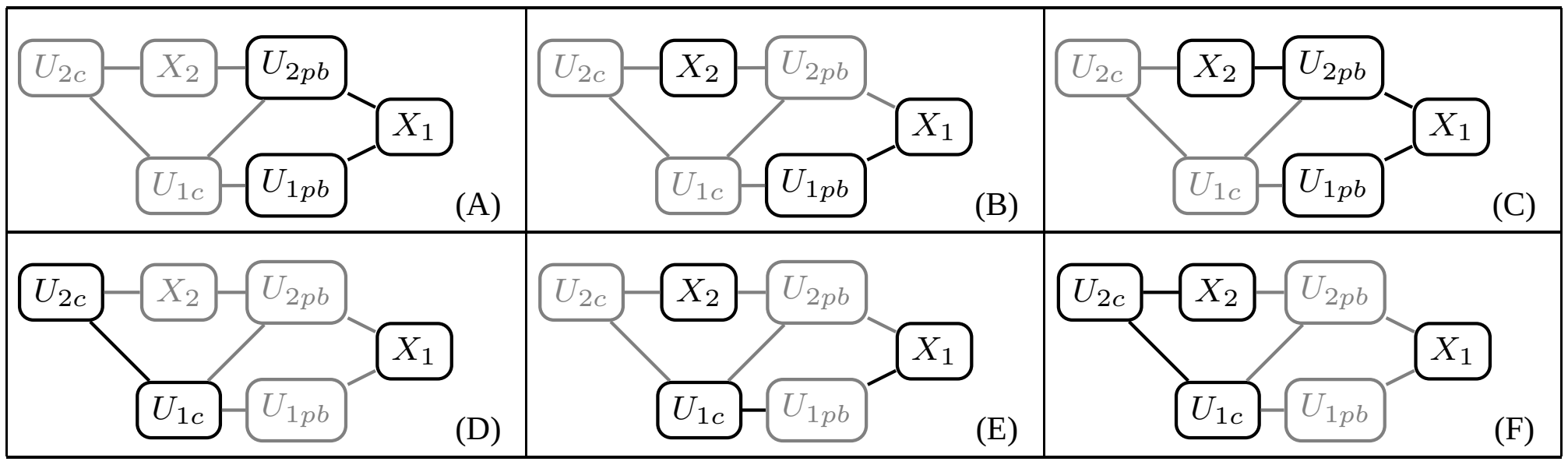

TABLE II

THE ROLE OF THE DIFFERENT ACHIEVABLE SCHEMES IN THE FOLLOWING SECTIONS

| Scheme | $U_{2c}$ | $U_{1c}$ | $X_2$ | $U_{1pb}$ | $U_{2pb}$ | Role | Where |
|---|---|---|---|---|---|---|---|
| (A) | | | | • | • | constant gap in a subspace of the parameter region | Thm. VI.4 |
| (B) | | | • | • | | capacity in “weak interference” | Thm II.1, Thm. VI.1, Thm. VI.3 |
| (C) | | | • | • | • | constant gap in the whole parameter region | Thm. VI.1 |
| (D) | • | • | | | | capacity in “very strong strong interference” | Thm. II.3 |
| | | | | | | constant gap in a subspace of the parameter region | Thm. VI.4 |
| (E) | | • | • | | | capacity in the “primary decodes cognitive” regime | Thm. V.1, Thm. V.3 |
| | | | | | | constant gap in a subspace of the parameter region | Thm. VI.3, Thm. VI.4 |
| (F) | • | • | | • | | Numerical results | Sec.VII |

taken over the union over of all $\alpha \in [0,1]$. When $|b| \geq 1$, $Y_1$ is a degraded version of $Y_2$ and the maximum achievable rate region when transmitter 2 is silent is

$$R_1 \leq \quad I(Y_1; U_{1pb}) \quad \overset{[38]}{=} \mathcal{C}\left(\frac{\bar{\alpha}P_1}{1+\alpha P_1}\right), \tag{21a}$$

$$R_2 \leq \quad I(Y_2; U_{2pb}|U_{1pb}) \quad \overset{[38]}{=} \mathcal{C}(|b|^2 \alpha P_1), \tag{21b}$$

taken over the union of all $\alpha \in [0,1]$.

### B. *Achievable scheme with $X_2$ and $U_{1pb}$: capacity achieving in the “weak interference” regime.*

*Motivation: Completeness.*

In this scheme both messages are private and receiver 2 treats the interference from transmitter 1 as noise while transmitter 1 performs perfect DPC against the interference from transmitter 2. This scheme achieves capacity in the “weak interference regime” of Th.II.1 [6].

### C. *Achievable scheme with $X_2, U_{1pb}$ and $U_{2pb}$: capacity achieving in the semi-deterministic DM-CIFC.*

*Motivation: Achieve the “strong interference” outer bound to within a constant gap in the whole parameter regime.*

This achievable strategy is obtained by combining the previous two transmission schemes, scheme (A) and (B), and it corresponds to the capacity achieving scheme for the semi-deterministic G-CIFC [11]. The broadcasting RV $U_{2pb}$ appears only in the $\mathcal{R}^{(\mathrm{RTD})}$ region and in [19], [25]. The achievable rate region is

$$
\begin{aligned}
R_1 &\leq I(Y_1;U_{1pb}) - I(U_{1pb};X_2) \\
&\overset{(23)}{=} \log(\sigma_{1pb}^2 + \alpha P_1) - \log\left(\sigma_{1pb}^2 + \frac{\mathrm{Var}[X_1 + aX_2]}{1 + \mathrm{Var}[X_1 + aX_2]}\right), && \text{(22a)}\\
R_2 &\leq I(Y_2;U_{2pb},X_2) \\
&\overset{(23)}{=} \log(1 + \mathrm{Var}[|b|X_1 + X_2]) - \log\left(1 + \frac{\sigma_{2pb}^2 \mathrm{Var}[|b|X_1|_{X_2}]}{\sigma_{2pb}^2 + \mathrm{Var}[|b|X_1|_{X_2}]}\right), && \text{(22b)}\\
R_1 + R_2 &\leq I(Y_2;U_{2pb},X_2) + I(Y_1;U_{1pb}) - I(U_{1pb};U_{2pb},X_2) \\
&\overset{(23)}{\leq} \text{(22a)} + \text{(22b)} + \log\left(1 - \frac{\left|[|b|P_1\alpha - \sqrt{\sigma_{1pb}^2\sigma_{2pb}^2}]^+\right|^2}{(|b|^2P_1\alpha + \sigma_{2pb}^2)(P_1\alpha + 1)}\right) && \text{(22c)}
\end{aligned}
$$

where

$$
\begin{aligned}
X_{1pb} &\sim \mathcal{N}_{\mathbb{C}}(0, \alpha P_1) \\
X_2 &\sim \mathcal{N}_{\mathbb{C}}(0, P_2), \ \text{independent of } X_{1pb}, \\
X_1 &= X_{1pb} + \sqrt{\frac{\bar{\alpha}P_1}{P_2}}X_2 \\
U_{1pb} &= X_1 + aX_2 + Z_{1pb} \\
U_{2pb} &= |b|X_1 + X_2 + Z_{2pb}, && \text{(23)}
\end{aligned}
$$

and

$$
\begin{bmatrix} Z_{1pb} \\ Z_{2pb} \end{bmatrix} \sim \mathcal{N}_{\mathbb{C}}\left(\mathbf{0}, \begin{bmatrix} \sigma_{1pb}^2 & \rho_{pb}\sqrt{\sigma_{1pb}^2\sigma_{2pb}^2} \\ \rho_{pb}^*\sqrt{\sigma_{1pb}^2\sigma_{2pb}^2} & \sigma_{2pb}^2 \end{bmatrix}\right),
$$

for $|\rho_{pb}| \leq 1$. The assignment in (23) is inspired by the capacity achieving scheme for the semi-deterministic CIFC of [11] where $U_{1pb}$ and $U_{2pb}$ are set to be equal to $Y_1$ and $Y_2$ respectively. The inequality in (22c) is obtained by optimizing $\rho_{pb}$ as detailed in Th. VI.1.

*D. Achievable scheme with $U_{1c}$ and $U_{2c}$: capacity achieving in "very strong interference" regime.*

*Motivation: Completeness.*

This scheme achieves the "strong interference" outer bound of Th. II.2 under the "very strong interference"

conditions of Th. II.3 [10]. The achievable rate region is

$$R_1 \le I(Y_1; X_1|X_2) \overset{(25)}{=} \mathcal{C}((1-|\rho|^2)P_1), \tag{24a}$$

$$R_1 \le I(Y_2; X_1|X_2) \overset{(25)}{=} \mathcal{C}((1-|\rho|^2)|b|^2 P_1), \tag{24b}$$

$$R_1 + R_2 \le I(Y_1; X_1, X_2) \overset{(25)}{=} \mathcal{C}(P_1 + |a|^2 P_2 + 2\mathrm{Re}\{a^* \rho\}\sqrt{P_1 P_2}), \tag{24c}$$

$$R_1 + R_2 \le I(Y_1; X_1, X_2) \overset{(25)}{=} \mathcal{C}(|b|^2 P_1 + P_2 + 2|b|\mathrm{Re}\{\rho\}\sqrt{P_1 P_2}), \tag{24d}$$

where the RHS of (24) is achieved with the assignment

$$\begin{aligned} X_{1c} &\sim \mathcal{N}_{\mathbb{C}}(0, (1-|\rho|^2)P_1) \\ X_2 &\sim \mathcal{N}_{\mathbb{C}}(0, P_2), \text{ independent of } X_{1c}, \\ X_1 &= X_{1c} + \rho\sqrt{\frac{P_1}{P_2}} X_2, \end{aligned} \tag{25}$$

for some $|\rho| \le 1$. This scheme was originally proposed for real-valued channels in [10]. Here we consider its extension to complex-valued valued channels.

### *E. Achievable scheme with $X_2, U_{1c}$: capacity achieving in "primary decodes cognitive" regime.*

*Motivation: Achieve capacity in the "primary decodes cognitive" regime.*

In this scheme the primary message is private while the cognitive message is public and binned against the interference created by the primary user at the cognitive decoder. This scheme can also be obtained as a special case of the scheme in [12] and [19]. The achievable rate region is

$$\begin{aligned} R_1 &\le I(Y_1; U_{1c}) - I(U_{1c}; X_2) \\ &\overset{(28)}{=} f\left(a + \sqrt{\frac{\bar{\alpha} P_1}{P_2}}, 1; \lambda\right), \end{aligned} \tag{26a}$$

$$\begin{aligned} R_2 &\le I(Y_2; U_{1c}, X_2) - (I(Y_2; U_{1c}) - I(U_{1c}; X_2)) \\ &\overset{(28)}{=} \mathcal{C}(P_2 + |b|^2 P_1 + 2\sqrt{\bar{\alpha}|b|^2 P_1 P_2}) - f\left(\frac{1}{|b|} + \sqrt{\frac{\bar{\alpha} P_1}{P_2}}, \frac{1}{|b|^2}; \lambda\right), \end{aligned} \tag{26b}$$

$$\begin{aligned} R_1 + R_2 &\le I(Y_2; U_{1c}, X_2) \\ &\overset{(28)}{=} \mathcal{C}(P_2 + |b|^2 P_1 + 2\sqrt{\bar{\alpha}|b|^2 P_1 P_2}), \end{aligned} \tag{26c}$$

for

$$\begin{aligned} f(h, \sigma^2; \lambda) &\triangleq I(X_{1c} + hX_2 + \sigma Z_1; U_{1c}) - I(U_{1c}; X_2) \\ &\overset{(28)}{=} \log\left(\frac{\sigma^2 + \alpha P_1}{\sigma^2 + \frac{\alpha P_1 |h|^2 P_2}{\alpha P_1 + |h|^2 P_2 + \sigma^2}\left|\frac{\lambda}{\lambda_{\mathrm{Costa}}(h, \sigma^2)} - 1\right|^2}\right), \end{aligned}$$

with

$$\lambda_{\text{Costa}}(h,\sigma^2) \triangleq \frac{\alpha P_1}{\alpha P_1+\sigma^2}\ h, \tag{27}$$

and where the RHS of (26) is achieved with the assignment

$$\begin{aligned} X_{1c} &\sim \mathcal{N}_{\mathbb{C}}(0,\alpha P_1) \\ X_2 &\sim \mathcal{N}_{\mathbb{C}}(0,P_2) \\ X_1 &\sim X_{1c}+\sqrt{\frac{\bar{\alpha}P_1}{P_2}}X_2 \\ U_{1c} &= X_{1c}+\lambda X_2, \end{aligned} \tag{28a}$$

for some $\alpha \in [0,1]$ and $\lambda \in \mathbb{C}$. Note that $f(h,\sigma^2;\lambda) \geq 0$ if $\left|\frac{\lambda}{\lambda_{\text{Costa}}(h,\sigma^2)}-1\right|^2 \leq 1+\frac{\alpha P_1+\sigma^2}{|h|^2P_2}$.

*F. Achievable scheme with $U_{2c}, X_2$ and $U_{1c}$.*

*<u>Motivation</u>: Achieve capacity in the largest subset of the "strong interference" regime.*

As for scheme (C), this scheme is obtained by combining the previous two schemes, scheme (D) and (E). The achievable rate region is

$$R_1 \leq I(Y_1;U_{1c}|U_{2c}) - I(X_2;U_{1c}|U_{2c}), \tag{29a}$$

$$R_1 \leq I(Y_2;U_{1c},X_2|U_{2c}), \tag{29b}$$

$$R_1+R_2 \leq I(Y_2;U_{2c},X_2,X_{1c}), \tag{29c}$$

$$R_1+R_2 \leq I(Y_2;X_2|U_{1c},U_{2c}) + I(Y_1;U_{1c},U_{2c}), \tag{29d}$$

$$2R_1+R_2 \leq I(Y_2;U_{1c},X_2|U_{2c}) + I(Y_1;U_{1c},U_{2c}) - I(U_{1c};X_2|U_{2c}). \tag{29e}$$

In particular, we consider the choice of RVs

$$X_{2c}, X_{2pa}, X_{1c} \sim \text{iid } \mathcal{N}_{\mathbb{C}}(0,1) \tag{30a}$$

$$X_2 = \sqrt{\beta P_2}X_{2c} + \sqrt{\bar{\beta}P_2}X_{2pa} \tag{30b}$$

$$X_1 = \sqrt{\alpha P_1}X_{1c} + \sqrt{\bar{\alpha}P_1}\left(\sqrt{\gamma}X_{2c}+\sqrt{\bar{\gamma}}X_{2pa}\right) \tag{30c}$$

$$U_{1c} = X_{1c} + \lambda X_{2pa} \tag{30d}$$

$$U_{2c} = X_{2c}. \tag{30e}$$

This scheme unifies the two schemes that achieve capacity in two different parameter regimes of of $|b|>1$ and hence is the scheme that achieves capacity in the largest subset of the "strong interference" regime.

## V. New capacity results

We now present two new capacity results for the G-CIFC. The first capacity result uses scheme (E) to achieve the "strong interference" outer bound in what we term the "primary decodes cognitive" regime, a subset of the "strong interference" regime that is not included in the "very strong interference" regime of Th. II.3, for which capacity is already known. The second capacity result focuses on the S-G-CIFC where we show that the BC-DMS-based outer bound of Th.III.2 is achieved by scheme (E) for a large set of parameters where capacity was previously unknown. Although the two results involve the same achievable scheme (E), in the first result the cognitive receiver performs Costa's "interference pre-cancellation" (or pre-coding) of the interference from the primary receiver while, in the second result, no pre-coding in necessary. In scheme (E) the pre-coding operation has an interesting effect on the rate region that we investigate in detail in Remark V.2.

Before presenting the new results, we describe scheme (E) in more detail. The achievable rate region is expressed in two parameters: $\alpha$ and $\lambda$. The parameter $\alpha$ denotes the fraction of power that encoder 1 employs to transmit its own message versus the power to broadcast $X_2$. For $\alpha = 0$, transmitter 1 uses all its power to broadcast $X_2$ as in a virtual Multiple Input Single Output (MISO) channel. When $\alpha = 1$, transmitter 1 utilizes all its power to transmit $X_{1c}$. The parameter $\lambda$ controls the amount of interference (created by $X_2$ at receiver 1) "pre-cancellation" achievable using DPC at transmitter 1. With $\lambda = 0$, no DPC is performed at transmitter 1 and the interference due to $X_2$ is treated as noise. On the other hand, with $\lambda = \lambda_{\rm Costa}$ for

$$\lambda_{\rm Costa}\left(a + \sqrt{\frac{\bar{\alpha}P_1}{P_2}},\ 1\right) \triangleq \lambda_{\rm Costa\ 1},$$

with $\lambda_{\rm Costa}(\cdot,\cdot)$ defined in (27), the interference due to $X_2$ at receiver 1 is completely "pre-canceled", thus achieving the maximum possible rate $R_1$. Different values of $\lambda$ are not usually investigated because, as long as the interference is a nuisance (i.e., no node in the network has information to extract from the interference), the best is to completely "pre-cancel" it by using $\lambda = \lambda_{\rm Costa}(h, \sigma^2)$.

However, $\lambda$ influences not only the rate $R_1$ in (26a), but also the rate $R_2$ in (26b). An interesting question is whether $\lambda \neq \lambda_{\rm Costa\ 1}$, although it does not achieve the largest possible $R_1$, would improve the achievable rate region by sufficiently boosting the rate $R_2$. We comment on this question later on in Section VII-D. At this point we make the following observation: $R_1$ is a concave function in $\lambda$, symmetric around $\lambda = \lambda_{\rm Costa\ 1}$ and with a global maximum at $\lambda = \lambda_{\rm Costa\ 1}$, while $R_2$ is a convex function in $\lambda$, symmetric around $\lambda = \lambda_{\rm Costa\ 2}$ and with a global minimum at $\lambda = \lambda_{\rm Costa\ 2}$, where

$$\lambda_{\rm Costa}\left(\frac{1}{|b|} + \sqrt{\frac{\bar{\alpha}P_1}{P_2}},\ \frac{1}{|b|^2}\right) \triangleq \lambda_{\rm Costa\ 2}.$$

Fig. 4 shows $R_1$ in (26a) and $R_2$ in (26b) as a function of $\lambda \in \mathbb{R}$, for $P_1 = P_2 = 6$, $b = \sqrt{2}$, $a = \sqrt{0.3}$, and $\alpha = 0.5$. For the chosen parameters, we observe a trade-off among the rates: $\lambda = \lambda_{\rm Costa\ 1}$ achieves the maximum for $R_1$, but it achieves close to the minimum for $R_2$. This observation will help in understanding why scheme (E) does not perform well in certain parameter regimes as will be pointed out in Remark V.2.

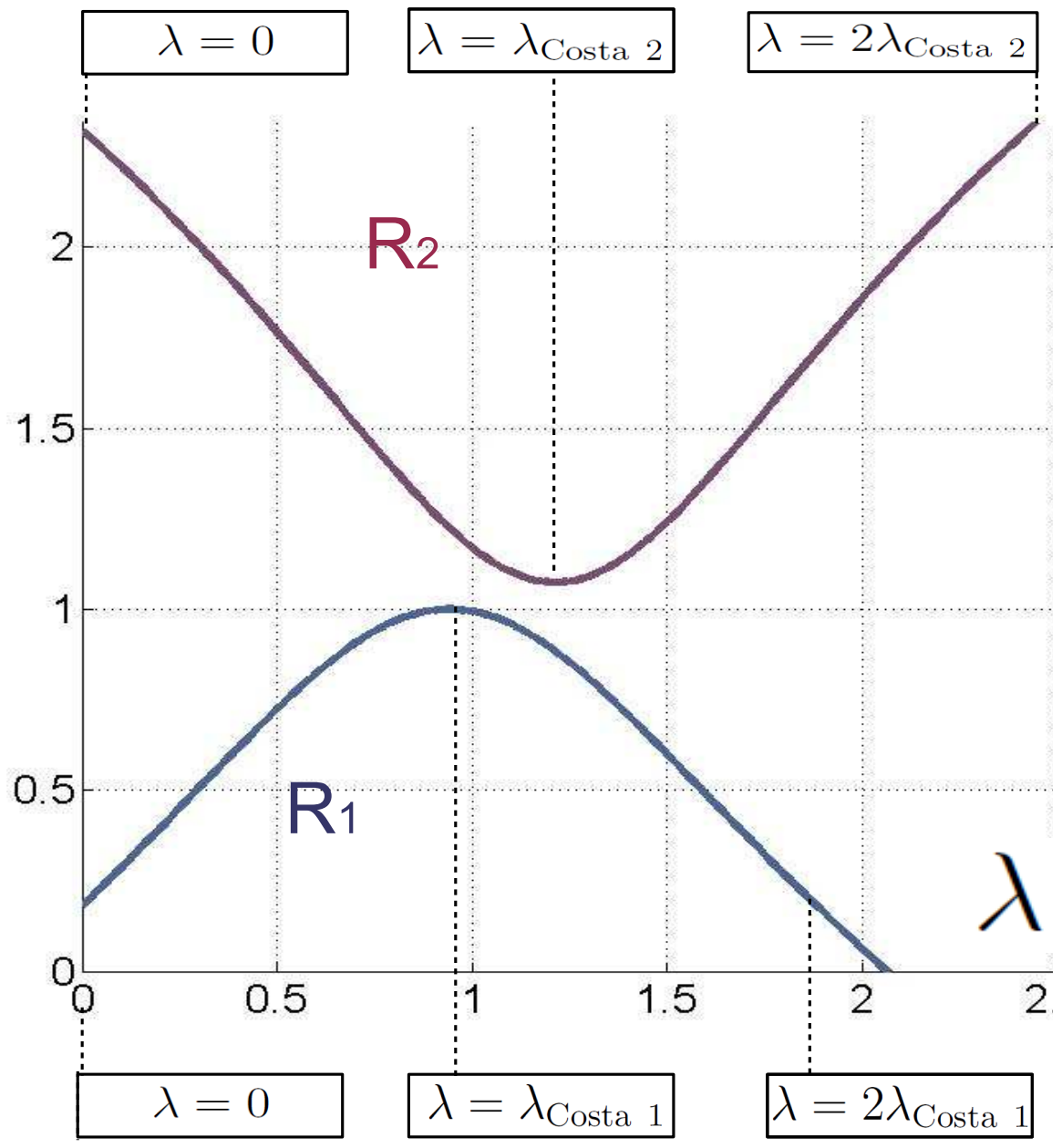

Fig. 4. The bound for $R_1$ in (26a) (bottom) and the bound for $R_2$ in (26b) (top) as a function of $\lambda \in \mathbb{R}$, for $P_1 = P_2 = 6$, $b = \sqrt{2}$, $a = \sqrt{0.3}$, $\alpha = 0.5$.

### *A. New capacity results for the C-CIFC.*

**Theorem V.1. Capacity in the "primary decodes cognitive" regime.** *When* $|b| > 1$ *and*

$$P_2 |1 - a|b||^2 \geq (|b|^2 - 1)(1 + P_1 + |a|^2 P_2) - P_1 P_2 \big|1 - a|b|\big|^2, \tag{31a}$$

$$P_2 |1 - a|b||^2 \geq (|b|^2 - 1)(1 + P_1 + |a|^2 P_2 + 2\mathrm{Re}\{a\}\sqrt{P_1 P_2}), \tag{31b}$$

*the "strong interference" outer bound of Th.II.2 is tight and achieved by scheme (E).*

The "primary decodes cognitive" regime, illustrated in Fig. 5 in the $(a, |b|)$-plane for $a \in \mathbb{R}$ and $P_1 = P_2 = 10$, covers parts of the "strong interference" regime $|b| > 1$ where capacity was not known. It also shows that the scheme in (26) (i.e., scheme (E)) is capacity achieving for part of the "very strong interference" region in (5), thus providing an alternative capacity achieving scheme to superposition coding [10] (i.e., scheme (D)).

*Proof:* We compare the achievable scheme (E) in Section IV-E with the "strong interference" outer bound of Th.II.2. Scheme (E) for $|b| > 1$, $\lambda = \lambda_{\text{Costa }1}$ and the assignment in (26). This achieves (6a)=(26a) and (6c)=(26c) (and (6b) is redundant). Therefore the "strong interference" outer bound is achievable when ((26a)+(26b))≥(6b),

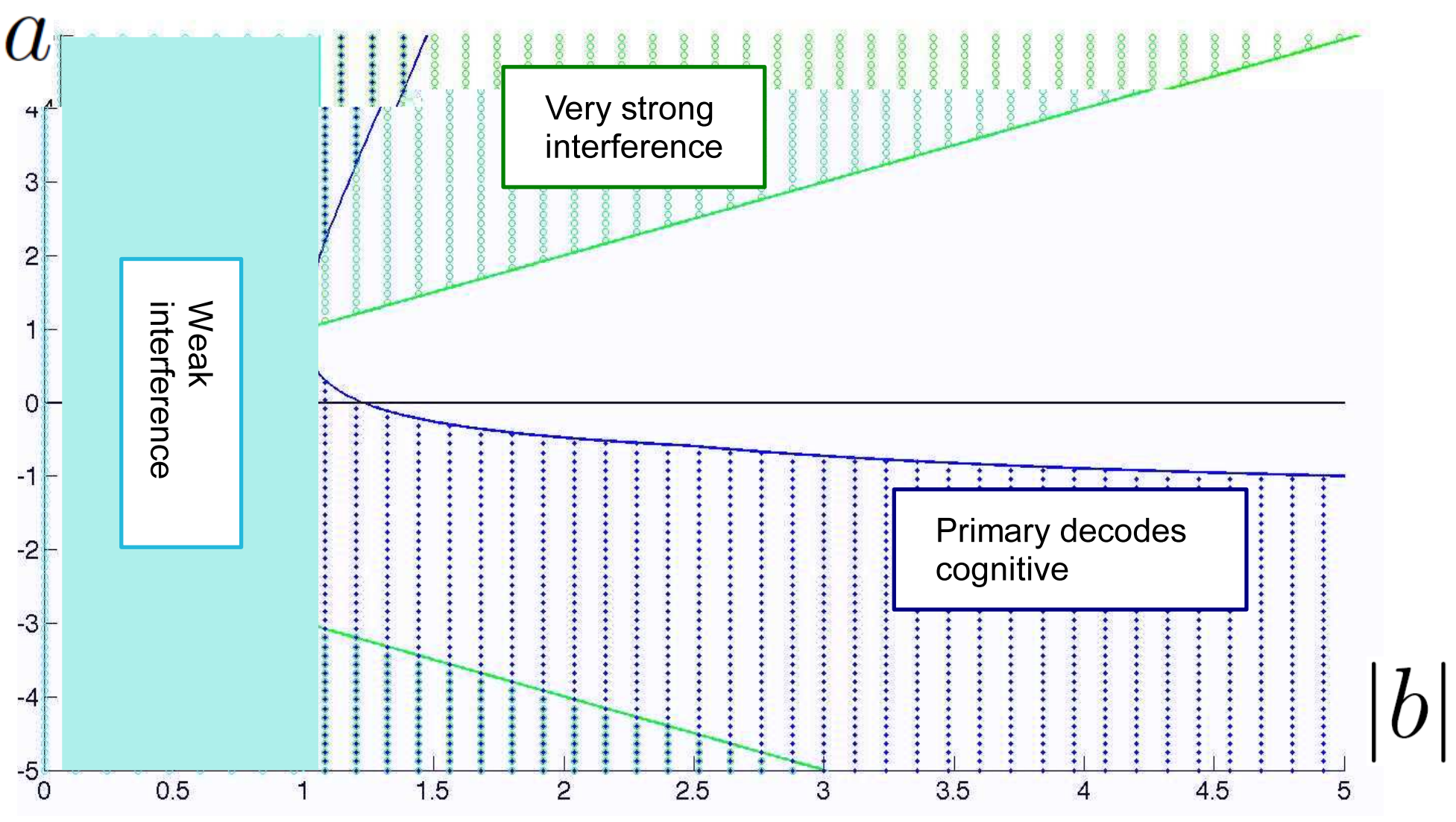

Fig. 5. A representation of the capacity result of Th.V.1 for a G-CIFC with $a \in \mathbb{R}$, $P_1 = P_2 = 10$ and $(a, |b|) \in [-5, 5] \times [0, 5]$.

i.e. when

$$\mathcal{C}(\alpha P_1) = f\left(a+\sqrt{\frac{\bar{\alpha} P_1}{P_2}}, 1; \lambda_{\text{Costa } 1}\right) \geq f\left(\frac{1}{|b|}+\sqrt{\frac{\bar{\alpha} P_1}{P_2}}, \frac{1}{|b|^2}; \lambda_{\text{Costa } 1}\right), \ \forall \alpha \in [0,1],$$

$$\iff \quad \alpha P_1 + |l_{\text{Costa}} a|^2 P_2 - \frac{\left||b|\alpha P_1 + \lambda\left(P_2+\sqrt{\bar{\alpha}|b|^2 P_1 P_2}\right)\right|^2}{|b|^2 P_1 + P_2 + 2\sqrt{\bar{\alpha}|b|^2 P_1 P_2}+1} \geq \frac{\alpha P_1}{\alpha P_1 + 1}, \ \forall \alpha \in [0,1],$$

$$\iff \quad \left(\frac{\alpha P_1}{\alpha P_1 + 1}\right)^2 \frac{Q(\alpha)}{1+|b|^2 P_1 + P_2 + 2\sqrt{\bar{\alpha}|b|^2 P_1 P_2}} \geq 0, \ \forall \alpha \in [0,1], \tag{32a}$$

where

$$Q(\alpha) \triangleq P_2\left|1 - a|b|\right|^2(\alpha P_1 + 1) - (|b|^2 - 1)\left(P_1 + |a|^2 P_2 + 2\text{Re}\{a\}\sqrt{\bar{\alpha} P_1 P_2} + 1\right).$$

Clearly the condition in (32a) is verified if for all $\alpha \in [0,1]$ we have $Q(\alpha) \geq 0$. $Q(\alpha)$ is a quadratic function in $x = \sqrt{1-\alpha}$ of the form $c_1 x^2 + c_2 x + c_3$ with $c_1 = -P_1 P_2 |1 - a|b||^2 \leq 0$, which implies that $Q(\alpha)$ is concave in $\alpha$. Hence, the inequality in (32a) is verified for every $\alpha \in [0,1]$ if it is verified for $\alpha = 1$ and $\alpha = 0$. The condition $Q(0) \geq 0$ corresponds to (31b) while the condition $Q(1) \geq 0$ corresponds to (31a). ■

*Remark* V.2. Previous capacity results for the G-CIFC imposed conditions on the channel parameters that lent themselves well to "natural" interpretations. For example, the "weak interference" condition $I(Y_1; X_1|X_2) \geq I(Y_2; X_1|X_2)$ of [6] in (1) suggests that decoding $X_1$ at receiver 2, even after having decoded the intended message in $X_2$, would constrain the rate $R_1$ too much, thus preventing it from achieving the interference-free rate in (6a). The "very strong interference" condition $I(Y_1; X_1, X_2) \geq I(Y_2; X_1, X_2)$ of [10] in (3) suggests that requiring receiver 1 to decode both messages should not prevent achieving the maximum sum-rate at receiver 2 given by (6c). A similar intuition about the new "primary decodes cognitive" capacity condition in (31) unfortunately does not emerge from the proof of Th.V.1.

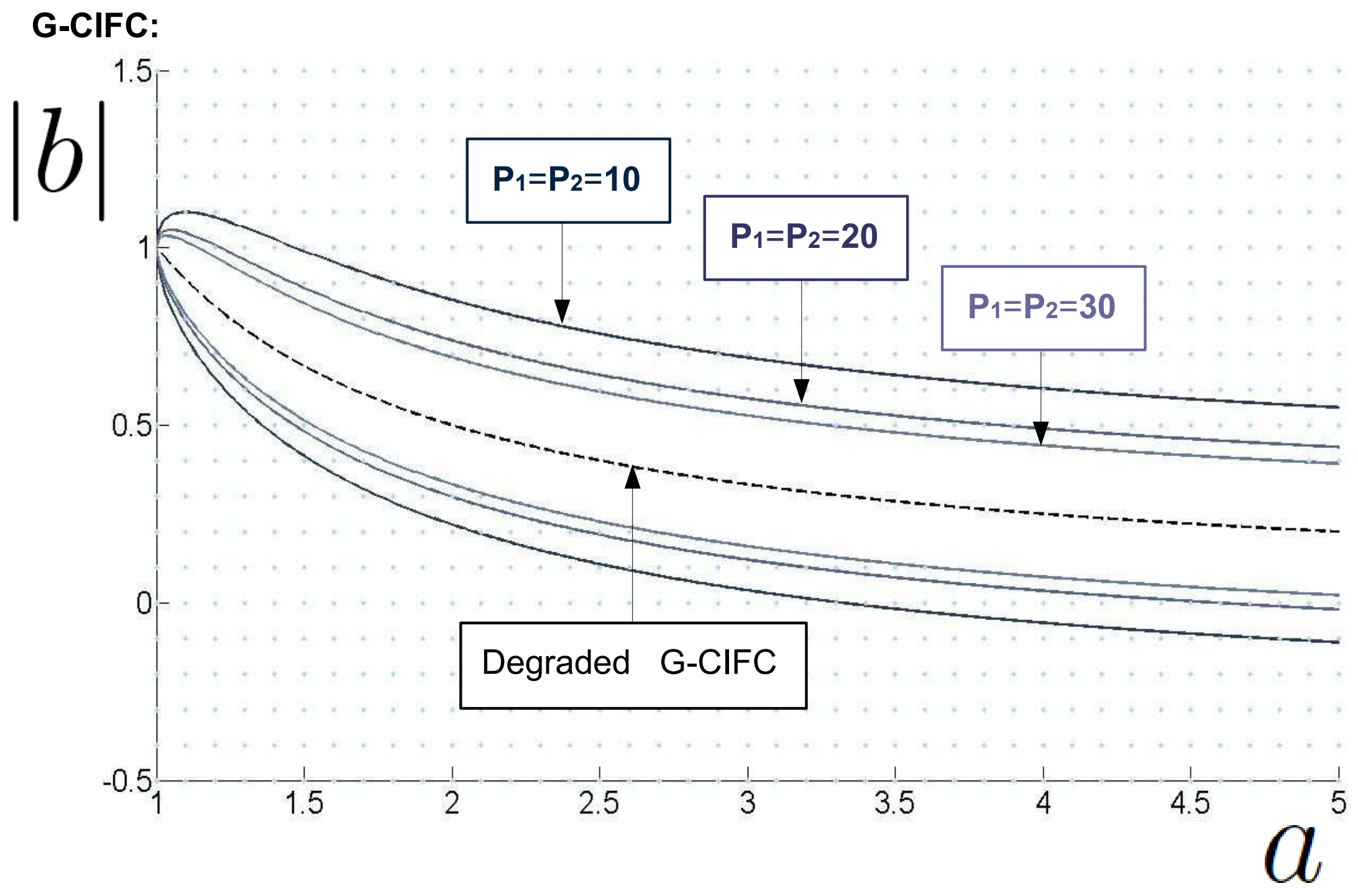

Fig. 6. Condition (33) for different values of $P_1 = P_2 = P$ for a G-CIFC with $a \in \mathbb{R}$ and $(a, |b|) \in [-0.5, 1, 5] \times [1, 5]$.

To provide some insight on the achievability conditions of Thm. V.1, we focus on the condition in (31a). When (31a) is verified, scheme (E) in Section IV-E achieves the "strong interference outer" bound point for $\alpha = 0$ in (4): to achieve more points on the "strong interference" outer bound Th. II.2 stricter conditions are necessary; to achieve all the points on the outer bound, both conditions (31a) and (31b) must be verified.

A representation of the region where the condition in (31a) holds is depicted in Fig. 6 for the case $a \in \mathbb{R}$ and $P_1 = P_2 = P$ with increasing $P$, in which case (31a) becomes

$$P(P+1)|1 - a|b||^2 \geq (|b|^2 - 1)(P + 1 + |a|^2 P). \tag{33}$$

We observe that, as $P$ increases, the region where the condition in (33) is not verified shrinks. Indeed, as $P \to \infty$, the condition in (33) is always verified unless the channel is degraded (i.e., $a|b| = 1$). For a degraded channel with "strong interference", the primary receiver is able to reconstruct $Y_1$ from $Y_2$ once $W_2$ has been decoded, as seen in (14). This means that $U_{1c}$ may be decoded at the primary receiver with no rate penalty for the cognitive user. Under this condition, the scheme with a common cognitive message and a private primary one seems a natural choice, reminiscent of the capacity achieving scheme in the degraded BC. Despite this intuition, in a degraded channel with large power $P$, $\lambda_{\text{Costa 1}}$ approaches $\lambda_{\text{Costa 2}}$ (similarly to the case depicted in Fig. 4) and thus the maximum of the rate $R_1$ in (26a) approaches the minimum of the rate $R_2$ in (26b). This rate penalty for the $R_2$-bound prevents us from achieving the "strong interference" outer bound point for $\alpha = 0$ in (4) when $a|b| = 1$.

Another consideration provides further insight on the condition in (31a): take a channel where

$$\frac{1}{|b|} = a\frac{P_1}{P_1+1}. \tag{34}$$

Then, as $P_1 \to \infty$ in (34) and for $\alpha = 0$, this condition approaches the degraded condition $a|b| = 1$. For this choice of $a$, $Y_2$ may be rewritten as $Y_2 = |b|U_{1c} + Z_2$, so that the $R_2$-bound of (26b) for $\alpha = 0$ becomes

$$R_2 \le I(Y_2, U_{1c}; X_2) = I(U_{1c}; X_2) = \mathcal{C}\left(\frac{P_2}{|b|^2 P_1}\right).$$

This observation reveals an interesting aspect of the RV $U_{1c}$. $U_{1c}$ is DPC coded against $X_2$ with the objective to remove (some of) the interference created by $X_2$ at $Y_1$. However, decoder 2 is not interested in removing $X_2$ from $Y_2$ (it must decode $X_2$!). Hence, for decoder 2, $U_{1c}$ acts as "side information" when decoding $X_2$. Now, both $U_{1c}$ and $Y_2$ contain $X_2$, but for this specific choice of parameters $Y_2$ is a noisy version of $U_{1c}$. This shows why the scheme performs poorly close to the degraded line: there is no gain for receiver 2 from having two observations (i.e., $Y_2$ and $U_{1c}$) of the intended message $X_2$ as they are noisy versions of each other.

*B. New capacity results for the S-G-CIFC.*

**Theorem V.3. Capacity for S-G-CIFC.** *For an S-G-CIFC (i.e., $a = 0$) with*

$$|b| \quad \le \quad \sqrt{1 + P_2\left(1 - \frac{P_1}{P_1+1}\right)} \tag{35}$$

*or with*

$$|b| \ge \sqrt{P_1P_2 + P_2 + 1} + \sqrt{P_1P_2} \tag{36}$$

*Th. III.6 is tight.*

*Proof:* When $|b| \le 1$, capacity is known so we focus only on the case $|b| > 1$. By setting $a = 0$ in Th.V.1 we obtain that scheme (E) with $\lambda = \lambda_{\text{Costa 1}}$ achieves the "strong interference" outer bound for

$$(|b|^2 - 1)(1 + P_1) \le \min\{P_2, P_2(1 + P_1)\} = P_2,$$

which is equivalent to (35).

Scheme (E) with $\lambda = 0$ achieves

$$\begin{aligned}
R_1 &\le I(Y_1; U_{1c}) - I(U_{1c}; X_2) = I(Y_1; U_{1c}) = \mathcal{C}\left(\tfrac{\alpha P_1}{1+\bar{\alpha}P_1}\right), \\
R_2 &\le I(Y_2, U_{1c}; X_2) = I(Y_2; X_2|U_{1c}) = \mathcal{C}\left((\sqrt{P_2} + \sqrt{\bar{\alpha}|b|^2 P_1})^2\right), \\
R_1 + R_2 &\le I(Y_2; X_2, U_{1c}) = \mathcal{C}\left(\alpha|b|^2 P_1 + (\sqrt{P_2} + \sqrt{\bar{\alpha}|b|^2 P_1})^2\right).
\end{aligned}$$

In this case the MIMO-BC outer bound may be achieved when the sum rate outer bound (17c) is redundant, that is, if

$$\begin{aligned}
&1 + P_2 + |b|^2 P_1 + 2\sqrt{\bar{\alpha}|b|^2 P_1 P_2} \ge \frac{1+P_1}{1+\bar{\alpha}P_1}(1 + P_2 + |b|^2 P_1 - \alpha|b|^2 P_1 + 2\sqrt{\bar{\alpha}|b|^2 P_1 P_2}) \qquad \forall \alpha \in [0,1] \\
&\iff \quad |b|^2 \ge 1 + P_2 + 2\sqrt{\bar{\alpha}|b|^2 P_1 P_2} \qquad \forall \alpha \in [0,1] \\
&\iff \quad |b|^2 \ge 1 + P_2 + 2\sqrt{|b|^2 P_1 P_2}
\end{aligned}$$

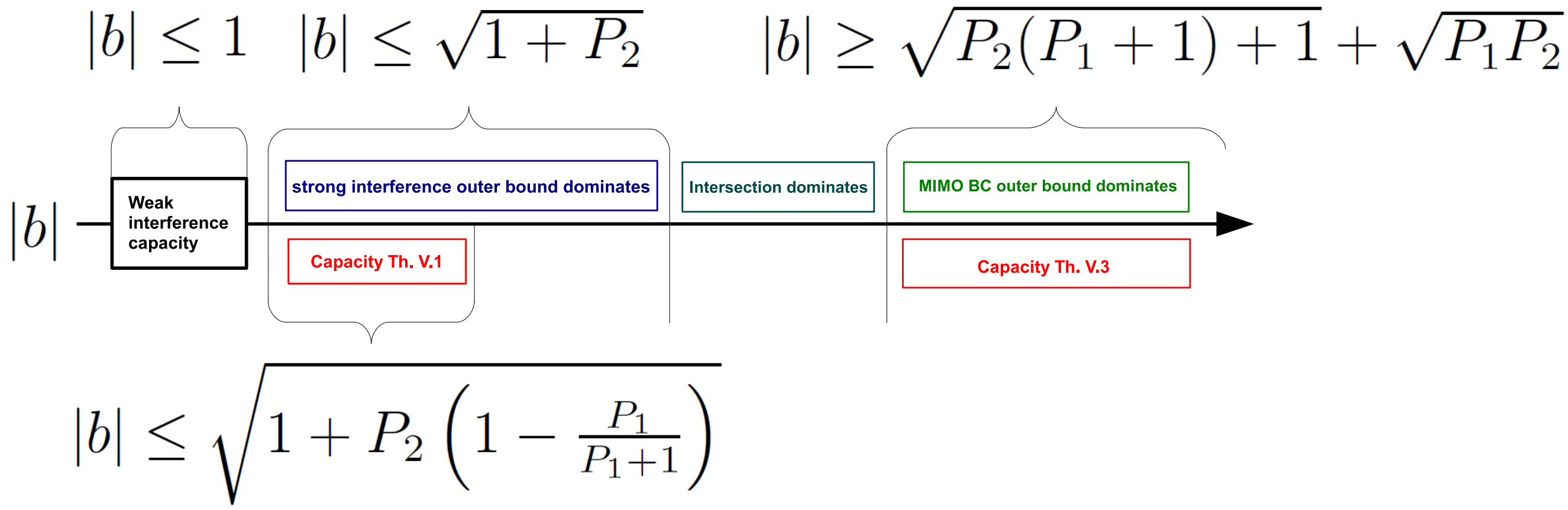

Fig. 7. A schematic representation of the capacity results for the S-G-CIFC in Th. V.3.

which corresponds to (36).

■

A representation of the region where capacity is known for the S-G CIFC is depicted in Fig. 7. Capacity remains unknown for

$$\sqrt{1+P_2\left(1-\frac{P_1}{P_1+1}\right)} \leq |b| \leq \sqrt{1+P_2+P_1P_2}+\sqrt{P_1P_2}$$

## VI. Capacity to within a constant gap

In the last couple of years a novel approach to the difficult task of determining the capacity region of a multi-user Gaussian network has been suggested. Rather than proving an equality between inner and outer bounds, the authors of [39] (and references therein) advocate a powerful new method for obtaining achievable rate regions that lie within a bounded distance from capacity region outer bounds, thereby determining the capacity region to *within a constant gap* for any channel configuration. Two measures are used to determine the distance between inner and outer bounds: the additive gap and the multiplicative gap. An additive gap corresponds to a finite difference between inner and outer bound, while a multiplicative gap corresponds to a finite ratio. The additive gap is useful at high SNR, where the difference between inner and outer bound is small in comparison to the magnitude of the capacity region, while the multiplicative gap is useful at low SNR, where the ratio between inner and outer bounds is a more indicative measure of their distance. In this section we show the capacity to within an additive gap of half a bit/s/Hz per real dimension and to within a multiplicative gap of a factor two. We also determine additional constant gap results that suggest which strategies approach the “strong interference” outer bound in different parameter regimes. Since the expressions of the BC-based outer bound of Th. III.2 and of the BC-DMS-based outer bound of Th. III.3 involve many parameters over which to optimize, it is not analytically straightforward to determine conditions for achievability; for this reason we restrict our attention to the “strong interference” outer bound of Th. II.2. These results are derived for the complex-valued channel and rather than for the real-valued channel as done in [7].

**Theorem VI.1.** ***Additive gap.*** *Capacity is known to within half a bit/s/Hz per real dimension.*

*Proof:* The capacity for weak interference ($|b| \leq 1$) was determined in [6], so we only need to concentrate on the strong interference regime ($|b| > 1$). We show the achievability of the "strong interference" outer bound in (4) to within a constant additive gap using the scheme (E) of Section IV-C with the assignment in (23). The assignment proposed in (23) is inspired by the capacity achieving scheme for deterministic channels in [11], where we showed that setting $U_c = Y_c$, $c \in \{1, 2\}$, is optimal. In a noisy channel, it is not possible to choose $U_c = Y_c$; we mimic this by setting $U_c \sim Y_c$, $c \in \{1, 2\}$.

Consider the achievable rate region in (22) and note that

$$\mathrm{Var}[X_1 + aX_2] = P_1 + |a|^2 P_2 + 2\mathrm{Re}\{a\}\sqrt{\bar{\alpha} P_1 P_2},$$
$$\mathrm{Var}[|b|X_1 + X_2] = |b|^2 P_1 + P_2 + 2\sqrt{\bar{\alpha}|b|^2 P_1 P_2}.$$

The inequality in (22c) follows by choosing

$$\begin{aligned}\rho_{pb} &= \arg\min_\rho I(U_1; U_2|X_2) \\ &= \arg\min_\rho |\mathbb{E}[U_1 U_2^* |X_2]|^2 \\ &= \arg\min_\rho \left| |b|P_1\alpha + \rho\sqrt{\sigma_{1pb}^2 \sigma_{2pb}^2} \right|^2 \\ &= -\min\left\{1, \frac{|b|P_1\alpha}{\sqrt{\sigma_{1pb}^2 \sigma_{2pb}^2}}\right\}.\end{aligned}$$

With $\sigma_{2pb}^2 = 0$ and $\sigma_{1pb}^2 = 1$ in (22) we have

$$R_1 \leq \log(1 + \alpha P_1) - \mathrm{GAP}(\alpha), \tag{37a}$$

$$R_1 + R_2 \leq \log(1 + \mathrm{Var}[|b|X_1 + X_2]) - \mathrm{GAP}(\alpha), \tag{37b}$$

with $\mathrm{GAP}(\alpha)$ bounded as

$$\mathrm{GAP}(\alpha) = \log\left(1 + \frac{\mathrm{Var}[X_1 + aX_2]}{1 + \mathrm{Var}[X_1 + aX_2]}\right) \leq \log(2) = 1,$$

as claimed. Notice that with $\sigma_{2pb}^2 = 0$, the $R_2$-bound in (22b) is equivalent to the sum-rate outer bound in (4b) and it is thus redundant. ■

To prove the multiplicative gap result, we utilize a looser version of Th. III.1 that we present in the next lemma.

**Lemma VI.2. "Piecewise linear strong interference" outer bound.** *The outer bound of Th.III.1 for $|b| > 1$ is contained in the region $\mathcal{R}^{(\mathrm{PL-SI})}$ defined as:*

$$R_1 \leq \mathcal{C}(P_1), \tag{38a}$$

$$R_1 + R_2 \leq \mathcal{C}\left((\sqrt{|b|^2 P_1} + \sqrt{P_2})^2\right). \tag{38b}$$

*Proof:* The bound in (38a) (respectively (38b)) is obtained by considering the maximum value of (6a) (respectively (6c)) over $\alpha \in [0,1]$. ∎

The region $\mathcal{R}^{(\mathrm{PL-SI})}$ in (38) has two Pareto optimal points:

$$A = \left(0, \mathcal{C}\left((\sqrt{|b|^2 P_1} + \sqrt{P_2})^2\right)\right), \tag{39a}$$

$$B = \left(\mathcal{C}(P_1), \mathcal{C}((\sqrt{|b|^2 P_1} + \sqrt{P_2})^2) - \mathcal{C}(P_1)\right). \tag{39b}$$

The point A is on the boundary of the "strong interference" outer bound region $\mathcal{R}^{(\mathrm{SI})}$ of Th.II.2 while Point B has the same $R_1$-coordinate as the point for $\alpha = 0$ in $\mathcal{R}^{(\mathrm{SI})}$, given by

$$C = \left(\mathcal{C}(P_1), \mathcal{C}(|b|^2 P_1 + P_2) - \mathcal{C}(P_1)\right), \tag{40}$$

but lies outside $\mathcal{R}^{(\mathrm{SI})}$. However the two points are no more than one bit away, i.e., $R_2^{(B)} \leq \log(2) + R_2^{(C)}$, as we will show later.

**Theorem VI.3. Multiplicative gap.** *For a Gaussian C-IFC, the capacity is known to within a factor two.*

*Proof:* The capacity for weak interference ($|b| \leq 1$) was determined in [6], thus we only need to concentrate on the strong interference regime ($|b| > 1$).

*Outer bound:*

We use the "piecewise linear strong interference" outer bound of Lemma VI.2, in particular we rewrite the outer bound as

$$\begin{aligned} R_2 &\leq \log\left(1 + |b|^2 P_1 + P_2 + 2\sqrt{|b|^2 P_1 P_2}\right) - R_1 \\ &\triangleq R_2^{(\mathrm{PL-SI})}(R_1), \end{aligned} \tag{41}$$

for $R_1 \in [0, \log(1+P_1)]$.

*Achievability to within a factor two:* Consider the following TDMA strategy. The rate-point

$$(R_1, R_2) = (\log(1+P_1), 0),$$

is achievable by silencing the primary transmitter, while the rate-point A in (39a) is achievable by beamforming. Hence, the following region is achievable by time sharing

$$\begin{aligned} R_2 &\leq \left(1 - \frac{R_1}{\log(1+P_1)}\right) \log(1 + (\sqrt{|b|^2 P_1} + \sqrt{P_2})^2) \\ &\triangleq R_2^{(\mathrm{tdma})}(R_1). \end{aligned} \tag{42}$$

The multiplicative gap is given by the smallest $M \geq 1$ for which

$$M R_2^{(\mathrm{tdma})}(R_1/M) \geq R_2^{(\mathrm{PL-SI})}(R_1), \tag{43}$$

that is

$$\left(1 - \frac{R_1}{M\log(1+P_1)}\right) M \log(1 + (\sqrt{|b|^2 P_1} + \sqrt{P_2})^2) - \log\left(1 + (\sqrt{|b|^2 P_1} + \sqrt{P_2})^2\right) + R_1 \geq 0. \tag{44}$$

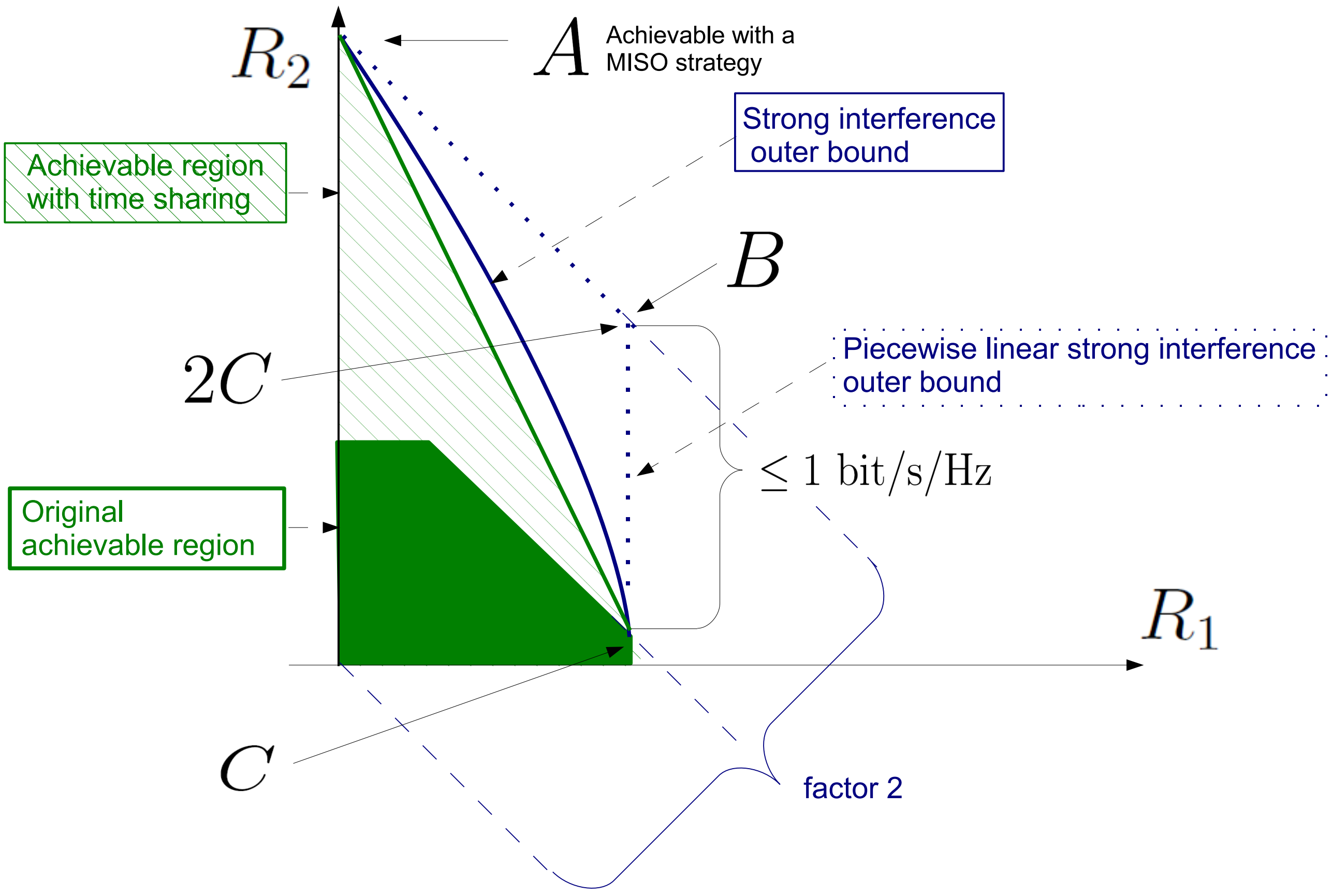

Fig. 8. A graphical representation of Th. VI.1 and Th. VI.3.

The LHS of (44) is a linear function of $R_1$ and thus has at most one zero. From this, it follows that the inequality in (44) is verified for every $R_1 \in [0, \log(1+P_1)]$ if it is verified at the boundary points of the interval. For $R_1 = 0$, the inequality is verified for $M \geq 1$ while for $R_1 = \log(1+P_1)$ it is verified if $M \geq 2$; thus the smallest $M$ for which (44) is verified for all channels is $M = 2$. ■

We remark here that we consider the multiplicative gap as the ratio of the outer bound over the inner bound; as originally introduced in [40] the multiplicative gap is defined as the ratio between the inner bound over the outer bound.

A schematic plot of the proofs of Th. VI.1, Th. VI.3 and Lemma VI.2 is provided in Fig. 8. The green hatched area represents the achievable rate region with scheme (E) in (22) which lies to within half a bit/s/Hz per real dimension from the "strong interference" outer bound of (4), illustrated by a solid blue line. The green cross-hatched area represents the achievable rate region with time sharing in (42) while the green dashed line is the region in (42) multiplied by a factor two, which contains the "piecewise linear strong interference outer bound" in (38), illustrated by a dotted blue line.

TABLE III
FURTHER CONSTANT GAP RESULTS

| What | Scheme | Regime | gap |
|---|---|---|---|
| perfect interference cancelation | (E) | $P_2(1+\|a\|^2-2\mathrm{Re}\{a\}\|b\|)\geq(b^2-1)(P_1+1)-P_1P_2\|1-\|a\|b\|^2$ | .5 |
| non perfect interference cancelation | (E) | $\|b\|>1$ and $\|b\|^2P_1\leq P_2$ | 1.87 |
| cognitive | (A) | $\|b\|\leq 1$ and $\|b\|^2P_1>P_2$ | 1 |
| broadcasting | (A) | $\|b\|>1$ and $\|b\|^2P_1>P_2$ | 1.5 |
| interference stripping | (D) | $\|a\|\geq 1$, $\|b\|>1$ and $\|b\|^2P_1\leq P_2$ | 1.5 |

### *A. Additional constant gap results*

In this section we provide additional additive gap results for specific subsets of the parameter region. Our aim is to provide insights on the relationship between inner and outer bounds for the region where capacity is still unknown.

**Corollary VI.4.** *The additive gaps between inner and outer bound in Table III are achievable under the prescribed conditions.*

*Proof:* See Appendix F. ■

In particular we consider four transmission strategies and show where they achieve capacity to within a constant gap:

- **Perfect interference cancelation.** Scheme (E) with Costa's DPC achieves the "strong interference" outer bound to within a constant gap in a larger parameter region than the "primary decodes cognitive" regime, where it achieves capacity.
- **Non perfect interference cancelation.** The scheme (E) with a specific DPC strategy achieves the "strong interference" outer bound to within a constant gap when the SNR is larger the INR at the primary receiver. The choice of DPC differs from Costa's and it favors the decoding of the common cognitive message at the primary decoder and enhances the performance for channel parameters close to the degraded G-CIFC.
- **Cognitive broadcasting.** When the INR is larger that the SNR the primary receiver, scheme (A) achieves a constant gap from the outer bound in both the "weak" and the "strong interference" regime. In this scheme the primary transmitter is silent and the cognitive transmitter acts as a broadcast trasmitter.
- **Interference stripping.** Scheme (D) achieves the "strong interference" outer bound to within a constant gap in a larger parameter region than the "very strong interference" regime, where it achieves capacity. In this scheme both decoders decode both messages as in a compound MAC.

## VII. NUMERICAL RESULTS

We now revisit each of the previous sections and provide numerical examples of the results therein. In the following we restrict ourselves to real-valued input/output G-CIFC so as to reduce the dimensionality of the search space for the optimal parameter values.

### *A. Section III: Outer bounds*

In Section III we introduced the tightest outer bound for a GCIFC in “strong intereference”, obtained as the intersection of the “strong interference” outer bound of Th. II.2 and the BC based outer bound of Th. III.2. This outer bound has a simple closed form expression for the degraded G-CIFC and the S-G-CIFC: Fig. 9 and Fig. 10 present the result of Corollaries III.5 and III.6 respectively, where the intersection of the “strong interference” outer bound and the BC-based outer bound for the degraded G-CIFC and the S-G-CIFC is derived. Note that we chose two channels where the two bounds intersect for some $R_1 \in (0, \mathcal{C}(P_1)]$ and neither bound strictly includes the other. The two outer bounds coincide at the point A in (39a). The maximum rate $R_1$ in the “strong interference” outer bound and the BC-based outer bound for the S-G-CIFC are the same: in this channel transmitter 2 does not influence the output at receiver 1 and hence full receiver cooperation does not increase the maximum attainable rate $R_1$.

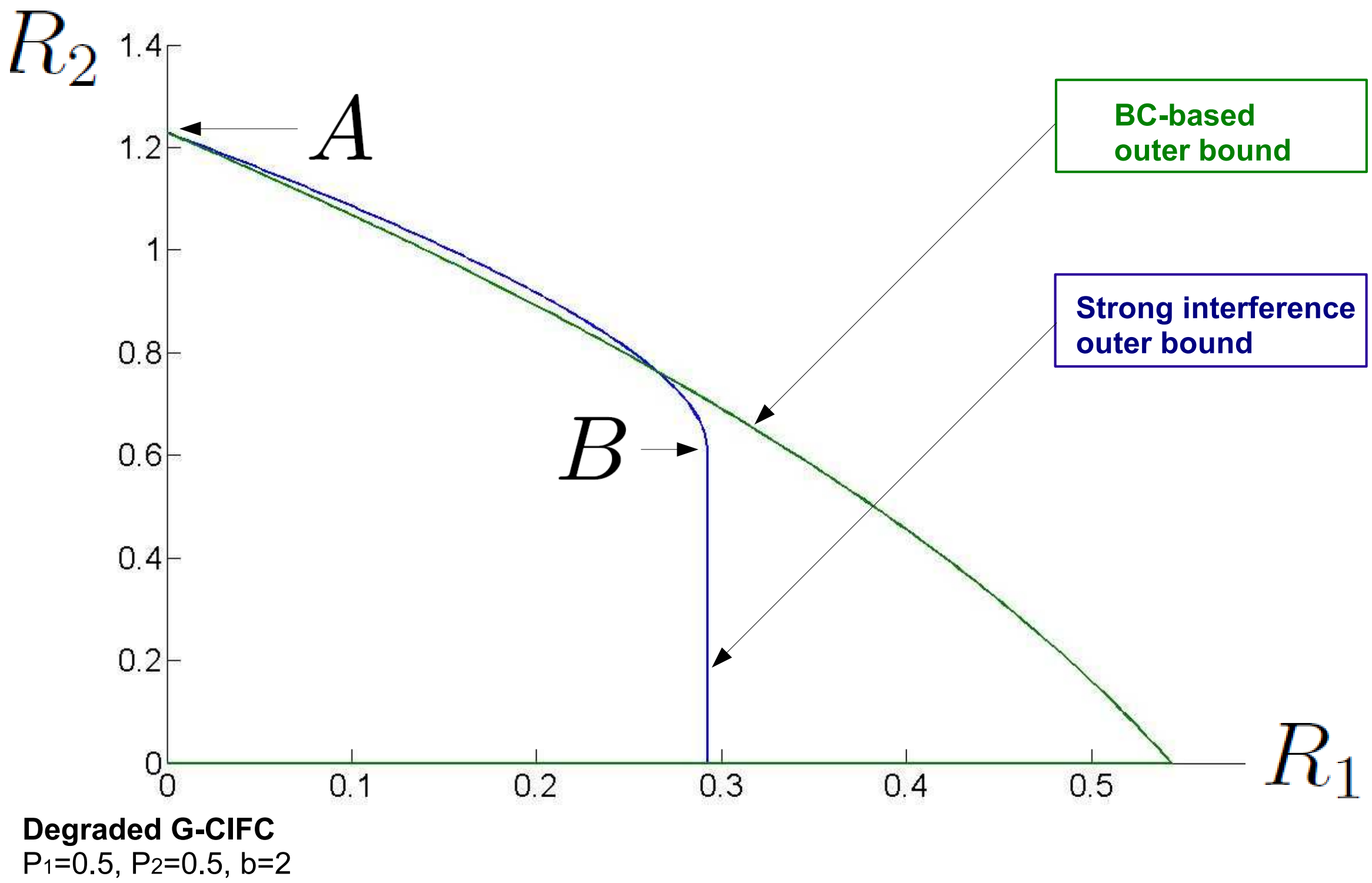

Fig. 9. The “strong interference” outer bound and the BC-based outer bound for the degraded G-CIFC.

For a general G-CIFC the intersection between the “strong interference” and the BC-based outer bound has no simple closed form expression. Consequently, it is difficult to determine where one dominates and find their intersection analytically. In Fig. 11 we show that the two bounds can intersect up to two times.

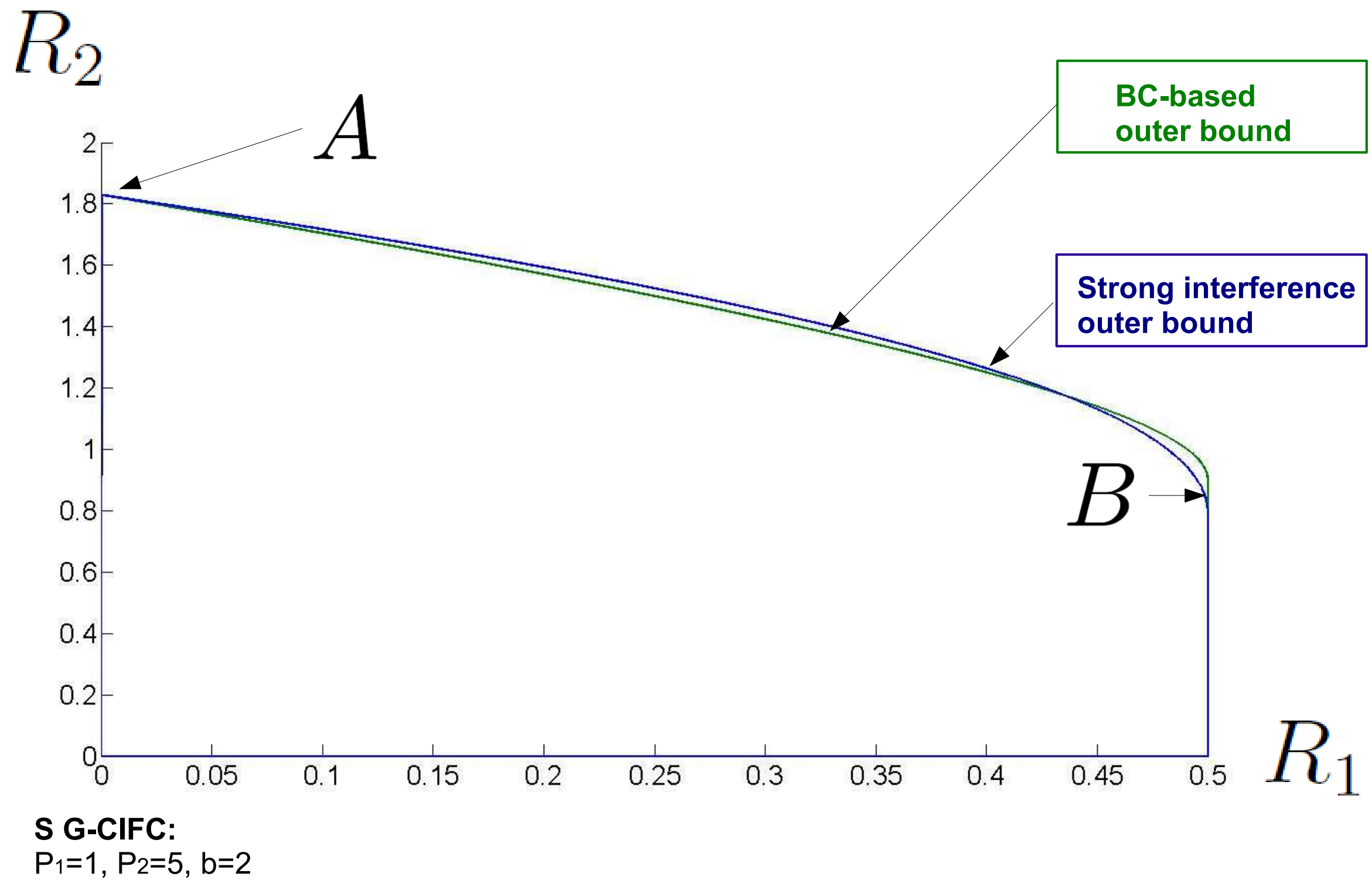

Fig. 10. The "strong interference" outer bound and the BC-based outer bound for theS-G-CIFC (right).

The outer bounds of Th. III.7 are presented in Fig. 12 which shows that these outer bounds may be tighter than either the "strong interference" or the BC-based outer bounds. Unfortunately, in the examples we considered, we did not find an instance where the outer bounds of Th. III.7 were tighter than the intersection of the "strong interference" and the BC-based outer bound. Despite this, we believe that our approach in transforming the channel provides a general, useful tool to derive outer bounds for channels with cognition.

### B. *Section IV: Inner bounds*

In Section IV we introduced the $\mathcal{R}^{(\mathrm{RTD})}$ achievable rate region and derived six sub-schemes from this general inner bound: in the following we plot these sub-schemes for the degraded channel, the S channel and a general G-CIFC. The "strong interference" and the "weak interference" outer bounds are provided for reference. Note that both the achievable rate regions and the outer bounds are expressed as a function of one parameter only, $\alpha \in [0,1]$, that controls the amount of cooperation between the cognitive and the primary transmitters.

We begin by considering the degraded G-CIFC in Fig. 13. The scheme that yields the largest achievable rate region in scheme (E) with the choice $\lambda = \lambda_{\mathrm{Costa}\ 1}$. Despite its superior performance (to other presented schemes) we may analytically show that this scheme cannot achieve either the "strong interference" or the BC-based outer bound.

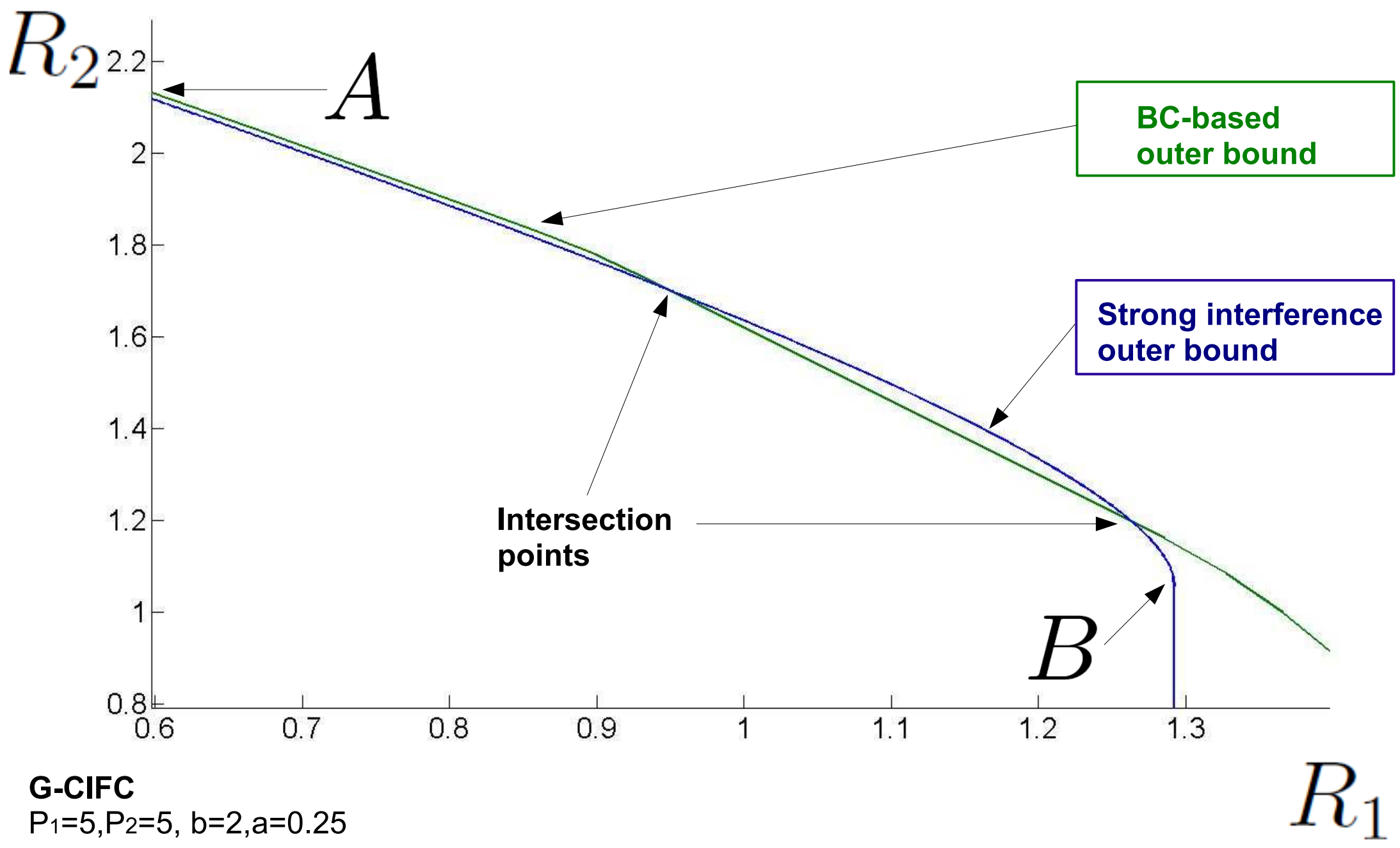

Fig. 11. The "strong interference" outer bound and the MIMO BC outer bound for a general G-CIFC in proximity to point C.

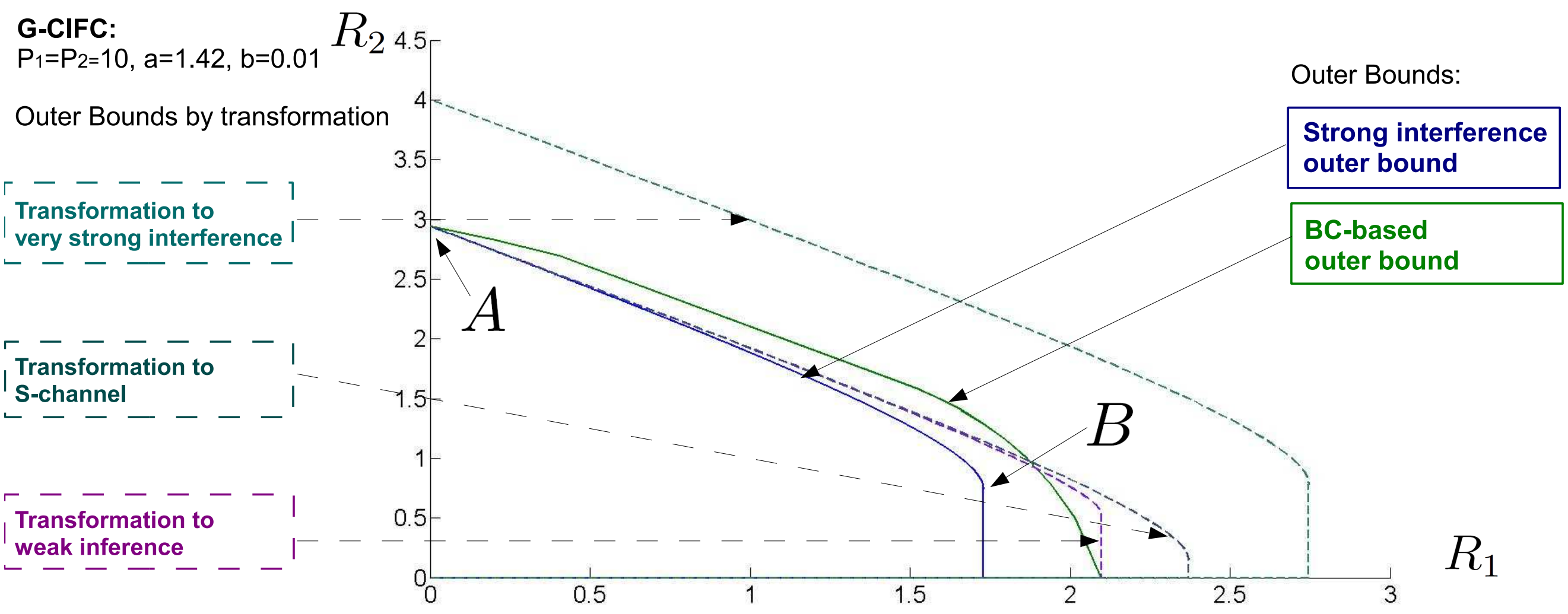

Fig. 12. The outer bounds of Lemma III.7 alongside the "strong interference" outer bound and the BC-based outer bound.

Both schemes (A) and (B) treat the interference at noise at receiver 1 and thus the maximum $R_1$ may be achieved only by silencing transmitter 2. For this reason $R_2 \to 0$ as $R_1 \to \mathcal{C}(P_1)$ for these two schemes.

The channels parameters are chosen to show how scheme (E) with the choice $\lambda = \lambda_{\text{Costa }1}$ achieves the "strong interference" outer bound for a subset of $R_1 \in (0, \mathcal{C}(P_1)]$ where the inequality in (32a) holds. The figure also shows how, in the S channel, it is possible to achieve the outer bound for $R_1 = \mathcal{C}(P_1)$ with scheme (E) without DPC. This is possible only in this channel, since $X_2$ does not influence $Y_1$ and no rate loss occurs at the cognitive receiver by treating the interference as noise. Note that scheme (D) performs the worst among all the achievable schemes: in this scheme the cognitive receiver is *required* to decode both messages – a very stringent constraint since $Y_1$ does not contain $X_2$. In particular, $R_2 \to 0$ when $R_1 \to \mathcal{C}(P_1)$ as in schemes (A) and (B): this is so because $R_1 = \mathcal{C}(P_1)$ may be achieved with scheme (D) only for $Y_1$ independent of $X_2$.

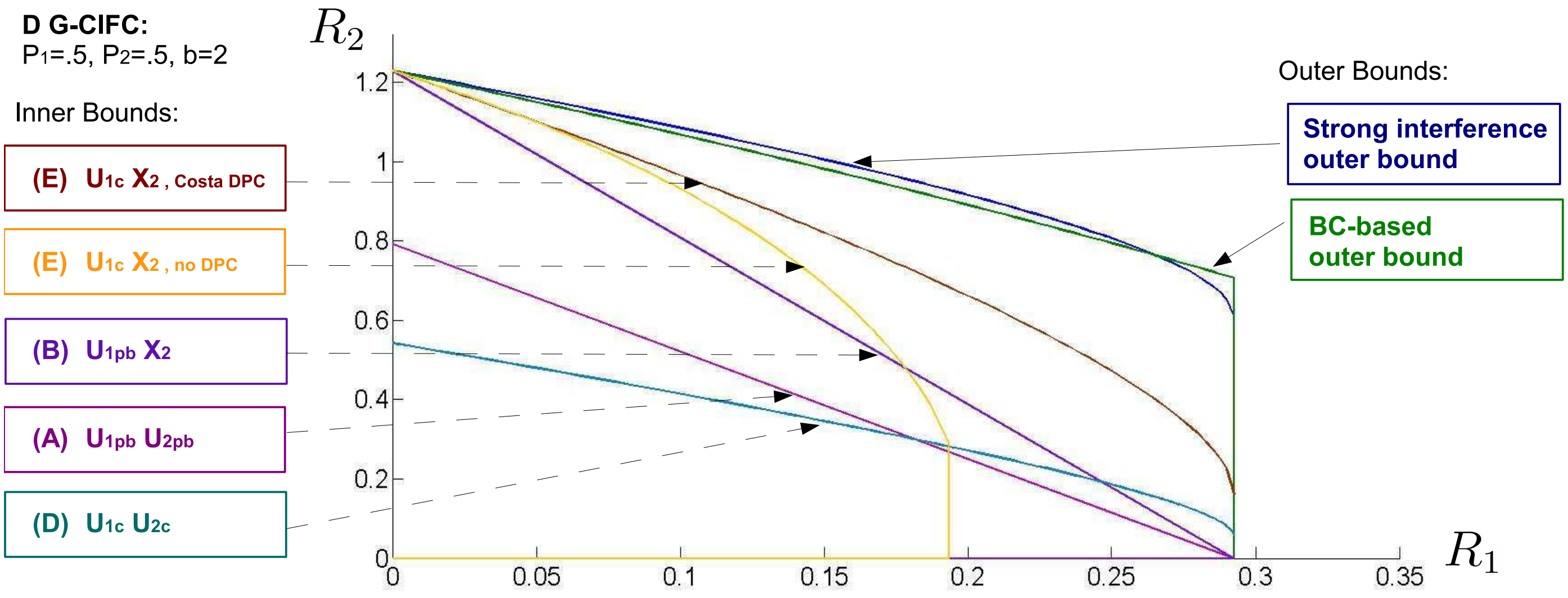

Fig. 13. The different schemes of Section IV for the degraded G-CIFC.

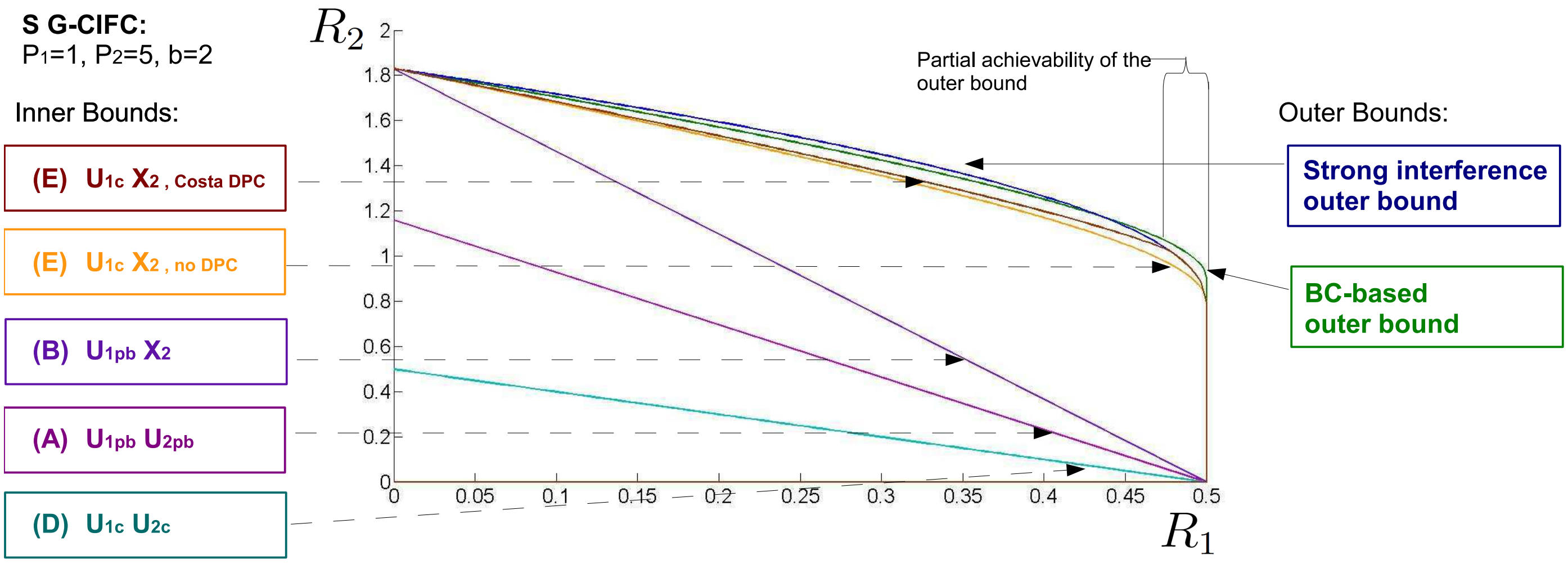

Fig. 14. The achievable schemes of Section IV for the S-G-CIFC.

A general G-CIFC in Fig. 15. In this example, scheme (E) with $\lambda = 0$ performs better than the scheme with $\lambda = \lambda_{\text{Costa 1}}$ for small $R_1$ while the opposite is true for large $R_1$. This is the first instance in which we see that a single choice of $\lambda$ does not yield the largest inner bound: for small INR, is better for the cognitive user to treat

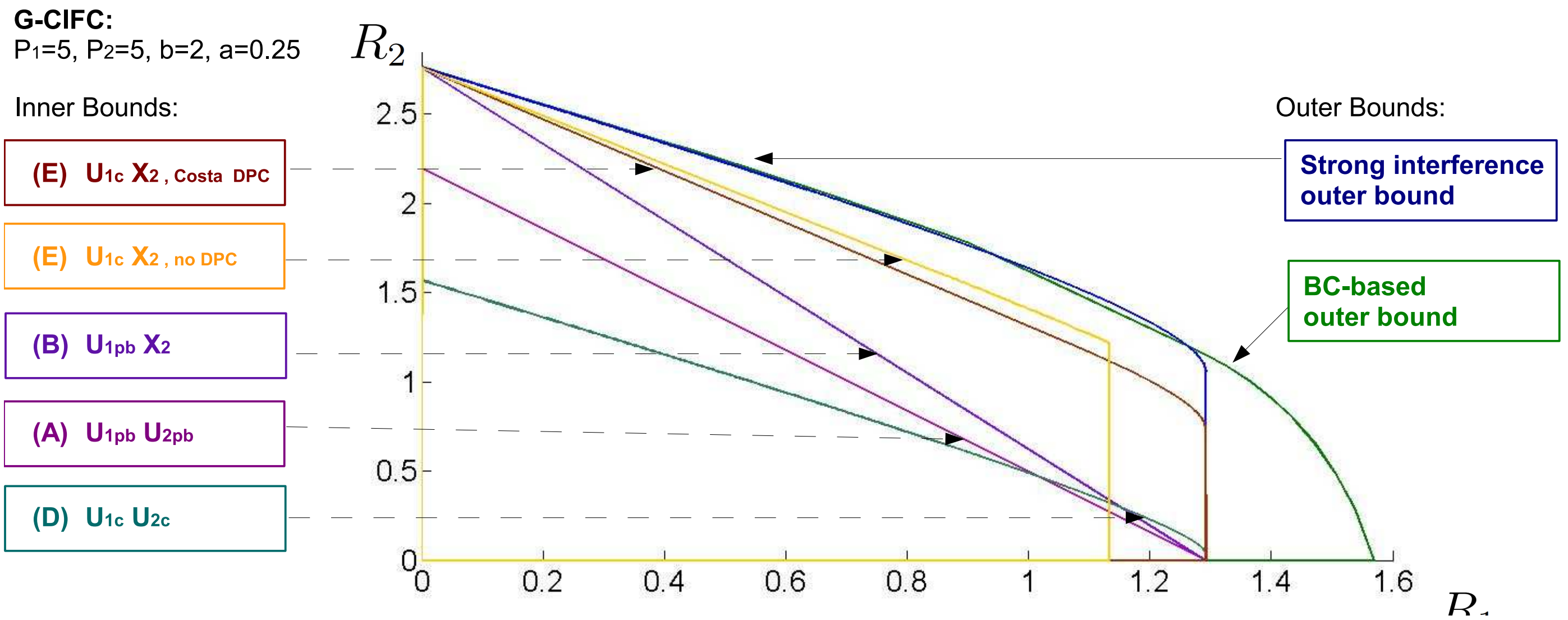

Fig. 15. The achievable schemes of Section IV for a general G-CIFC.

the interference as noise, while for large INR it is more advantageous to perform Costa's DPC.

From Section III-B we know that, for $|b| > 1$, the cognitive receiver can decode the primary message with no additional rate penalties; this may be observed by comparing scheme (E) with Costa's DPC and scheme (B). The primary message is private in both schemes while the cognitive message is common in scheme (E) and private in scheme (B). Since the primary receiver can decode the cognitive message at no cost, scheme (E) with Costa's DPC achieves larger rates than scheme (B). When no DPC is used ($\lambda = 0$) in scheme (E), the cognitive receiver observes an equivalent additive Gaussian noise noise of variance $1 + |a|^2 P_2$: for this region rate $R_1$ is always bounded by $R_1 \leq \mathcal{C}(P_1/(1 + |a|^2 P_2))$ and thus scheme (B) outperforms scheme (E) with no DPC in the interval $R_1 \in [\mathcal{C}(P_1/(1 + |a|^2 P_2), \mathcal{C}(P_1)]$.

The scheme (F) in Section IV-F unifies capacity achieving schemes in the "very strong interference" and the "primary decodes cognitive" regimes. It is possible that by unifying the two schemes, we may show achievability in a larger region than the union of the two regimes. Unfortunately determining the achievability conditions in closed form is not straightforward as it requires the optimization of the four parameters in (30). In Fig. 16 we show through numerical evaluation that scheme (F) indeed achieves a larger region of than the union of the schemes (E) and (D). Whether this scheme achieves capacity for a larger parameter region remains an open question.

### C. *Section V: New capacity results*

In Section V we determine new capacity results for the "primary decodes cognitive" regime both for a general G-CIFC and the S-G-CIFC. In Fig. 17 we plot the "primary decodes cognitive" regime in (31) for different transmitter powers $P_1 = P_2 = P$. Note that the "weak interference" and the "very strong interference" regimes do not depend on $P$ so their plot does not vary. As the power $P$ increases, the "primary decodes cognitive" region expands from the line $|b| = 1$ to cover a larger region around the degraded line. Interestingly the "primary decodes cognitive" regime intersects with the "very strong interference" regime, thus showing that the "strong interference" outer bound may be achieved with two different transmission schemes for some channels.

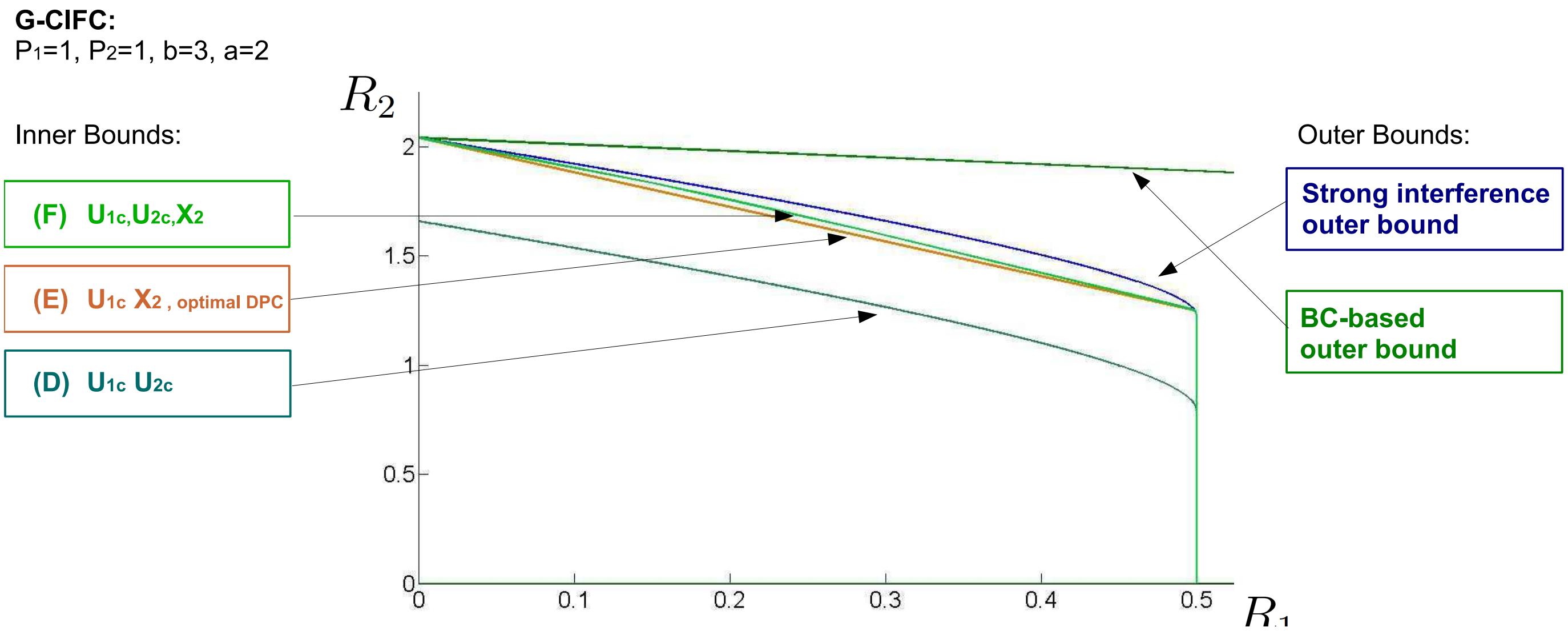

Fig. 16. The achievable region of schemes (D), (E) and (F) for a general G-CIFC.

In a similar fashion, Fig. 18 shows the capacity results of Th. V.3 for the case $P_1 = P_2 = P$ on the plane $P \times |b|$. For equal transmitter powers, the conditions in (35) and in (36) become

$$|b|^2 \leq \frac{2P+1}{P+1} \approx 2 \tag{45a}$$

$$|b|^2 \geq P + \sqrt{P^2+P+1} \approx 2P \tag{45b}$$

and these two asymptotic behaviors are clearly visible in Fig. 18.

## D. *Section VI: Capacity to within a constant gap*

In Th. VI.1 we established the capacity of a general G-CIFC to within half a bit/s/Hz with a specific assignment in the region of (22). This specific assignment was chosen to mimic the capacity achieving scheme in the deterministic CIFC of [11] and partially optimized to yield the smallest gap between inner and outer bound. A larger achievable rate region could be obtained by considering the scheme C with the assignment of RV

$$U_{1pb} = X_1 + c_1 X_2 + Z_{1pb} \tag{46a}$$

$$U_{2pb} = c_2 X_1 + X_2 + Z_{2pb}, \tag{46b}$$

in (23). The region in (22) considers only the case $c_1 = a$, $c_2 = |b|$ while the assignment in (46) parameterizes any covariance matrix $\mathrm{Cov}\left([X_1\ X_2\ U_{1pb}\ U_{2pb}]\right)$. Unfortunately, this scheme is parametrized by five coefficients and the algebraic optimization of the additional parameters is quite involved.

Instead, in Figure 19, we may use numerical evaluations to investigate the rate improvements that may be obtained with the more general achievable scheme of (46). We consider a degraded G-CIFC with high power and show that this choice of RVs greatly improves on the result in Thm. VI.1. With the assignment in (22) it is not possible to approach the outer bound of Thm. III.5 for large $R_1$. On the other hand, with the more general formulation of the auxiliary RVs in (46), it is possible to greatly reduce the distance between inner and outer bounds.

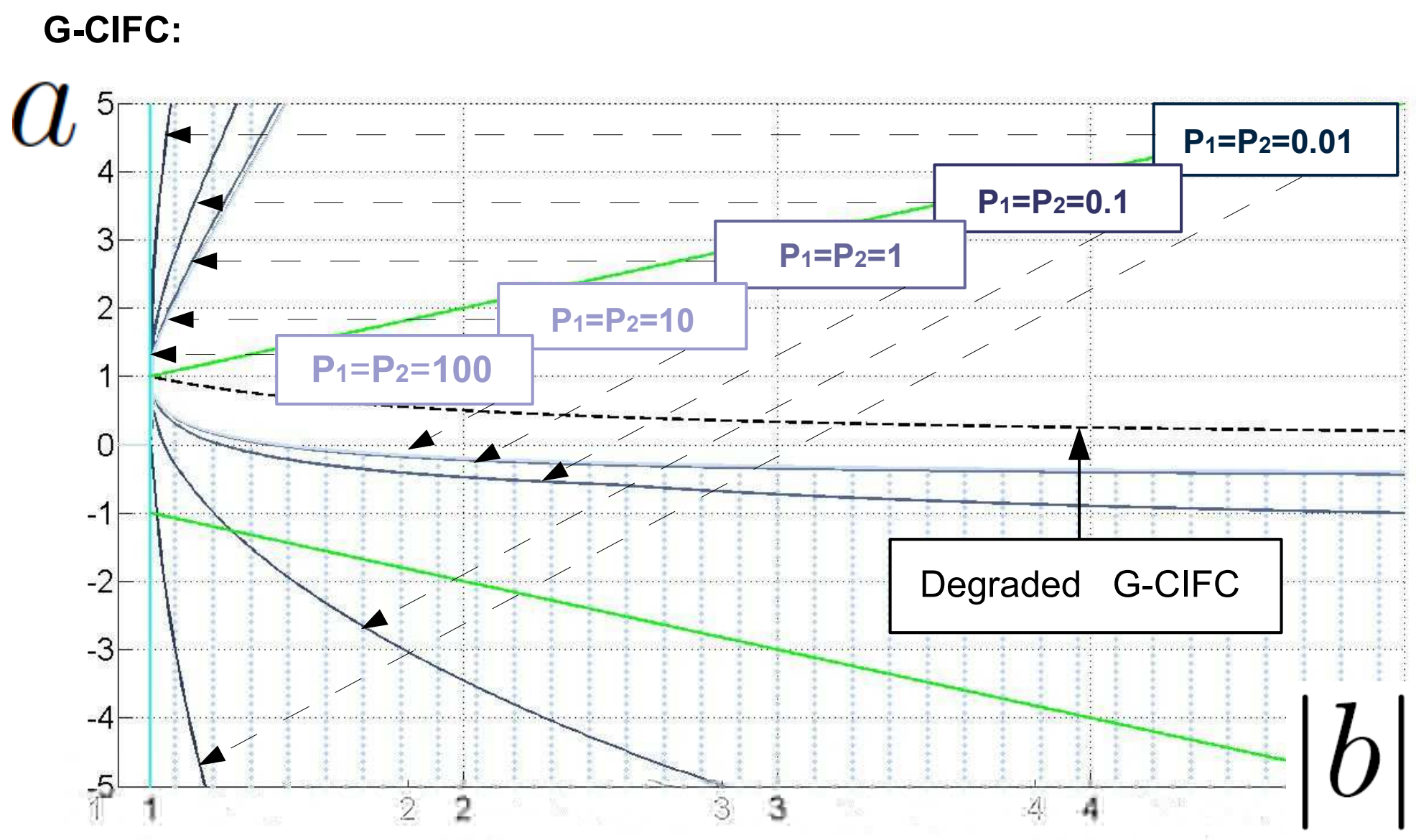

Fig. 17. The "primary decodes cognitive" region for different powers $P_1 = P_2 = P$ for a G-CIFC with $a \in \mathbb{R}$ and $(a, |b|) \in [-5, 5] \times [0, 5]$.

Although the scheme (E) in Section IV-E does not achieve capacity outside the "primary decodes cognitive" regime, we next show by numerical evaluation that scheme (E) is close to optimal for a general channel in "strong interference", especially when considering the union over all $\lambda \in \mathbb{C}$ instead of the choice $\lambda = \lambda_{\text{Costa 1}}$. Fig. 20 shows the position of point

$$D(\lambda) = \Big( (26\text{a}), \min\{(26\text{b}), (26\text{c}) - (26\text{a})\} \Big)$$

in the range $\lambda \in [0, 2\lambda_{\text{Costa 1}}]$, for a fixed $\alpha^{(\text{in})}$, together with the outer bound point C for $\alpha^{(\text{out})} = \alpha^{(\text{in})}$. Under the "primary decodes cognitive" condition, $D(\lambda_{\text{Costa 1}}) = C$ for every $\alpha \in [0, 1]$. However, here we show a channel where the condition in (31a) is not satisfied. In this case the choice $\lambda = \lambda_{\text{Costa 1}}$ minimizes the distance of the $R_1$-coordinate between D and C, but it does not minimize the Euclidean distance between the two points.

The rate improvements that may be obtained with any $\lambda \in \mathbb{C}$ are exemplified in Fig. 21. In this figure we plot the achievable rate regions of (26) obtained for $\lambda = 0$, $\lambda = \lambda_{\text{Costa 1}}$ and any $\lambda \in [0, 2\lambda_{\text{Costa 1}}]$. Unlike Fig. 15, the scheme for $\lambda = \lambda_{\text{Costa 1}}$ strictly outperforms the scheme for $\lambda = 0$; the choice $\lambda \in [0, 2\lambda_{\text{Costa 1}}]$ not only includes the previous regions but improves on the case $\lambda = \lambda_{\text{Costa 1}}$ as well. The inner bound point for $R_1 = 0$ corresponds to point A in (39a) and is always achievable; the inner bound point for $R_1 = \mathcal{C}(P_1)$ may be achieved only for $\lambda = \lambda_{\text{Costa 1}}$.

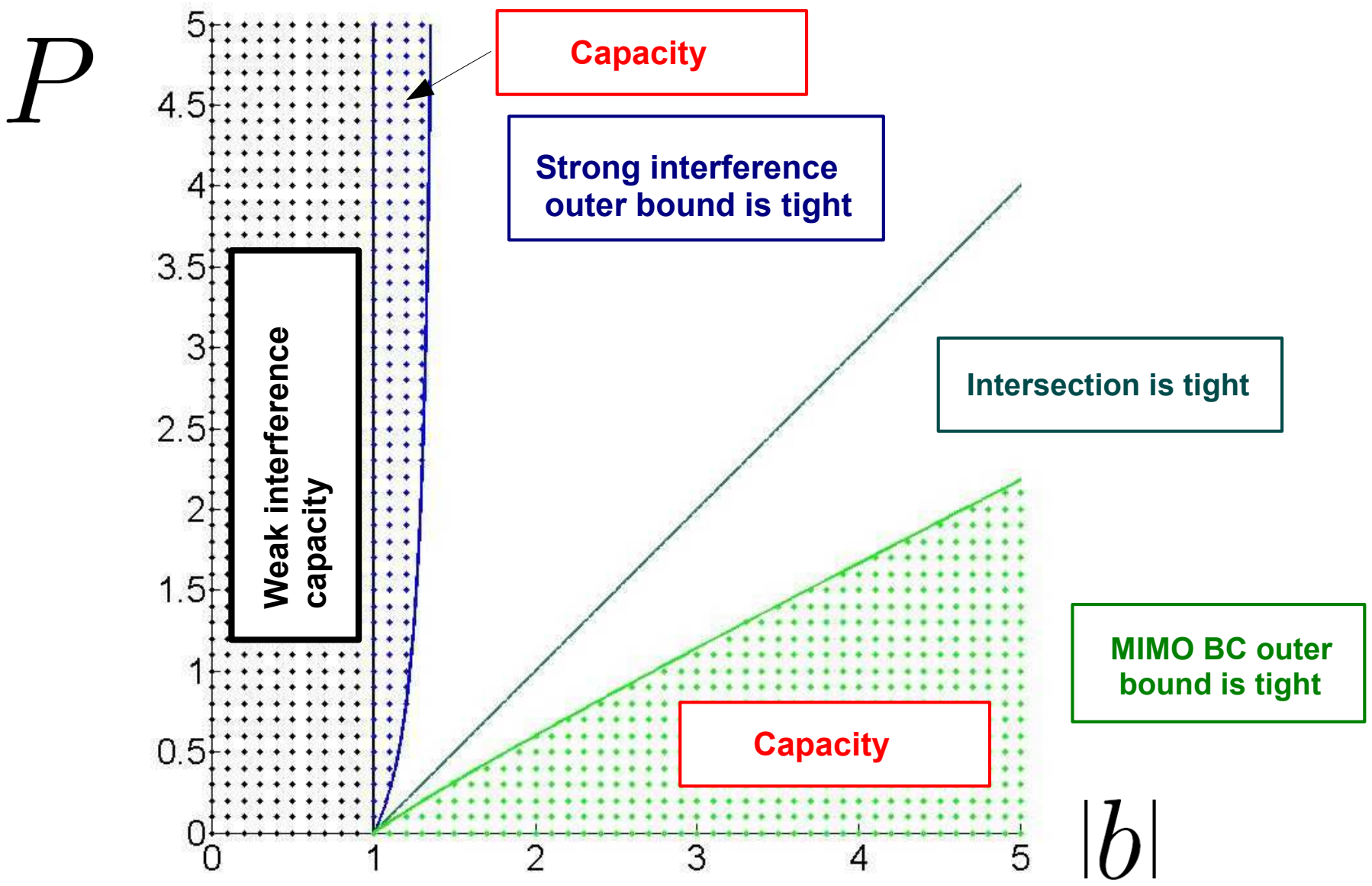

Fig. 18. The capacity results for the S-G-CIFC for the case $P_1 = P_2 = P$ for $(P, |b|) \in [0, 5] \times [0, 5]$.

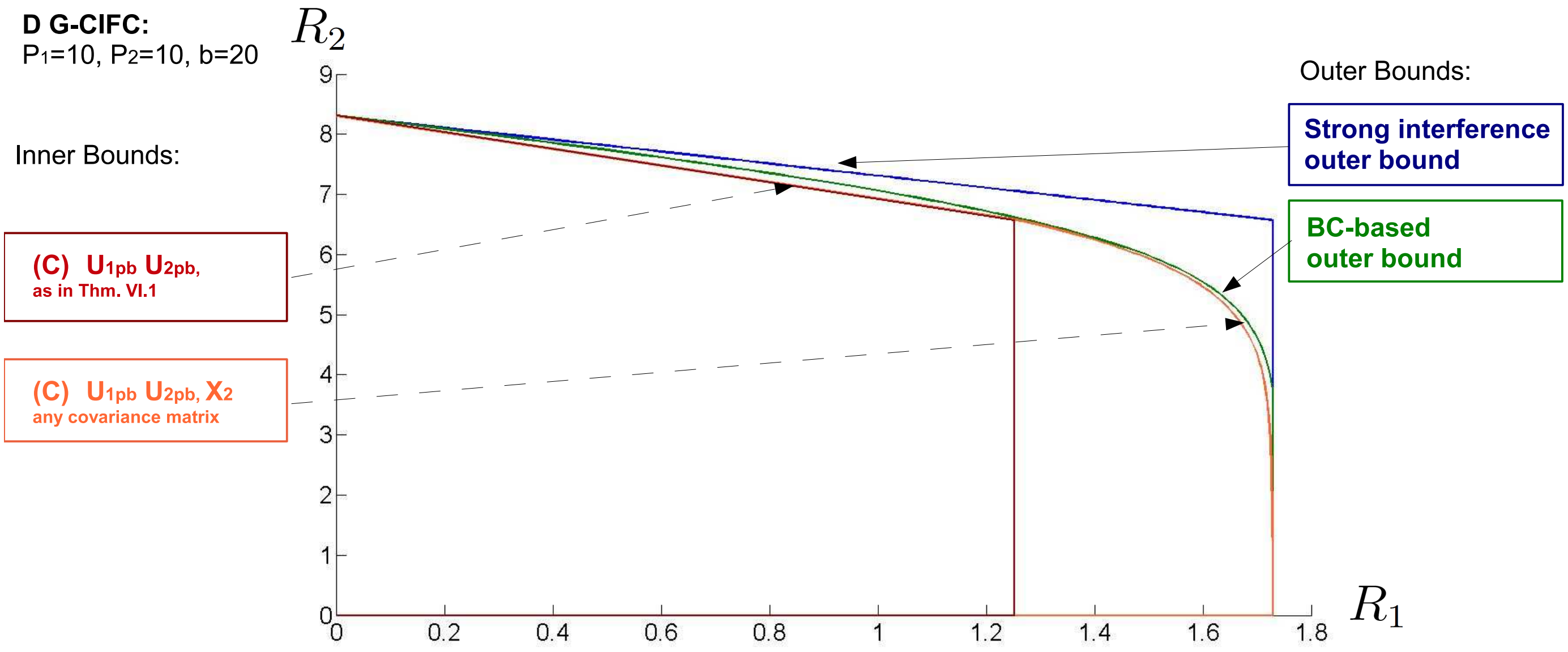

Fig. 19. The achievable region of scheme C with the assignment of RVs in (23) and in (46).

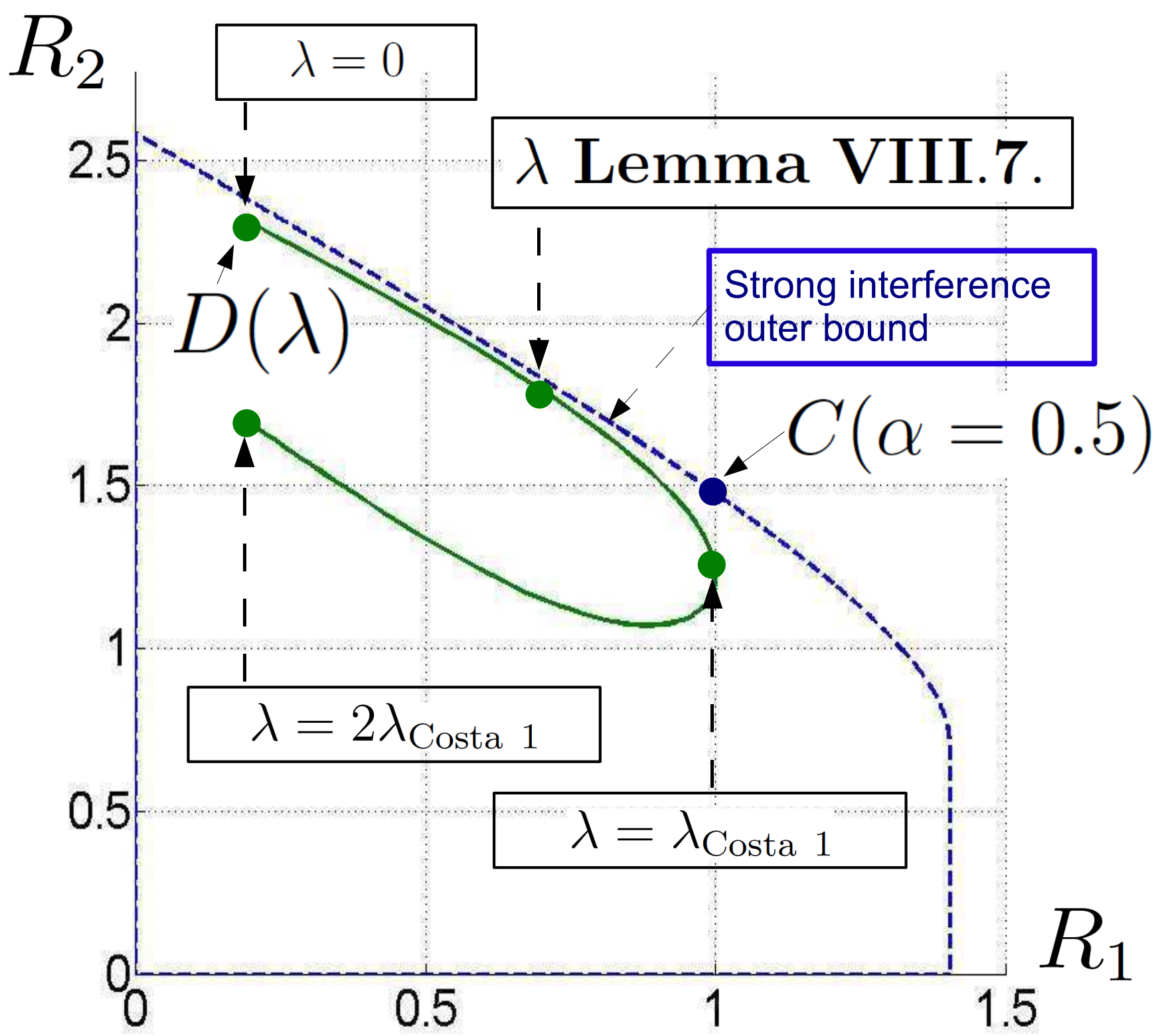

Fig. 20. A plot of points $C(\lambda)$ and $D(\lambda)$ for $\alpha = 0.5$, together with the "strong interference" bound for the G-CIFC with parameters $P_1 = P_2 = 6, b = \sqrt{2}$ and $a = \sqrt{0.3}$.

## VIII. Conclusion and Future Work

In this paper we presented outer bounds, inner bounds, and new capacity results for the Gaussian cognitive interference channel. We derived the tightest known outer bound for the cognitive interference channel in "strong interference", which is based on the capacity of the MIMO BC with degraded message sets. We showed the achievability of this outer bound in the subset of the channel parameter space which we term the "primary decodes cognitive" regime. We also proved capacity to within both an additive and a multiplicative gap, thus providing a characterization of the capacity region in both high and low SNR.

Despite the new results presented, the capacity of the Gaussian cognitive interference channel remains unknown in general. The achievable rate region of [1] provides a comprehensive inner bound that may yield new capacity results: only some specific choices of parameters for this region have been considered so far and we expect that new results may be derived from this region. We have shown that the tightest outer bound for the Gaussian cognitive interference channel in "strong interference" is obtained as the intersection of different bounds. The expression of this outer bound does not have a simple closed form expression except in some special cases like the S and the degraded channels. Even in these two subcases, capacity is not known in general. Another interesting open question is how much rate improvement is attainable with binning at the cognitive encoder: we have shown how dirty paper coding may be used to boost the rate of both the primary and the cognitive user; whether non perfect interference cancellation achieves capacity is still unknown.

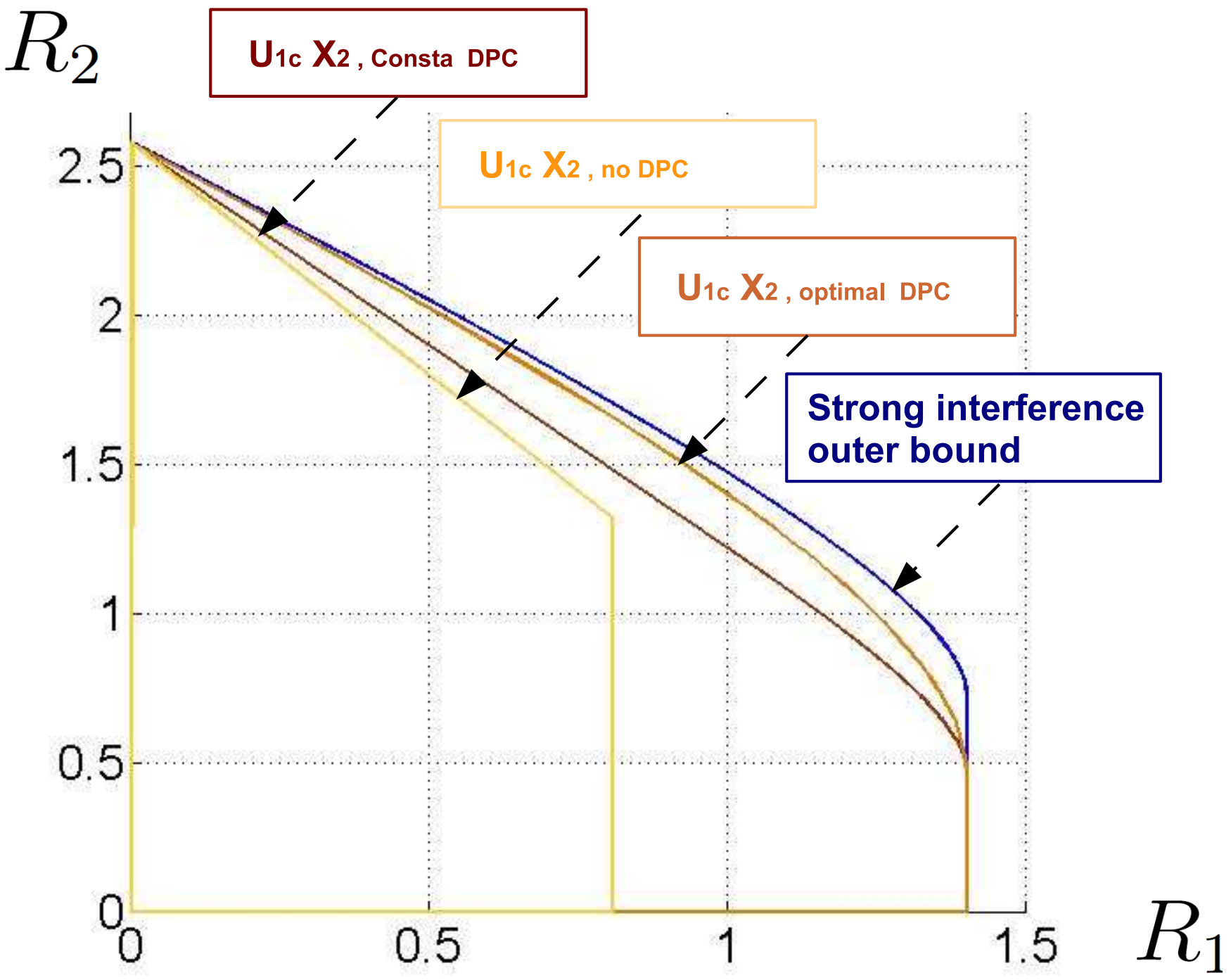

Fig. 21. The achievable region of (26) for $\lambda = 0$, $\lambda = \lambda_{\text{Costa 1}}$ and any $\lambda \in [0, 2\lambda_{\text{Costa 1}}]$ for the G-CIFC with parameters $P_1 = P_2 = 6$, $b = \sqrt{2}$ and $a = \sqrt{0.3}$ for $P_1 = P_2 = 6$, $b = \sqrt{2}$ and $a = \sqrt{0.3}$.

**Acknowledgment** The authors would like to thank Prof. Shlomo Shamai and Prof. Gerhard Kramer for insightful discussions on the problem during IZS 2010 and ITA 2010 respectively.

# Appendix A
# The G-CIFC in standard form

A general G-CIFC has outputs

$$\widetilde{Y}_1 = h_{11}\widetilde{X}_1 + h_{12}\widetilde{X}_2 + \widetilde{Z}_1$$
$$\widetilde{Y}_2 = h_{21}\widetilde{X}_1 + h_{22}\widetilde{X}_2 + \widetilde{Z}_2$$

where

$$\widetilde{Z}_i \sim \mathcal{N}_{\mathbb{C}}(0,\sigma_i^2), \qquad \sigma_i^2 > 0, \qquad i \in \{1,2\},$$

and the inputs are subject to the power constraint

$$\mathbb{E}[|\widetilde{X}_i|^2] \le \widetilde{P}_i, \qquad \widetilde{P}_i \ge 0, \qquad i \in \{1,2\}.$$

When $h_{11} \neq 0$ and $h_{22} \neq 0$ , we may scale each channel output by the standard deviation (assumed strictly positive) of the corresponding additive Gaussian noise and change the phase as

$$\begin{aligned}
Y_1 &\triangleq \frac{\widetilde{Y}_1}{\sigma_1} \\
Y_2 &\triangleq \frac{\widetilde{Y}_2}{\sigma_2} \mathrm{e}^{\mathrm{j}(\angle h_{11} - \angle h_{12})} \\
X_1 &\triangleq \frac{h_{11}}{\sigma_1}\widetilde{X}_1 \quad \text{such that} \quad \mathbb{E}[|X_1|^2] \le P_1 \triangleq \frac{|h_{11}|^2}{\sigma_1^2}\widetilde{P}_1 \\
X_2 &\triangleq \frac{h_{22}}{\sigma_2} \mathrm{e}^{\mathrm{j}(\angle h_{11} - \angle h_{12})}\widetilde{X}_2 \quad \text{such that} \quad \mathbb{E}[|X_2|^2] \le P_2 \triangleq \frac{|h_{22}|^2}{\sigma_2^2}\widetilde{P}_2 \\
a &\triangleq \frac{h_{12}}{\sigma_1}\frac{\sigma_2}{h_{22}} \mathrm{e}^{\mathrm{j}(-\angle h_{11} + \angle h_{12})} \in \mathbb{C} \\
b &\triangleq \frac{|h_{21}|}{\sigma_2}\frac{\sigma_1}{|h_{11}|} \in \mathbb{R}^+,
\end{aligned} \tag{47}$$

to obtain the equivalent channel outputs have additive noise of unit variance, unit gain on the direct link, as claimed in Section II-C. To remind the reader that $b$ is always real-valued and non-negative we use the notation $|b|$.

When $h_{22} = 0$, transmitter 2 can only create interference at receiver 1 and thus the channel reduces to a BC where the cognitive transmitter is sending both messages to both receivers. When $h_{22} = 0$ in (47), we have $a = \infty$ and $P_2 = 0$ corresponds to the scenario above; the same in not true when $h_{11} = 0$.

If $h_{11} = 0$, the channel reduces to a MISO, point-to-point channel since decoder 1 can only receive interference from transmitter 2. For $h_{11} = 0$ the transformation in (47) does not yield a MISO channel, since in this case $P_1 = 0$ and $b = \infty$. In [41, Sec. II.B], this fact is overlooked and the transformation in (47) is considered to be without loss of generality.

Note that the equivalent transformation in standard form for a classical interference channel does not require $h_{11} > 0$, since the transmitters cannot cooperate.

## APPENDIX B
## PROOF OF THEOREM II.3

For a complex-valued G-CIFC with $|b| > 1$, the outer bound of Th.II.2 is achievable by the superposition-only (scheme (D) of Section IV-D) if $I(Y_1; X_1, X_2) \geq I(Y_2; X_1, X_2)$ for all input distributions [10], that is, if

$$\begin{aligned}\mathbb{E}[|Y_1|^2] - \mathbb{E}[|Y_2|^2] = (|a|^2 - 1)P_2 - (|b|^2 - 1)P_1 + \\ + 2\sqrt{P_1 P_2}(\mathrm{Re}\{a^* \rho\} - |b|\mathrm{Re}\{\rho\}) \geq 0, \quad \forall |\rho| \leq 1. \end{aligned} \tag{48}$$

Let $\rho = |\rho| \mathrm{e}^{\mathrm{j}\phi_\rho}$ and $a = |a| \mathrm{e}^{\mathrm{j}\phi_a}$. We have

$$\begin{aligned}\mathrm{Re}\{a^* \rho\} - |b|\mathrm{Re}\{\rho\} &= |\rho||a| \cos(\phi_\rho - \phi_a) - |\rho||b| \cos(\phi_\rho) \\ &= |\rho| \Big[|a| \cos(\phi_a) - |b|\Big] \cos(\phi_\rho) + |\rho| \Big[|a| \sin(\phi_a)\Big] \sin(\phi_\rho) \\ &= |\rho| \sqrt{(\Big[|a| \cos(\phi_a) - |b|\Big])^2 + \Big[|a| \sin(\phi_a)\Big]^2} \cos(\phi) \\ &= \big|a - |b|\big| \cdot |\rho| \cos(\phi),\end{aligned}$$

for some angle $\phi$. The condition in (48) is thus verified for all $|\rho| \cos(\phi) \in [-1, +1]$ if it is verified for $|\rho| \cos(\phi) = -1$ as claimed in Th.II.3. ■.

## APPENDIX C
## CLOSED FORM EXPRESSION FOR $\mathcal{R}^{(\mathrm{BC-PR})}$

The closed form expression of $\mathcal{R}^{(\mathrm{BC-PR})}$ was obtained in [32] and is presented here for completeness.

Consider an input covariance matrix defined as follows

$$\boldsymbol{S} \triangleq \begin{bmatrix} P_1 & \rho\sqrt{P_1 P_2} \\ \rho^* \sqrt{P_1 P_2} & P_2 \end{bmatrix} : \quad \rho = \sqrt{1 - \alpha}\, \mathrm{e}^{\mathrm{j}\theta}, \ \theta \in \mathbb{R}, \ \alpha \in [0, 1]. \tag{49}$$

The capacity region of a Gaussian *MIMO BC with private rates only* with a per-antenna power constraint is given by [32]

$$\mathcal{R}^{(\mathrm{BC-PR})} = \mathcal{CH} \bigcup_{\mathbf{S}} \mathcal{R}^{(\mathrm{BC-PR})}(\mathbf{S})$$

where $\mathcal{CH}$ denotes the convex-hull operation, $\bigcup_{\mathbf{S}}$ denotes the union over all input covariance matrices $\mathbf{S}$ that satisfy the per-antenna power constraint, and where

$$\mathcal{R}^{(\mathrm{BC-PR})}(\mathbf{S}) = \bigcup_{u \in \{1,2\}} \mathcal{R}^{(\mathrm{DPC}\ u)}(\mathbf{S})$$

where $\mathcal{R}^{(\mathrm{DPC}\ u)}(\mathbf{S})$ is the DPC region for the encoding order where user $u$ is pre-coded against the interference created by the other user at its intended receiver, which is given by

$$\mathcal{R}^{(\mathrm{DPC}\ u)}(\mathbf{S}) = \bigcup_{0 \preceq \mathbf{B}_1,\ 0 \preceq \mathbf{B}_2,\ \mathbf{B}_1 + \mathbf{B}_2 = \mathbf{S}} \mathcal{R}^{(\mathrm{DPC}\ u)}(\mathbf{B}_1, \mathbf{B}_2), \quad u \in \{1, 2\},$$

and where, for

$$\mathbf{B}_1 = \begin{bmatrix} \alpha_1 P_1 & \rho_1\sqrt{\alpha_1 P_1\,\alpha_2 P_2} \\ \rho_1^*\sqrt{\alpha_1 P_1\,\alpha_2 P_2} & \alpha_2 P_2 \end{bmatrix}, \quad \mathbf{B}_2 = \begin{bmatrix} \bar{\alpha}_1 P_1 & \rho_2\sqrt{\bar{\alpha}_1 P_1\,\bar{\alpha}_2 P_2} \\ \rho_2^*\sqrt{\bar{\alpha}_1 P_1\,\bar{\alpha}_2 P_2} & \bar{\alpha}_2 P_2 \end{bmatrix},$$

with

$$(\alpha_1, \alpha_2, |\rho_1|, |\rho_2|) \in [0,1]^4 : \quad \rho_1\sqrt{\alpha_1\,\alpha_2} + \rho_2\sqrt{\bar{\alpha}_1\,\bar{\alpha}_2} = \rho,$$

the region $\mathcal{R}^{(\text{DPC }1)}(\mathbf{B}_1, \mathbf{B}_2)$ is given by

$$R_1 \leq \mathcal{C}(\alpha_1 P_1 + |a|^2\alpha_2 P_2 + 2\text{Re}\{a^*\rho_1\}\sqrt{\alpha_1\alpha_2 P_1 P_2}), \tag{50a}$$

$$R_2 \leq \mathcal{C}\left(\frac{\bar{\alpha}_1|b|^2 P_1 + \bar{\alpha}_2 P_2 + 2\text{Re}\{\rho_2\}\sqrt{\bar{\alpha}_1\bar{\alpha}_2|b|^2 P_1 P_2}}{1 + |b|^2\alpha_1 P_1 + \alpha_2 P_2 + 2\text{Re}\{\rho_1\}\sqrt{\alpha_1\alpha_2|b|^2 P_1 P_2}}\right), \tag{50b}$$

and $\mathcal{R}^{(\text{DPC }2)}(\mathbf{B}_1, \mathbf{B}_2)$ is given by

$$R_1 \leq \mathcal{C}\left(\frac{\alpha_1 P_1 + |a|^2\alpha_2 P_2 + 2\text{Re}\{a^*\rho_1\}\sqrt{\alpha_1\alpha_2 P_1 P_2}}{1 + \bar{\alpha}_1 P_1 + |a|^2\bar{\alpha}_2 P_2 + 2\text{Re}\{a^*\rho_2\}\sqrt{\bar{\alpha}_1\bar{\alpha}_2 P_1 P_2}}\right), \tag{51a}$$

$$R_2 \leq \mathcal{C}(\bar{\alpha}_1|b|^2 P_1 + \bar{\alpha}_2 P_2 + 2\text{Re}\{\rho_2\}\sqrt{\bar{\alpha}_1\bar{\alpha}_2|b|^2 P_1 P_2}). \tag{51b}$$

The quantity $\alpha_u$, $u \in \{1,2\}$, represents the fraction of power $P_u$ used to send the cognitive message $W_1$ on antenna $u$. The requirement $(\alpha_1, \alpha_2) \in [0,1]^2$ guarantees that the per-antenna power constraints are verified.

# APPENDIX D
# PROOF OF COROLLARIES III.5 AND III.6

## A. *Proof of Corollary III.5*

When allowing full transmitter cooperation for a channel with $a|b| = 1$ and $|b| > 1$, we obtain an equivalent degraded BC with input $X_{\text{eq}} = |b|X_1 + X_2$ and outputs

$$Y_2 = (|b|X_1 + X_2) + Z_2 = X_{\text{eq}} + Z_2,$$

$$|b|Y_1 = (|b|X_1 + X_2) + |b|Z_1 \sim Y_2 + \sqrt{|b|^2 - 1}\,Z_0,$$

with $Z_0 \sim \mathcal{N}_{\mathbb{C}}(0,1)$ and independent of everything else. The input of the equivalent BC is subject to the power constraint

$$\mathbb{E}[|X_{\text{eq}}|^2] \leq (\sqrt{|b|^2 P_1} + \sqrt{P_2})^2 \triangleq P_{\text{eq}}.$$

For this order of degradedness among the users, the capacity region of the degraded BC with private rates equals the capacity with degraded message sets. In general $\mathcal{R}^{(\text{BC}-\text{DMS})} \subseteq \mathcal{R}^{(\text{BC}-\text{PR})}$, but since here $Y_1$ is a degraded version of $Y_2$, decoder 2 can decode the message of decoder 1 without imposing any rate penalty to user 1, thus $\mathcal{R}^{(\text{BC}-\text{PR})}$ is achievable. This implies $\mathcal{R}^{(\text{BC}-\text{DMS})} = \mathcal{R}^{(\text{BC}-\text{PR})}$.

The capacity region of the equivalent BC is [30]

$$R_1 \;\leq\; R_1^{(\text{BC}-\text{deg})}(\alpha') = \mathcal{C}\left(\frac{\alpha' P_{\text{eq}}}{(1-\alpha')P_{\text{eq}} + |b|^2}\right), \tag{52a}$$

$$R_2 \;\leq\; R_2^{(\text{BC}-\text{deg})}(\alpha') = \mathcal{C}\left((1-\alpha')P_{\text{eq}}\right), \tag{52b}$$

taken over the union of all $\alpha' \in [0,1]$, i.e., that is $\rho_1 = \rho_2 = 1,\ \alpha_1 = \alpha_2 = \alpha'$ and $\mathcal{R}^{(\mathrm{BC-PR})} = \mathcal{R}^{(\mathrm{DPC\ 2})}$ in (51).

To intersect the region in (52) with the "strong interference" outer bound of Th.II.2 we equate the $R_1$-bounds in (52a) and (4a) to obtain

$$\alpha' = \frac{\alpha P_1}{1+\alpha P_1}\left(1 + \frac{|b|^2}{(\sqrt{|b|^2 P_1} + \sqrt{P_2})^2}\right). \tag{53}$$

Notice that $\alpha'$ in (53) satisfies $\alpha' \leq 1$ (the maximum value of 1 is obtained for $P_2 = 0$ and $\alpha = 1$). By substituting $\alpha'$ from (53) in (52b), we obtain the bound in (16b).

The BC-based outer bound is more stringent than the "strong interference" outer bound if

$$\begin{aligned}
& R_1^{(\mathrm{BC-deg})}(\alpha) + R_2^{(\mathrm{BC-deg})}(\alpha) \leq R_{\mathrm{sum}}^{(\mathrm{SI})}(\alpha) \quad \forall \alpha \in [0,1] \\
\iff\ & \alpha P_1 + P_2 + (1-\alpha)|b|^2 P_1 + 2\sqrt{|b|^2 P_1 P_2} \leq P_2 + |b|^2 P_1 + 2\sqrt{\bar{\alpha}|b|^2 P_1 P_2} \quad \forall \alpha \in [0,1] \\
\iff\ & 2\sqrt{|b|^2 P_1 P_2}(1-\sqrt{\bar{\alpha}}) \leq \alpha P_1(|b|^2 - 1) \quad \forall \alpha \in [0,1] \\
\iff\ & 2\frac{\sqrt{|b|^2 P_1 P_2}}{P_1(|b|^2-1)} \leq 1 + \sqrt{\bar{\alpha}} \quad \forall \alpha \in [0,1] \quad \big(\text{since } \alpha = (1-\sqrt{\bar{\alpha}})(1+\sqrt{\bar{\alpha}})\big) \\
\iff\ & 2\frac{\sqrt{|b|^2 P_1 P_2}}{P_1(|b|^2-1)} \leq \min_{\bar{\alpha} \in [0,1]} \{1 + \sqrt{\bar{\alpha}}\} = 1 \\
\iff\ & 1 + \frac{P_2}{P_1} \leq \left(|b| - \sqrt{\frac{P_2}{P_1}}\right)^2 \\
\iff\ & |b| \geq \sqrt{1 + \frac{P_2}{P_1}} + \sqrt{\frac{P_2}{P_1}},
\end{aligned}$$

as claimed.

*Remark* D.1. The capacity of the equivalent degraded BC may be achieved both by using superposition coding and binning. An achievable scheme inspired by the degraded BC and employing superposition coding is scheme (E) with $\lambda = 0$. An achievable scheme inspired by the degraded BC and employing binnig coding is scheme (B). Both schemes achieve the outer bound only in point A in (39a). The capacity region of the degraded CIFC in therefore unknown in general it remains an interesting open problem.

### B. *Proof of Corollary III.6*

To establish the result in Corollary III.6 we proceed as follows: first we prove that the capacity region of the Gaussian BC-DMS may be obtained form the region in [31] by considering Gaussian inputs and auxiliary RV. Successively we perform a partial optimization of the region in [31] in the Gaussian case and obtain a looser outer bound that may be expressed as a function of a single parameter. Finally we intersect this outer bound with the "strong interference" outer bound of (4) to obtain the expression in (17).

The capacity region of the general BC-DMS is found in [31] and is expressed as the union over all possible distributions of the input and one auxiliary RV. A closed form expression of the capacity region of the Gaussian BC-DMS is derived in [42] and is expressed as the intersection of the capacity region of a general BC and an additional sum rate constraint. We derive another simpler expression of the capacity region of the Gaussian BC-DMS and we do so by showing that we may restrict the union in [31] over all Gaussian inputs and auxiliary RV.

Consider the BC-DMS defined as:

$$Y_i = \boldsymbol{H}_i X + Z_i \qquad \forall\ i \in [1,2] \tag{54}$$

where:

- $X$ is a real valued input vector of size $n \times 1$ subject to the second moment constraint $\mathrm{Cov}[X] = \boldsymbol{K}_X \preceq \boldsymbol{S}$ for some $\boldsymbol{S} \succeq 0$,
- $Y_i$ is a real valued output vector of size $m_i \times 1$ received by user $i \in [1,2]$,
- $\boldsymbol{H}_i$ is a fixed real valued gain matrix imposed on user $i \in [1,2]$. This is a matrix of size $m_i \times n$,
- $Z_i$ is a real valued Gaussian random vector with zero mean and covariance matrix $\mathrm{Cov}[Z_i] = \boldsymbol{K}_Z \succ 0$.

As for the BC of [42], we consider real valued channels; the extension to complex valued channels is easily obtained by doubling the real dimensions. We first derive the capacity of a Gaussian BC-DMS for the case where $\boldsymbol{H}_i$ is square and invertible, we than argue that the case for a general $\boldsymbol{H}_i$ may be obtained by series of channel transformations originally devised for the BC in [32].

**Theorem D.2.** *The capacity region of the Gaussian BC-DMS in (54) is*

$$R_1 \le I(U; Y_1), \tag{55a}$$

$$R_2 \le I(X; Y_2|U), \tag{55b}$$

$$R_1 + R_2 \le I(X; Y_2). \tag{55c}$$

*taken over the union of all Gaussian $U$ and $X$ vectors of size $n$ such that $\boldsymbol{K}_X \preceq \boldsymbol{S}$.*

*Proof:*

The region in (55) was originally obtained in [31] for a general BC-DMS but considering the union over any distribution $P_{UX}$. To prove the theorem we need to show that only Gaussian $U$ and $X$ need to be considered.

First, we notice that (55c) is always maximized by having $X$ Gaussian by the "Gaussian maximizes entropy" of [43]. Since (55c) is maximized by Gaussian inputs, we have to show that the region obtained by considering (55a) and (55b) only is optimized by Gaussian inputs as well. To this end we write the region with (55a) and (55b) as

$$\begin{aligned}\text{(55a)} + (1-\mu)\text{(55b)} &= \max_{P_{X|U}: Cov[X] \preceq \mathbf{S}} \mu I(U; Y_1) + (1-\mu) I(X; Y_2|U) \\ &\le \mu h(\boldsymbol{H}_1 X_G + Z_1) - (1-\mu) h(Z_2) + (1-\mu) \max_{P_{X|U}:\ Cov[X|U] \preceq \mathbf{S}} \left( h(\boldsymbol{H}_1 X + Z_1|U) - \frac{\mu}{(1-\mu)} h(\boldsymbol{H}_2 X + Z_2|U) \right),\end{aligned} \tag{56}$$

for any $\mu \in [1/2, 1]$ and where $X_G$ is a Gaussian vector with $\boldsymbol{K}_X \preceq \boldsymbol{S}$.

We need not consider $\mu \in [0, 1/2]$ because the region in (55) is convex and contained in the triangular region

$$R_1, R_2 \geq 0, \tag{57a}$$

$$R_1 + R_2 \leq I(\boldsymbol{H}_2 X_G + Z_2; X_G), \tag{57b}$$

see [31].

For these reasons, the region in (55) cannot contain any rate point with tangent greater than $-1$ and thus there is no loss of generality in restricting $\mu$ in (56) to the interval $[1/2, 1]$.

We now show that solution of the optimization problem

$$\max_{P_{X|U}:\ \mathrm{Cov}[X|U]\preceq \mathbf{S},\ \mu\in[1/2,1]} \quad h(\boldsymbol{H}_1 X + Z_1|U) - \tfrac{\mu}{1-\mu} h(\boldsymbol{H}_2 X + Z_2|U).$$

must be Gaussian by using the extremal inequality of [44]. We first focus on channels where $\boldsymbol{H}_i,\ i \in [1, 2]$ is square and invertible, then show how this result may be used to establish a general channel using the perturbation techniques of [32].

If $\boldsymbol{H}_i,\ i \in [1, 2]$, are square we may write

$$\begin{aligned}
&\max_{P_{X|U}:\ \mathrm{Cov}[X|U]\preceq \mathbf{S}} \quad h(\boldsymbol{H}_1 X + Z_1|U) - \frac{\mu}{(1-\mu)} h(\boldsymbol{H}_2 X + Z_2|U) \\
&= \frac{\mu}{(1-\mu)} \left(\log|\boldsymbol{H}_2|\right)^{-1} - \left(\log|\boldsymbol{H}_1|\right)^{-1} + \max_{P_{X|U}:\ \mathrm{Cov}[X|U]\preceq \mathbf{S}} \quad h(X + \boldsymbol{H}_1^{-1} Z_1|U) - \frac{\mu}{(1-\mu)} h(X + \boldsymbol{H}_2^{-1} Z_2|U).
\end{aligned} \tag{58}$$

Th. 8 in [44] grants that the solution of the optimization problem in (58) is Gaussian since $\mu/(1-\mu) > 1$ for $\mu \in [0, 1/2]$. Since we have established that both (56) and (55c) are maximized by Gaussian $X$ and $U$, we conclude that (55) is also maximized by Gaussian $X$ and $U$ as well.

Finally the perturbation technique in [32, Section V.B] allows us to extend this result to a general channel where $\boldsymbol{H}_i$ in not necessarily square and invertible. The derivation in [32, Section V.B] was originally devised for the general BC but it extends in a straight-forward manner to the BC-DMS, since it solely relies on the channel matrix and the covariance of the noise and not on the message set.

■

We can now evaluate the outer bound of Th. D.2 for the S-G-CIFC.

Note that (55a) and (55b) correspond the region $\mathcal{R}^{(\mathrm{DPC}\ 2)}$ in (51); moreover (55c) depend on the parameter $\rho$ in (49). For these reasons we may write:

$$\begin{aligned}
\mathcal{R}^{\mathrm{BC-DMSG-S-CIFC}} &= \bigcup_{|\rho|\leq 1\ \ \alpha_1\in[0,1]} \left(\mathcal{R}^{DPCG-S-CIFC}(\rho,\alpha_1) \cap \mathcal{R}^{SUM-RATE\ G-S-CIFC}(\rho)\right) \\
&\subseteq \bigcup_{|\rho|\leq 1, \alpha_1\in[0,1]} \mathcal{R}^{\mathrm{DPCG-S-CIFC}}(\rho,\alpha_1) \\
&= \bigcup_{\alpha_1\in[0,1]} \mathcal{R}^{\mathrm{DPCG-S-CIFC}}(\sqrt{\bar{\alpha}_1},\alpha_1)
\end{aligned} \tag{59}$$

where

$$\mathcal{R}^{\rm SUM-RATEG-S-CIFC}(\rho) = \left\{R_1 + R_2 \le \mathcal{C}(|b|^2 P_1 + P_2 + 2{\rm Re}\{\rho\}\sqrt{|b|^2 P_1 P_2})\right\}$$

and where

$$\mathcal{R}^{DPCG-S-CIFC}(\rho, \alpha_1) = \left\{ \begin{array}{ll} R_1 & \le R_1^{(\rm DPC\ 2\ S-G-CIFC)}(\alpha_1) \\ R_2 & \le R_2^{(\rm DPC\ 2\ S-G-CIFC)}(\rho, \alpha_1) \end{array} \right\}$$

with

$$R_1 \le \mathcal{C}\left(\frac{\alpha_1 P_1}{1+\bar{\alpha}_1 P_1}\right) \triangleq R_1^{(\rm DPC\ 2\ S-G-CIFC)}(\alpha_1), \tag{60a}$$

$$\begin{aligned} R_2 \le & \max_{\rho_1,\rho_2,\alpha_2\ {\rm s.t.}\rho=\rho_1\sqrt{\alpha_1\alpha_2}+\rho_2\sqrt{\bar{\alpha}_1\bar{\alpha}_2}} \mathcal{C}(|b|^2\bar{\alpha}_1 P_1 + \bar{\alpha}_2 P_2 + 2{\rm Re}\{\rho_2\}\sqrt{\bar{\alpha}_1\bar{\alpha}_2|b|^2 P_1 P_2}) \\ & \triangleq R_2^{(\rm DPC\ 2\ S-G-CIFC)}(\rho, \alpha_1). \end{aligned} \tag{60b}$$

In (59) we have used the fact that, for a fixed $\alpha_1$, we have

$$R_2^{(\rm DPC\ 2\ S-G-CIFC)}(\rho,\alpha_1) \le \mathcal{C}(|b|^2\bar{\alpha}_1 P_1 + P_2 + 2\sqrt{\bar{\alpha}_1|b|^2 P_1 P_2}) = R_2^{(\rm DPC\ 2\ S-G-CIFC)}(\sqrt{\bar{\alpha}_1}, \alpha_1).$$

Equation (60a) corresponds to (51a) when $a = 0$ and equation (60b) corresponds to (51b) optimized over $\rho_1, \rho_2, \alpha_2$.

This proves that for the S-G-CIFC, $\mathcal{R}^{(\rm BC-PR)}$ is contained in the region

$$R_1 \le R_1^{(\rm BC-DMS-S)}(\alpha) = \mathcal{C}(\alpha P_1), \tag{61a}$$

$$R_2 \le R_2^{(\rm BC-DMS-S)}(\alpha) = \mathcal{C}\left(\left(\sqrt{|b|^2\frac{(1-\alpha)P_1}{1+\alpha P_1}} + \sqrt{P_2}\right)^2\right), \tag{61b}$$

taken over the union of all $\alpha \in [0, 1]$.

The BC-based outer bound of (61) is more stringent than the "strong interference" outer bound in (4) if

$$\begin{aligned} & R_1^{(\rm BC-DMS-S)}(\alpha) + R_2^{(\rm BC-DMS-S)}(\alpha) \le R_{\rm sum}^{(\rm SI)}(\alpha) \quad \forall\alpha \in [0,1] \\ \iff\ & \alpha P_1 + \left(\sqrt{|b|^2(1-\alpha)P_1} + \sqrt{P_2(1+\alpha P_1)}\right)^2 \le P_2 + |b|^2 P_1 + 2\sqrt{(1-\alpha)|b|^2 P_1 P_2} \quad \forall\alpha \in [0,1] \\ \iff\ & \alpha P_1(1 + P_2 - |b|^2) + 2\sqrt{(1-\alpha)|b|^2 P_1 P_2}(\sqrt{1+\alpha P_1} - 1) \le 0 \quad \forall\alpha \in [0,1] \\ \iff\ & 1 + P_2 - |b|^2 \le 0 \quad \text{(condition for } \alpha = 1\text{; the one for } \alpha = 0 \text{ is always verified).} \end{aligned}$$

## APPENDIX E
## PROOF OF LEMMA III.7

Let $X_1^N(W_1, W_2), X_2^N(W_2)$ be a good code for the channel $(a, |b|, P_1, P_2)$. Consider now the inputs

$$X_1' = AX_1 + BX_2,$$
$$X_2' = CX_2,$$

on a channel with parameters $(a', |b'|, P_1', P_2')$ resulting in the outputs

$$Y_1' = X_1' + a'X_2' + Z_1 \propto X_1 + \frac{B + a'C}{AX_2} + \frac{Z_1}{A},$$

and

$$Y_2' = |b'|X_1' + X_2' + Z_2 \propto \frac{|b'|}{|b'|B + C} X_1 + X_2 + \frac{Z_1}{|b'|B + C}.$$

If

$$\begin{aligned} a &= \frac{B + a'C}{A}, \\ |b| &= \frac{|b'|}{b'B + C}, \\ |A|^2 &\geq 1, \\ ||b'|B + C|^2 &\geq 1, \\ P_1' &\geq (\sqrt{|A|^2 P_1} + \sqrt{|B|^2 P_2})^2, \\ P_2' &\geq |C|^2 P_2, \end{aligned} \tag{62}$$

the output of the channel $(a, b, P_1, P_2)$ may be reconstructed in the channel $(a', b', P_1', P_2')$. This implies

$$\boldsymbol{C}(a, b, P_1, P_2) \subseteq \bigcap_{A,B,C: |A| \geq 1, |C/(1-B|b|)| \geq 1} \boldsymbol{C}\left(\frac{aA - B}{C}, \frac{C|b|}{1 - B|b|}, (\sqrt{|A|^2 P_1} + \sqrt{|B|^2 P_2})^2, |C|^2 P_2\right).$$

**S-G-CIFC.**

By considering the transformation in (62) with

$$\begin{aligned} A &= 1 \\ B &= a \\ C &= 1 - a|b| \end{aligned}$$

we see that the capacity of a general G-CIFC $\boldsymbol{C}(a, |b|, P_1, P_2)$ is contained in the capacity region of S-G-CIFC $\boldsymbol{C}(0, |b|, |\sqrt{P_1} + a\sqrt{P_2}|^2, |1 - a|b||^2 P_2)$.

**G-CIFC in “weak interference”.**

By considering the transformation in (62) with

$$\begin{aligned} A &= |b| \\ B &= \tfrac{a(1-|b|)}{a-1} \\ C &= \tfrac{a|b|-1)}{a-1} \end{aligned}$$

we have that the capacity of a general G-CIFC $\boldsymbol{C}(a, |b|, P_1, P_2)$ is contained in the capacity region of G-CIFC in “weak interference” $\boldsymbol{C}\left(a, 1, \left|\sqrt{|b|^2 P_1} + \frac{a(1-|b|)}{a-1}\sqrt{P_2}\right|^2, \left|\frac{a|b|-1}{a-1}\right|^2 P_2\right)$.

**G-CIFC in “very strong interference”.**

By considering the transformation in (62) with

$$\begin{aligned} A &= |b| \\ B &= |b|\tfrac{1-a|b|}{|b|^2-1} - a \\ C &= \tfrac{1-a|b|}{|b|^2-1} \end{aligned}$$

we have that the capacity of a G-CIFC $\boldsymbol{C}(a, |b|, P, P)$ is contained in the capacity region of G-CIFC $\boldsymbol{C}\left(|b|, |b|, P', P'\right)$, $P' = \frac{P}{(|b|^2-1)^2} \max\{||b|^2 - 1 + |b| - a|^2, |1 - a|b||^2\}$.

# Appendix F
## Proof of Corollary VI.4

In the following we use the fact that point $B$ in (39b) is to within one bits/s/Hz and a factor two from point C in (40). This is the case as, for the additive gap, $R_1^{(B)} = R_1^{(C)}$ and

$$R_2^{(B)} - R_2^{(C)} = \mathcal{C}\left(\frac{2\sqrt{|b|^2 P_1 P_2}}{1+|b|^2 P_1 + P_2}\right) \leq \mathcal{C}\left(\sqrt{\frac{P_2}{P_2+1}}\right) \leq \log(2) = 1, \quad (63)$$

where we use the fact that $R_2^{(B)} - R_2^{(C)}$ has a maximum in $|b|^2 P_1 = P_2 + 1$.

A representation of "strong interference" outer bound and the "piecewise linear strong interference" outer bound is shown in Fig. 22. The "strong interference" outer bound coincides with the "piecewise linear strong interference" outer bound at point A and the largest distance between the two outer bounds is attained between points B and C. This figure also introduces a new corner point of the inner bound: point D, the inner bound point with the largest $R_2$ rate when $R_1 = \mathcal{C}(P_1) - \Delta_1$.

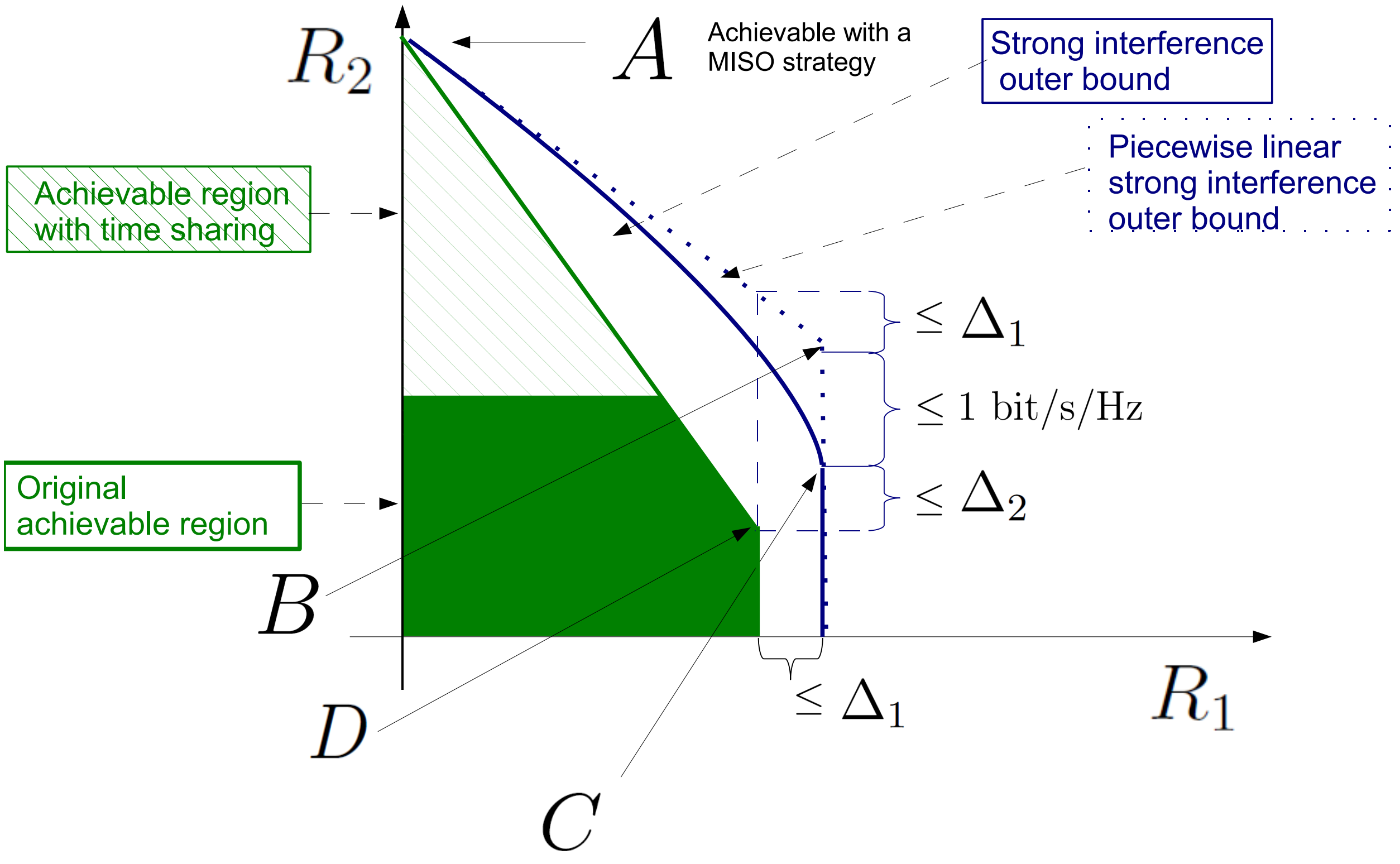

Fig. 22. A graphical representation of the relationship between inner and outer bound for Th. F.2, F.5 and F.5.

*1) Perfect interference cancellation:* In the proof of Th. V.1 we have seen that under condition (31a) it is possible achieve point C in (40) with scheme (E) with Costa's DPC. This result may be used to show achievability of the "strong interference" outer bound to within half a bit/s/Hz per real dimension.

**Theorem F.1.** *If condition in (31a) holds, the "strong interference" outer bound of Th.II.2 is achievable to within half a bit/s/Hz per real dimension.*

*Proof:* Under the condition in (31a), point C is achievable. This point lies to within half a bit/s/Hz per real dimension from the outer bound. ■

*2) Non perfect interference cancellation:* Although it is not possible to achieve point C using scheme (E) and perfect interference cancellation, it is possible to achieve this point to within a bounded distance using non perfect interference cancellation in the strong interference ($|b| > 1$) and strong signal ($P_2 \geq |b|^2 P_1$) regimes.

**Theorem F.2.** *When* $|b| > 1$ *and* $P_2 \geq |b|^2 P_1$*, the outer bound of Th. II.2 may be achieved to within* $1.87$ *bits/s/Hz per real dimension.*

*Proof:* To prove this theorem we show the achievability of a point D in Fig. 22 which lies at a bounded distance from point C using scheme (E) in (26) for $\alpha = 0$. Fig. 22 shows the different additive gaps between inner and outer bound points in the following proof. If equation (26a) is tight there are two possible scenarios: the corner point D is determined by 1) the intersection between (26c) and (26a) or by 2) the intersection of (26b) and (26a). We choose $\lambda$ so that both (26a) and (26b) lie within a finite distance from $R_1^{(B)}$ and $R_2^{(B)}$ respectively. The sum rate bound (26c) does not depend on the choice of $\lambda$ and is always equal to $R_1^{(C)} + R_2^{(C)}$. We divide the proof in two subcases $\mathrm{Re}\{a\} \gtrless |b|^{-1}$.

*Sub-case* $\mathrm{Re}\{a\} \leq |b|^{-1}$*:* When $P_1 \leq 1$ a gap of 1 bit per dimension is achievable by having both transmitters transmit to receiver 2 at rate $R_2^{(C)}$. In this case the distance along the rate $R_2$ is zero and on the rate $R_1$ is $R_1^{(C)} - 0 \leq \log(1 + 1) < 2$. For $P_1 \geq 1$ let $\lambda = \frac{P_1 - \sqrt{P_1}}{P_1 + 1} a$, in (26). The distance between inner and outer bound for $R_1$ is

$$\Delta_1 \triangleq R_1^{(C)} - R_1^{(D)} \quad = \log\left(\frac{1+P_1+2|a|^2 P_2}{1+P_1+|a|^2 P_2}\right) \leq 1,$$

where we have used the inequality $P_2 \geq |b|^2 P_1$. Similarly letting (26b) hold with equality, we obtain

$$\begin{aligned}
\Delta_2 &\triangleq R_2^{(C)} - R_2^{(D)} \\
&\leq \max_{a:\mathrm{Re}\{a\} \leq |b|^{-1}} \log\left(\frac{\frac{1+2P_2}{1+P_1}}{1+P_2\left|1-\frac{(P_1-\sqrt{P_1})a\,|b|}{1+P1}\right|^2}\right) \\
&\leq \log\left(\frac{(1+P_1)(1+2P_2)}{(1+P_1)(1+P_2+P_1)+2P_2\sqrt{P_1}}\right) \\
&\leq \log\left(\frac{1+2P_2}{1+P_2+P_1}\right) \leq 1,
\end{aligned}$$

where we have used that the expression has a global maximum in $a^* > \frac{1}{|b|}$. The largest gap between the inner bound and B is thus bounded by $\max\{1 + \Delta_1, \Delta_1 + \Delta_2\} = 2$, and so the overall gap between the specified achievable scheme of (26) and the outer bound is within $1 + 2 = 3$ bits/s/Hz for a complex valued channel.

*Sub-case* $\mathrm{Re}\{a\} > |b|^{-1}$*:* When $P_1 \leq 3$ a gap of 1 bit per dimension is achievable by having transmitter 1 remain silent (rate $R_1 = 0$) since in this case $R_1^{(B)} - 0 \leq \log(3 + 1)$. When $P_1 > 3$ let $\lambda = \frac{P_1 + 2\sqrt{P_1}}{P_1 + 1}$ in (26). The gap for

$R_1$ may be bounded as

$$\Delta_1 \triangleq R_1^{(C)} - R_1^{(D)} = \log\left(\frac{1+P_1+5|a|^2P_2}{1+P_1+|a|^2P_2}\right) \le \log(5),$$

while that for the rate $R_2$ of transmitter 2 may be bounded as

$$\begin{aligned}
\Delta_2 &\triangleq R_2^{(C)} - R_2^{(D)} && \text{(64a)}\\
&\le \max_{a:\mathrm{Re}\{a\}\le|b|^{-1}} \log\left(\frac{1+2P_2}{(1+P_1)\left(P_2\left|1-\frac{P_1+2a|b|\sqrt{P_1}}{1+P1}\right|^2+1\right)}\right) && \text{(64b)}\\
&\le \log\left(\frac{(1+P_1)(1+2P_2)}{P_2-4P_2\sqrt{P_1}+4P_2P_1+(1+P_1)^2}\right) && \text{(64c)}\\
&\le \log\left(\frac{(1+P_1)(1+2P_2)}{2P_1P_2+(1+P_1)^2}\right) && \text{(64d)}\\
&\le \log\left(\frac{P_1+1}{P_1}\right) && \text{(64e)}\\
&\le \log\left(\frac{4}{3}\right), && \text{(64f)}
\end{aligned}$$

where (64c) follows since the expression has a global maximum for $a^{(opt)} < \frac{1}{|b|}$ and (64d) follows since $4P_1 - 4\sqrt{P_1}+1 > 2P_1$ for $P_1 > 3$. Finally (64e) and (64f) follow since the expression is monotonically increasing in $P_2$ and decreasing in $P_1$. As in the sub-case $\mathrm{Re}\{a\} \le |b|^{-1}$, the maximum distance between points C and D is bounded by $\max\{1+\Delta_1, \Delta_1+\Delta_2\} \le \log\left(\frac{20}{3}\right)$ so that the overall gap is bounded by $\log\left(\frac{40}{3}\right) \approx 3.74$ bits/s/Hz for a complex valued channel.

■

*3) Cognitive broadcasting:* The outer bound Thm II.1 is achievable in "weak interference": the capacity achieving scheme in this regime is scheme (B) in Section IV-B and it employs Costa's DPC at the cognitive transmitter to "pre-cancel" the known interference generated by the primary user. While capacity is known in this regime, we show that the very simple broadcast strategy of scheme (A) in Section IV-A achieves capacity to within a constant gap from the outer bound when the INR is larger than the SNR at the primary receiver ( i.e $|b|^2P_1 > P_2$). When the INR is larger than the SNR at the primary receiver, scheme (A) achieves a constant gap from the outer bound in "strong interference" as well. Although the resulting gap does not improve on the result of Th. VI.1, this result suggests that, in a general scheme, rate improvement may be obtained by having the cognitive transmitter send part of the primary message.

**Theorem F.3.** *When $|b| < 1$ and $|b|^2P_1 \ge P_2$, the outer bound of Th. II.1 may be achieved within 1 bit/s/Hz per real dimension.*

Consider the strategy (A) in Section IV-A for $|b| \le 1$. Then since (2a) and (20a) are the same for every $\alpha$ there is zero gap for the rate $R_1$. By considering the difference between (2b) and (20b), the gap for the rate $R_2$ is bounded

as

$$
\begin{aligned}
(2b) - (20b) &\leq \mathcal{C}\left(|b|^2P_1 + P_2 + 2\sqrt{\bar{\alpha}|b|^2P_1P_2}\right) - \mathcal{C}\left(|b|^2P_1\right)\\
&\leq \mathcal{C}\left(\frac{P_2 + 2\sqrt{|b|^2P_1P_2}}{1+|b|^2P_1}\right)\\
&\leq \mathcal{C}\left(\frac{3|b|^2P_1}{1+|b|^2P_1}\right)\\
&\leq \log(4) = 2.
\end{aligned}
$$

**Theorem F.4.** *When $|b| > 1$ and $|b|^2P_1 \geq P_2$, the outer bound of Th. II.2 may be achieved within 1.5 bits/s/Hz per real dimension.*

*Proof:* Consider scheme (A) in Section IV-A for $|b| > 1$ and $\alpha = \min\{1, 1/P_1\}$ in (21): the gap for user 1 is

$$\Delta_1 \triangleq R_1^{(B)} - R_1^{(C)} = \mathcal{C}(\min\{1, P_1\}) \leq \log(2) = 1,$$

while the gap for user 2 (using $P_2 \leq |b|^2P_1$ and $|b|^2 \geq 1$) is

$$
\begin{aligned}
\Delta_2 \triangleq R_2^{(B)} - R_2^{(C)} &\leq \mathcal{C}\left(\frac{1+2|b|^2P_1}{(1+P_1)(1+|b|^2\min\{1,P_1\})}\right)\\
&\leq \max\left\{\log\left(\frac{2|b|^2}{1+|b|^2}\right), \log\left(\frac{2}{1+P_1}\right)\right\}\\
&\leq \log(2) = 1.
\end{aligned}
$$

As shown in Fig. 22, the achievable point C in (40) is at most at $1 + \Delta_1 + \Delta_2 \leq 3$ bits from the outer bound. By time sharing between points A and C, we have an achievable rate region that is at most at $\max\{1, 3\} = 3$ bits/Hz/s from the outer bound for complex valued channel.

■

*4) Interference stripping:*

**Theorem F.5.** *When $|a| \geq 1$ , $|b| \geq 1$ and $|b|^2P_1 \leq P_2$, the outer bound of Th. II.2 may be achieved within 1.5 bits/s/Hz per real dimension.*

*Proof:* We consider scheme (D)'s performance in the "strong interference" regime when $|b^2| > 1, |a|^2 \geq 1$. When we set $\alpha = 1$, it achieves the rate

$$
\begin{aligned}
R_1 &\leq \mathcal{C}\left(P_1\right)\\
R_1 + R_2 &\leq \mathcal{C}\left(\min\{|a|^2P_2 + P_1, P_2 + |b^2|P_1\}\right).
\end{aligned}
$$

Referring again to Fig. 22, the gap between points B and C may be bounded as

$$\Delta_1 \triangleq R_1^{(B)} - R_1^{(C)} \leq \log(2) = 1,$$

and

$$\begin{aligned}\Delta_2 \triangleq R_2^{(B)} - R_2^{(C)} &\leq \mathcal{C}\left(\frac{1+|b^2|P_1+P_2}{1+\min\{|a|^2P_2+P_1, P_2+|b|^2P_1\}}\right)\\ &\leq \mathcal{C}\left(\max\left\{1, \frac{1+|b|^2P_1+P_2}{1+|a|^2P_2+P_1}\right\}\right)\\ &\leq \mathcal{C}\left(\max\left\{1, \frac{1+2P_2}{1+|a|^2P_2+P_1}\right\}\right)\\ &\leq \mathcal{C}\left(\max\left\{1, \frac{1+2P_2}{1+|a|^2P_2+P_1}\right\}\right)\\ &\leq \log(2) = 1.\end{aligned}$$

We thus achieve a rate pair that lies within $1+\Delta_1+\Delta_1 = 3$ bits/s/Hz of the outer bound for complex valued channel .

■